\documentclass[12pt]{article}
\usepackage{amssymb}
\usepackage{epsf}
\usepackage{epsfig}
\usepackage{psfrag}
\newcommand{\lsim}{
\mathrel{\hbox{\rlap{\hbox{\lower4pt\hbox{$\sim$}}}\hbox{$<$}}}}

\newcommand{\gsim}{
\mathrel{\hbox{\rlap{\hbox{\lower4pt\hbox{$\sim$}}}\hbox{$>$}}}}

\newcommand{\vcb}{|V_{cb}|}

\newcommand{\vub}{|V_{ub}/V_{cb}|}

\def\epe{\varepsilon'/\varepsilon}

\newcommand{\mev}{\, {\rm MeV}}
\newcommand{\bsi}{B_6^{(1/2)}}
\newcommand{\bei}{B_8^{(3/2)}}
\newcommand{\Lms}{\Lambda_{\overline{\rm MS}}}

\newcommand{\be}{\begin{equation}}
\newcommand{\ee}{\end{equation}}
\newcommand{\bi}{\begin{itemize}}
\newcommand{\ei}{\end{itemize}}
\newcommand{\ord}{{\cal O}}

\newcommand{\RE}{{\rm Re}}
\newcommand{\IM}{{\rm Im}}

\def\kpn{K^+\rightarrow\pi^+\nu\bar\nu}

\def\klpn{K_{\rm L}\rightarrow\pi^0\nu\bar\nu}

\newcommand{\Ctilde}{\tilde{C}}

\newcommand{\kmm}{K_{\rm L} \to \mu^+ \mu^-}

\begin{document}
\begin{titlepage}
\vspace*{-0.5truecm}

\begin{flushright}
CERN-PH-TH/2004-020\\
TUM-HEP-540/04\\
MPP-2004-14\\
hep-ph/0402112
\end{flushright}

\vspace*{0.3truecm}

\begin{center}
\boldmath
{\Large{\bf Anatomy of Prominent $B$ and $K$ Decays and 

\vspace{0.3truecm}

Signatures of CP-Violating New Physics in the 

\vspace{0.4truecm}

Electroweak Penguin Sector}}
\unboldmath
\end{center}

\vspace{0.5truecm}

\begin{center}
{\bf Andrzej J. Buras,${}^a$ Robert Fleischer,${}^b$ 
Stefan Recksiegel${}^a$ and Felix Schwab${}^{c,a}$}
 
\vspace{0.4truecm}

${}^a$ {\sl Physik Department, Technische Universit\"at M\"unchen,
D-85748 Garching, Germany}

\vspace{0.2truecm}

${}^b$ {\sl Theory Division, Department of Physics, CERN, 
CH-1211 Geneva 23, Switzerland}

\vspace{0.2truecm}

 ${}^c$ {\sl Max-Planck-Institut f{\"u}r Physik -- Werner-Heisenberg-Institut,
 D-80805 Munich, Germany}

\end{center}

\vspace{0.7cm}
\begin{abstract}
\vspace{0.2cm}\noindent
The recent observation of $B_d\to\pi^0\pi^0$ at the $B$ factories with a
surprisingly large branching ratio represents a challenge for theory, and 
complements the amazingly small $B_d\to\pi^+\pi^-$ rate. We point 
out that all puzzling $B\to\pi\pi$ features can be accommodated 
in the Standard Model (SM) through non-factorizable hadronic interference 
effects, extract the relevant parameters, and predict the CP asymmetries of 
$B_d\to\pi^0\pi^0$. Using then $SU(3)$ flavour-symmetry and plausible 
dynamical assumptions, we fix the hadronic $B\to\pi K$ parameters through 
their $B\to\pi\pi$ counterparts, and determine the CKM angle $\gamma$, 
with a result in remarkable accordance with the usual fits for the unitarity 
triangle. We may then analyse the $B\to\pi K$ system in the SM, where we find 
agreement with the experimental picture, with the exception of those 
observables that are significantly affected by electroweak (EW) penguins, 
thereby suggesting new physics (NP) in this sector. Indeed, a moderate 
enhancement of these topologies and a large CP-violating NP phase allow us 
to describe any currently observed feature of the $B\to\pi K$ modes, and to 
predict the CP-violating $B_d\to\pi^0K_{\rm S}$ observables. If we then 
restrict ourselves to a specific scenario where NP enters only through $Z^0$ 
penguins, we obtain a link to rare $K$ and $B$ decays, where the most 
spectacular NP effects are an enhancement of the $K_{\rm L}\to\pi^0\nu\bar\nu$ 
rate by one order of magnitude with $\mbox{BR}(\klpn)\approx 4 
\mbox{BR}(\kpn)$, $\mbox{BR}(K_{\rm L}\to\pi^0e^+e^-)=
\ord(10^{-10})$, $(\sin2\beta)_{\pi\nu\bar\nu}<0$, and a large 
forward--backward CP asymmetry in $B_d\to K^*\mu^+\mu^-$. We address also 
$\epe$ and other prominent decays, including $B\to \phi K$ 
and $B\to J/\psi K$ modes.
\end{abstract}

\vspace*{0.5truecm}
\vfill
\noindent
February 2004

\end{titlepage}

\thispagestyle{empty}
\vbox{}
\newpage

\setcounter{page}{1}
\pagenumbering{roman}

\tableofcontents

\newpage

\setcounter{page}{1}
\pagenumbering{arabic}

\section{Introduction}\label{sec:intro}
\setcounter{equation}{0}
In this decade, dedicated $B$- and $K$-decay experiments aim at stringent 
tests of the flavour dynamics of the Standard Model (SM) and in particular
of the Kobayashi--Maskawa mechanism of CP violation \cite{KM}. 
The central target of these studies, which will hopefully
shed light on new physics (NP), is the well-known unitarity triangle (UT) 
of the Cabibbo--Kobayashi--Maskawa (CKM) matrix, with its three angles 
$\alpha$, $\beta$ and $\gamma$ (for  detailed reviews, see 
\cite{RF-Phys-Rep,schladming}). Thanks to the SLAC and KEK $B$ factories 
with their detectors BaBar and Belle, respectively, mixing-induced CP 
violation is now an established effect in the $B$-meson system. The 
corresponding determination of $\sin\phi_d$ through the ``golden'' mode 
$B_d\to J/\psi K_{\rm S}$, where $\phi_d$ denotes the 
$B^0_d$--$\bar B^0_d$ mixing phase ($\phi_d=2\beta$ in the SM),
agrees remarkably well with the CKM fits \cite{CKM-Book} that include in
particular the size of the well-known ``indirect'' CP violation in 
$K_{\rm L}\to\pi\pi$ decays. In spite of this tremendous success of the 
SM, one should realize that only a handful of CP-violating decays and rare 
$B$ and $K$ decays have been measured and it is to be seen whether some 
modifications of the SM picture of flavour dynamics and CP violation will 
be required in the future when the data improve. For instance, the BaBar 
and Belle data on $\phi_d$ still admit two solutions with 
$\phi_d\sim 47^\circ$, in accordance with the SM, and $133^\circ$, pointing 
towards NP contributions. This ambiguity can be resolved by measuring the
the sign of $\cos\phi_d$ \cite{ambig}, which is in progress 
at the $B$ factories, even though it is a challenging task \cite{itoh}. 

While testing the SM and its possible extensions it is essential to consider
simultaneously as many processes as possible. Only in this manner can the
parameters of a given theory be fully determined; having them at hand,
predictions for other observables can be made. In this enterprise
correlations between various observables play an important r\^ole, as they 
may exclude or pinpoint a given extension of the SM even without a detailed
knowledge of the parameters specific to this theory.

Interestingly, the current $B$-factory data for a number of processes 
indicate potential inconsistencies with the SM description of CP violation
and flavour dynamics that may suggest the presence of NP contributions 
and/or the deficiencies in our understanding of hadron dynamics that 
necessarily enters the analyses of non-leptonic $B$ decays. In particular: 

\begin{itemize}
\item
The BaBar and Belle collaborations have very recently reported the observation 
of $B_d\to\pi^0\pi^0$ decays with CP-averaged branching ratios of 
$(2.1\pm0.6\pm0.3)\times10^{-6}$ and $(1.7\pm0.6\pm0.2)\times10^{-6}$, 
respectively \cite{Babar-Bpi0pi0,Belle-Bpi0pi0}. These measurements represent
quite a challenge for theory. For example, in a recent state-of-the-art 
calculation \cite{Be-Ne} within QCD factorization \cite{BBNS1}, a branching 
ratio that is about six times smaller is favoured, whereas the calculation of
$B_d\to\pi^+\pi^-$ points towards a branching ratio about two times 
larger than the current experimental average. On the other hand, the
calculation of $B^+\to\pi^+\pi^0$ reproduces the data rather well. This
``$B\to\pi\pi$ puzzle'' is reflected by the following quantities:
\begin{eqnarray}
R_{+-}^{\pi\pi}&\equiv&2\left[\frac{\mbox{BR}(B^+\to\pi^+\pi^0)
+\mbox{BR}(B^-\to\pi^-\pi^0)}{\mbox{BR}(B_d^0\to\pi^+\pi^-)
+\mbox{BR}(\bar B_d^0\to\pi^+\pi^-)}\right]\frac{\tau_{B^0_d}}{\tau_{B^+}}=
2.12\pm0.37~~\mbox{}\label{Rpm-def}\\
R_{00}^{\pi\pi}&\equiv&2\left[\frac{\mbox{BR}(B_d^0\to\pi^0\pi^0)+
\mbox{BR}(\bar B_d^0\to\pi^0\pi^0)}{\mbox{BR}(B_d^0\to\pi^+\pi^-)+
\mbox{BR}(\bar B_d^0\to\pi^+\pi^-)}\right]=0.83\pm0.23.\label{R00-def}
\end{eqnarray}
In order to calculate the numerical values, we have used 
$\tau_{B^+}/\tau_{B^0_d}=1.086\pm0.017$ and the most recent compilation 
of the Heavy Flavour Averaging Group (HFAG) \cite{HFAG}, adding the errors
in quadrature. The central values calculated within QCD factorization 
\cite{Be-Ne} give $R_{+-}^{\pi\pi}=1.24$ and $R_{00}^{\pi\pi}=0.07$.
As was recently pointed out \cite{BFRS-II}, these data indicate important 
non-factorizable contributions rather than NP contributions, and can be
perfectly accommodated in the SM. 
\item
In the $B\to\pi K$ system, the CLEO, BaBar and Belle collaborations have 
measured the following ratios of 
CP-averaged branching ratios \cite{BF-neutral1}:
\begin{eqnarray}
R_{\rm c}&\equiv&2\left[\frac{\mbox{BR}(B^+\to\pi^0K^+)+
\mbox{BR}(B^-\to\pi^0K^-)}{\mbox{BR}(B^+\to\pi^+ K^0)+
\mbox{BR}(B^-\to\pi^- \bar K^0)}\right]=1.17\pm0.12\label{Rc-def}\\
R_{\rm n}&\equiv&\frac{1}{2}\left[
\frac{\mbox{BR}(B_d^0\to\pi^- K^+)+
\mbox{BR}(\bar B_d^0\to\pi^+ K^-)}{\mbox{BR}(B_d^0\to\pi^0K^0)+
\mbox{BR}(\bar B_d^0\to\pi^0\bar K^0)}\right]=0.76\pm0.10,\label{Rn-def}
\end{eqnarray}
with numerical values following from \cite{HFAG}. As noted in 
\cite{BF-neutral2}, the pattern of $R_{\rm c}>1$ and -- in particular -- 
$R_{\rm n}<1$, which
is now consistently favoured by the separate BaBar, Belle and CLEO data, 
is actually very puzzling. On the other hand, the quantity \cite{FM}
\begin{equation}\label{R-def}
R\equiv\left[\frac{\mbox{BR}(B_d^0\to\pi^- K^+)+
\mbox{BR}(\bar B_d^0\to\pi^+ K^-)}{\mbox{BR}(B^+\to\pi^+ K^0)+
\mbox{BR}(B^-\to\pi^- \bar K^0)}
\right]\frac{\tau_{B^+}}{\tau_{B^0_d}}
=0.91\pm0.07
\end{equation}
does not show any anomalous behaviour. Since $R_{\rm c}$ and $R_{\rm n}$ are 
affected significantly by colour-allowed electroweak (EW) penguins, whereas 
these topologies may only contribute to $R$ in colour-suppressed form, this 
``$B\to\pi K$ puzzle'' may be a manifestation of NP in the EW penguin sector 
\cite{BFRS-II,BF-neutral2,BFRS-I}, offering an attractive avenue for 
physics beyond the SM to enter the $B\to\pi K$ system 
\cite{FM-NP}--\cite{gr-ewp}.

\item Another potential discrepancy with the SM expectation is indicated
by the decay $B_d\to\phi K_{\rm S}$. Within the SM, this transition is 
governed by QCD penguins \cite{BphiK-old} and receives sizeable EW penguin 
contributions \cite{RF-EWP,DH-PhiK}, so that it may well be affected by NP. 
The current experimental status of the CP-violating $B_d\to\phi K_{\rm S}$ 
observables is given as follows \cite{browder-talk,Belle-BphiK}:
\begin{equation}\label{aCP-Bd-phiK-dir}
{\cal A}_{\rm CP}^{\rm dir}(B_d\to \phi K_{\rm S})=
\left\{\begin{array}{ll}
-0.38\pm0.37\pm0.12 &\mbox{(BaBar)}\\
+0.15\pm0.29\pm0.07 &\mbox{(Belle)}
\end{array}\right.
\end{equation}
\begin{equation}\label{aCP-Bd-phiK-mix}
{\cal A}_{\rm CP}^{\rm mix}(B_d\to \phi K_{\rm S})=
\left\{\begin{array}{ll}
-0.45\pm0.43\pm0.07 &\mbox{(BaBar)}\\
+0.96\pm0.50^{+0.11}_{-0.09} &\mbox{(Belle),}
\end{array}\right.
\end{equation}
where we have employed the same notation for the direct and mixing-induced 
CP asymmetries ${\cal A}_{\rm CP}^{\rm dir}(B_d\to \phi K_{\rm S})$ and 
${\cal A}_{\rm CP}^{\rm mix}(B_d\to \phi K_{\rm S})$, respectively, 
as in \cite{RF-Phys-Rep}.\footnote{This notation may differ in signs from the
one used by some authors. However, the explicit example in (\ref{rate-asym}) 
and (\ref{obs}) should allow for an easy comparison of signs.}
In the SM, the direct asymmetry is 
$\ord(\lambda^2)$, i.e.\ tiny, whereas the mixing-induced CP asymmetry 
is --  up to corrections of $\ord(\lambda^2)$ -- equal to the one of 
the ``golden'' decay $B_d\to J/\psi K_{\rm S}$ \cite{growo}--\cite{GLNQ}, 
which is measured to be
\begin{equation}\label{ACPmix-gol}
{\cal A}_{\rm CP}^{\rm mix}(B_d\to J/\psi K_{\rm S})=-0.736\pm0.049.
\end{equation}
Consequently, we may well arrive at a discrepancy with the SM description 
of CP violation, although the experimental situation is of course very 
unclear at present.
\end{itemize}

In view of significant experimental uncertainties, none of these 
exciting results is conclusive at the moment, but it is legitimate and 
interesting to take them seriously and to search for possible origins of 
these ``signals'' for deviations from the SM expectations. As we are 
dealing here with non-leptonic decays, the natural question arises of whether 
these signals originate in the NP contributions or/and result from our 
insufficient understanding of the hadron dynamics. 
The purpose of the present paper is to develop a strategy that would allow 
us to address this question in a systematic manner once the experimental 
data on the relevant non-leptonic $B$ decays and rare $B$ and $K$ decays 
improve. In order to illustrate this strategy in explicit terms, we shall 
consider a simple extension of the SM in which NP enters dominantly through 
enhanced $Z^0$ penguins involving a CP-violating weak phase. As we will see 
below, this choice is 
dictated by the pattern of the data on the $B\to\pi K$ observables and 
the great predictivity of this scenario. It was first considered in 
\cite{Buras:1998ed}--\cite{Buras:1999da} to study correlations between rare 
$K$ decays and the ratio $\epe$ measuring direct CP violation
in the neutral kaon system, and was generalized to rare $B$ decays in 
\cite{Buchalla:2000sk}. Here we extend these considerations to non-leptonic
$B$-meson decays, which allows us to confront this extension of the SM with
many more experimental results. Our strategy consists of three interrelated 
steps, and has the following logical structure:

\vspace*{0.3truecm}

\noindent
{\bf Step 1:}

\noindent
Since $B\to\pi\pi$ decays and the usual analysis of the UT are only 
insignificantly affected by EW penguins, the $B\to\pi\pi$ system can be 
described as in the SM. Employing the $SU(2)$ isospin flavour symmetry
of strong interactions and the information on $\gamma$ from the
UT fits, we may extract the relevant hadronic parameters, and find large 
non-factorizable contributions, which are in particular reflected by 
large CP-conserving strong phases. Having these parameters at hand, we 
may then also predict the direct and mixing-induced CP asymmetries
of the $B_d\to\pi^0\pi^0$ channel. A future measurement of one of these
observables allows a determination of $\gamma$.

\vspace*{0.3truecm}

\noindent
{\bf Step 2:}

\noindent
If we use the $SU(3)$ flavour symmetry and plausible dynamical 
assumptions, we may determine the hadronic $B\to\pi K$ parameters 
through the $B\to\pi\pi$ analysis, and may calculate the $B\to\pi K$ 
observables in the SM. Interestingly, we find agreement with the pattern 
of the $B$-factory data for those observables where EW penguins play only 
a minor r\^ole. On the other hand, the observables receiving significant
EW penguin contributions do {\it not} agree with the experimental picture, 
thereby suggesting NP in the EW penguin sector. Indeed, a detailed analysis
shows that we may describe all the currently available data through sizeably 
enhanced EW penguins with a large CP-violating NP phase around $-90^\circ$,
in the spirit of the NP scenario considered here. A crucial future test of 
this scenario will be provided by the CP-violating $B_d\to \pi^0 K_{\rm S}$ 
observables, which we may predict. Moreover, we may obtain valuable insights 
into $SU(3)$-breaking effects, which support our working assumptions, and 
may also determine the UT angle $\gamma$, with a result in remarkable 
agreement with the well-known UT fits.  

\vspace*{0.3truecm}

\noindent
{\bf Step 3:}

\noindent
In turn, the sizeably enhanced EW penguins with their large CP-violating 
NP phase have important implications for rare $K$ and $B$ decays as well as 
$\epe$, where the new weak phase plays a particularly important r\^ole.
Interestingly, several predictions differ significantly from the SM 
expectations and should easily be identified once the data improve. 
Similarly, we may explore specific NP patterns in other non-leptonic 
$B$ decays such as $B_d\to\phi K_{\rm S}$.

\vspace*{0.3truecm}

The most interesting results of this study have recently been summarized 
in \cite{BFRS-II}. Here we discuss the details of our analysis, present 
several additional results, and propose other methods that will be 
useful for the confrontation of the forthcoming data with the SM and 
the search for possible indications of NP. The outline of this paper is 
as follows: in Section~\ref{sec:Basic}, we discuss in detail our scenario 
of NP, using low-energy effective Hamiltonians as the starting point. 
In the subsequent three sections, we execute the three steps described 
above. In Section~\ref{sec:Bpipi}, we discuss the $B\to\pi\pi$ system
in detail. We then move on to the $B\to\pi K$ system in Section~\ref{sec:BpiK},
and explore the impact of this study on rare $K$ and $B$ decays and
$\epe$, as well as the implications of the current data on rare decays on 
the $B\to\pi K$ observables, in 
Section~\ref{sec:rare}. As the last element of our analysis, we 
investigate in Section~\ref{sec:B-prom} the effects of our NP scenario
on the prominent $B$-meson decays of the kind $B\to\phi K$
and $B\to J/\psi K$. Finally, we summarize our 
conclusions in Section~\ref{sec:concl}.
A compendium of the most relevant formulae for $B\to\pi\pi$, 
$B_{(s)}\to\pi K$ and $B_s\to K^+K^-$ observables and some 
technical details can be found in Appendices A--D.

Our paper discusses a large number of observables and it is useful already
now to list them, as well as the parameters they depend on, and indicate 
where the explicit expressions for them can be found in our paper. We do this 
in Table~\ref{guide}, where we also present our predictions (TH), list the 
input values and the present experimental values (EXP) for the observables 
in question. It should be emphasized that, within our approach, the 
hadronic $B_{(s)}\to\pi K$ and $B_s\to K^+K^-$ parameters can be 
calculated in terms of the parameters of the $B\to\pi\pi$ system. 
The relevant formulae are given in (\ref{r-rho-rel}), (\ref{rho-n-det}), 
(\ref{x-rel}) and (\ref{r-det}). A compendium of the formulae for all 
observables of Table~\ref{guide} is given in Appendix~\ref{sec:comp}.

\begin{table}[hbt]
\vspace{0.4cm}
\begin{center}
\begin{tabular}{|c||c|c|c|c|}
\hline
{Quantity} & Eq.\ &  Parameters & 
TH & EXP
 \\ \hline
$R_{+-}^{\pi\pi}$ & (\ref{Rpm-expr}) &  $x,~\Delta,~d,~\theta $ 
& input & $2.12\pm0.37$
\\ \hline
$R_{00}^{\pi\pi}$ & (\ref{R00-expr}) &  $x,~\Delta,~d,~\theta $ 
&input & $0.83\pm0.23$
\\ \hline
${\cal A}_{\rm CP}^{\rm dir}(B_d\!\to\! \pi^+\pi^-)$ & (\ref{Adir-Bpipi})  & 
$d,~\theta $ & input & $-0.38\pm0.16$
\\ \hline
${\cal A}_{\rm CP}^{\rm mix}(B_d\!\to\! \pi^+\pi^-)$ & (\ref{Amix-Bpipi})  & 
 $d,~\theta,~\phi_d$ & input & $0.58\pm0.20$
\\ \hline
${\cal A}_{\rm CP}^{\rm dir}(B_d\!\to\! \pi^0\pi^0)$ & (\ref{Adir-Bpipi0})  & 
 $x,~\Delta,~d,~\theta $ & $-0.41^{+0.35}_{-0.17}$ \rule{0em}{1.05em} & $-$
\\ \hline
${\cal A}_{\rm CP}^{\rm mix}(B_d\!\to\! \pi^0\pi^0)$ & (\ref{Amix-Bpipi0})  & 
 $x,~\Delta,~d,~\theta,~\phi_d$ & $-0.55^{+0.43}_{-0.45}$ \rule{0em}{1.05em} 
& $-$
\\ \hline
 \hline
$R$ & (\ref{R-expr})  &  $r,~\delta $ 
& $0.943^{+0.033}_{-0.026}$ \rule{0em}{1.05em} & $0.91\pm0.07$
\\ \hline
$R_{\rm c}$ & (\ref{Rc-expr})  &  $r_{\rm c},~\delta_{\rm c},~q,~\omega,~\phi$ 
& $1.00^{+0.12}_{-0.08}$ \rule{0em}{1.05em} & $1.17\pm0.12$
\\ \hline
$R_{\rm n}$ & (\ref{Rn-expr})  &  
$r,\delta,r_{\rm c},\delta_{\rm c},q,\omega,\phi,\rho_{\rm n},\theta_{\rm n}$ 
& $0.82^{+0.12}_{-0.11}$ \rule{0em}{1.05em} & $0.76\pm0.10$
\\ \hline
${\cal A}_{\rm CP}^{\rm dir}(B_d\!\to\!\pi^\mp K^\pm)$ &(\ref{ACPdir-r}) & 
$r,~\delta $ & $0.140^{+0.139}_{-0.087}$ \rule{0em}{1.05em} & $0.095\pm 0.028$
\\ \hline
${\cal A}_{\rm CP}^{\rm dir}(B^\pm\!\to\!\pi^\pm K) $ &(\ref{ACP-B+pi+K}) & 
$\rho_{\rm c},~\theta_{\rm c}$ & $\sim0$ & $-0.02\pm 0.06$
\\ \hline
${\cal A}_{\rm CP}^{\rm dir}(B^\pm\!\to\!\pi^0 K^\pm) $ 
&(\ref{ACP-Bpi0K-defa}) & 
$r_{\rm c},~\delta_{\rm c},~q,~\omega,~\phi$  
& $0.03^{+0.32}_{-0.24}$ \rule{0em}{1.05em} & $0.00\pm0.07$
\\ \hline
${\cal A}_{\rm CP}^{\rm dir}(B_d\!\to\!\pi^0 K_{\rm S})$ 
&(\ref{ACPdir-Bpi0K0}) & 
$r,\delta,r_{\rm c},\delta_{\rm c},q,\omega,\phi,\rho_{\rm n},\theta_{\rm 
n}$   & $0.08^{+0.18}_{-0.22}$ \rule{0em}{1.05em} & $0.40^{+0.29}_{-0.30}$
\\ \hline 
${\cal A}_{\rm CP}^{\rm mix}(B_d\!\to\!\pi^0 K_{\rm S})$ 
&(\ref{ACPmix-Bpi0K0}) & 
$r,\delta,r_{\rm c},\delta_{\rm c},q,\omega,\phi,\rho_{\rm n},\theta_{\rm 
n},\phi_d$ & $-0.98^{+0.05}_{-0.02}$ \rule{0em}{1.05em} 
& $-0.48_{-0.40}^{+0.48}$
\\ \hline
${\cal A}_{\rm CP}^{\rm dir}(B_s\!\to\! K^+K^-)$ &(\ref{ACPdir-BsKK}) & 
$d,~\theta~$ & $0.14^{+0.14}_{-0.09}$ \rule{0em}{1.05em} & $-$
\\ \hline
${\cal A}_{\rm CP}^{\rm mix}(B_s\!\to\! K^+K^-)$ &(\ref{ACPmix-BsKK}) & 
$d,~\theta,~\phi_s$ & $-0.18^{+0.08}_{-0.07}$ \rule{0em}{1.05em} & $-$ 
\\ \hline
${\cal A}_{\rm CP}^{\rm dir}(B_s\!\to\! \pi^\pm K^\mp)$ & 
(\ref{ACPdir-BspiK}) & $r,~\delta~$ & $-0.38^{+0.16}_{-0.16}$ 
\rule{0em}{1.05em} & $-$
\\ \hline
\end{tabular}
\end{center}
\caption[]{\small A guide to master formulae. We suppress the dependence on 
$\gamma$ as it enters all quantities. For those quantities that depend on
the EW penguin parameters $q$ and $\phi$, the rare decays constraints as 
discussed in Section~\ref{sec:rare} have been employed in the calculation 
of our theoretical values (TH).\label{guide}}
\end{table}

\boldmath
\section{A Simple Scenario for New Physics}
\label{sec:Basic}
\setcounter{equation}{0}
\unboldmath
\subsection{General Structure}
The scenario of NP with the dominant $Z^0$-penguin contributions presented 
here was first considered in \cite{Buras:1998ed}--\cite{Buras:1999da}, 
where correlations between rare $K$ decays and $\epe$ were studied. It 
was generalized to rare $B$ decays in \cite{Buchalla:2000sk}. The new 
feature of our analysis is a simultaneous study of rare $K$ and $B$ decays, 
in addition to non-leptonic $B$ decays, in particular 
$B\to\pi\pi,\pi K$ modes.

It should be noted that in
\cite{Buras:1998ed}--\cite{Buchalla:2000sk} 
model-independent analyses and studies within particular supersymmetric 
scenarios were presented. As we will see below, in the present analysis, 
we determine the size of the enhancement of the $Z^0$-penguin function 
$C$ and the magnitude of its complex phase by the $B\to\pi K$ data and, 
consequently, our predictions are more specific than was possible in 
these papers. In what follows, we will recall the basic ingredients 
of this scenario in a notation suitable to our analysis.

In order to discuss the non-leptonic $B$-meson decays and rare $K$ and $B$
decays in a particular extension of the SM, and to investigate correlations 
between the NP contributions to different observables, it is essential to
formulate the theory with the help of an effective Hamiltonian. The effective 
weak Hamiltonian for $\Delta F=1$ decays with $F=S,B$ is generally 
given as follows 
\cite{Buchalla:1995vs}:
\be\label{b1}
{\cal H}_{\rm eff}=\frac{G_{\rm F}}{\sqrt{2}}
\sum_i V^i_{\rm CKM}C_i(\mu)Q_i,
\ee
where $G_{\rm F}$ is the Fermi constant and $V^i_{\rm CKM}$ the relevant CKM 
factors. Next, the $Q_i$ are local operators, which govern the decays in 
question, and the $C_i(\mu)$ are the corresponding Wilson coefficients, 
which describe the strength with which a given operator enters the 
Hamiltonian. The latter coefficients can be calculated in 
renormalization-group-improved perturbation theory and carry, in particular, 
the information about the physics contributions at scales higher than 
$\mu$, which is usually chosen to be $\ord(m_b)$ and 
$\ord((1\mbox{--}2)~{\rm GeV})$ 
for $B$ decays and $K$ decays, respectively. Thus, the $C_i$ include the 
top-quark contributions and contributions from other heavy particles if 
extensions of the SM are considered. Consequently, the $C_i(\mu)$ generally 
depend on the top-quark mass $m_t$ and also on the masses and couplings 
of the new particles. This dependence can be found by evaluating box and 
penguin diagrams with $W$-, $Z$-, top- and new-particle exchanges and properly 
including short-distance QCD effects. 

The amplitude for the decay of a given meson $M= K, B, ...$\ into a final 
state $F=\pi\pi, \pi K, \pi\nu\bar\nu, ...$\ is then simply given by
\be\label{amp5}
A(M\to F)=\langle F|{\cal H}_{\rm eff}|M\rangle=\frac{G_{\rm F}}{\sqrt{2}}
\sum_i V^i_{\rm CKM}C_i(\mu)\langle F|Q_i(\mu)|M\rangle,
\ee
where the $\langle F|Q_i(\mu)|M\rangle$ are the hadronic matrix elements of 
the operators $Q_i$ between $M$ and $F$, evaluated at the renormalization 
scale $\mu$. As demonstrated in \cite{PBE,BH92}, the formula (\ref{amp5}) 
can be cast into the following useful expression: 
\be\label{mmaster}
{A({\rm Decay})}= 
P_0({\rm Decay}) + \sum_r P_r({\rm Decay} ) \, F_r(v).
\end{equation}
To this end, we choose $\mu=\mu_0=\ord(M_W,m_t)$ in (\ref{amp5}) and 
rewrite the $C_i(\mu_0)$ as linear combinations of the so-called master
functions $F_r(v)$, which result from various penguin and box diagrams 
with heavy particle exchanges and $v$ denoting collectively the 
parameters of a given model. In the SM, the functions $F_r(v)$ 
reduce to  the  Inami--Lim functions \cite{IL}, with $v=m_t^2/M_W^2$. 
The term $P_0$ summarizes the contributions stemming from light internal 
quarks, such as the charm and up quarks, and the sum incorporates the 
remaining contributions.
The general properties of the $P_0$, $P_r$ and $F_r$ have recently been 
discussed at length in \cite{Cracow}. We recall here only 
the following features:
\begin{itemize}
\item
$F_r(v)$ are {\it process-independent, universal} functions that depend 
on the particular model considered. NP enters
the decay amplitudes only through these functions.
\item
$P_0$ and $P_r$ are {\it process-dependent} quantities. In particular, they 
depend on the hadronic matrix elements of the operators 
$Q_i$.\footnote{These quantities should not be confused with the 
penguin parameters in Sections~\ref{sec:Bpipi} and \ref{sec:BpiK}.}
\end{itemize}
In the models with ``minimal flavour violation'' (MFV), 
as defined in \cite{Cracow,UUT}, the set of the $F_r(v)$ consists 
of seven functions
\be\label{masterf}
S(v),~X(v),~Y(v),~Z(v),~E(v),~D'(v),~E'(v),
\ee
which are discussed in detail in \cite{Cracow}. The important property of 
the functions in (\ref{masterf}) is that they are real-valued, so that the 
CP-violating effects are governed entirely by the complex CKM phase 
hiding in the parameters $P_r$.
Other definitions of MFV can be found in \cite{MFV2,BOEWKRUR}.
\subsection{Going Beyond MFV}
In the present paper, we would like to investigate a scenario along the 
lines of \cite{Buras:1998ed}--\cite{Buchalla:2000sk}, which goes beyond
the usual MFV framework. In particular, we consider the presence of a 
single weak complex phase in the functions $F_r(v)$, leading to additional 
CP-violating effects that are not present in models with MFV. In order 
to make the discussion as simple as possible, we will choose the functions 
$S$, $E$, $D'$ and $E'$ to be real-valued, as in the MFV case. Moreover,
we will assume that the functions $E$, $D'$ and $E'$ are not affected 
significantly by NP contributions, so that $D'$ and $E'$, which result from 
magnetic penguin diagrams, describe with their SM values automatically 
the $B\to X_s\gamma$ decay. The function $E$ resulting from QCD penguins 
and entering the Wilson coefficients of penguin operators at scales 
$\ord(M_W)$ is phenomenologically not very relevant and can safely be set 
to its SM value. The point is that the contributions of QCD penguins to 
essentially all decays are 
dominated by renormalization-group effects for scales below $M_W$ and are
included in the coefficient $P_0$ in (\ref{mmaster}).

Concerning the $\Delta F=2$ box function $S(v)$, it enters only the CP 
violation parameter $\varepsilon_K$ and the $B^0_{d,s}$--$\bar B^0_{d,s}$ 
mass differences $\Delta M_{d,s}$. Being real, it does not introduce any 
new complex phases in the standard analysis of the UT that is based on 
$\vcb$, $\vub$, $\varepsilon_K$ and $\Delta M_{d,s}$. On the other hand, 
the enhanced function $C$ and its large complex phase can, in principle, 
affect the analysis of the UT through the double $Z^0$-penguin contributions 
to $\varepsilon_K$ and $\Delta M_{s,d}$. However, we have checked that
these effects are at the few-percent level. In particular, the phases of 
the $B^0_{d,s}$--$\bar B^0_{d,s}$ mixing amplitudes are essentially unchanged.

Next the decay amplitude of the ``golden'' mode $B_d\to J/\psi K_{\rm S}$ 
is only insignificantly, i.e.\ at the level of the current experimental 
and theoretical uncertainties, affected by the NP in the EW penguin sector 
in our scenario. Consequently, as discussed in detail in 
Subsection~\ref{ssec:BpsiK}, we may still convert -- to a good approximation --
the mixing-induced CP asymmetry 
${\cal A}_{\rm CP}^{\rm mix}(B_d\to J/\psi K_{\rm S})$ into the angle 
$\beta$ of the UT, as in the SM. In particular, a ``universal unitarity 
triangle'' \cite{UUT} can also be constructed in our NP scenario by using 
only $\vcb$, $\Delta M_{d}/\Delta M_s$ and $\sin 2\beta$. This simplifies our
phenomenological analysis significantly, since the determination of the 
CKM parameters can be separated from the study of the hadronic and NP 
effects in non-leptonic $B$ decays and rare decays. However, as 
advertised in Step 2 of our approach, one can also determine $\gamma$ by 
complementing the $B\to\pi\pi$ system with those $B\to\pi K$ modes 
that are insensitive to the NP effects discussed here, i.e.\ to EW 
penguin contributions, and subsequently compare the result with the 
one obtained by using the conventional UT fits. Following these lines
and using the experimental information on the side $R_b$ of the UT,
we may also determine the remaining two angles $\alpha$ and $\beta$ in 
an alternative manner, and eventually arrive at a remarkably consistent 
overall picture. 

Most interesting for us are the functions $X$, $Y$ and $Z$, which are given by 
\begin{equation}\label{XYZ} 
X(v)=C(v)+B^{\nu\bar\nu}(v),\quad  
Y(v)  =C(v)+B^{\mu\bar\mu}(v),\quad
Z(v)  =C(v)+\frac{1}{4}D(v),
\end{equation}
with $C(v)$, $D(v)$, $B^{\nu\bar\nu}(v)$ and $B^{\mu\bar\mu}(v)$ resulting 
from $Z^0$-penguin diagrams, $\gamma$-penguin diagrams and $\Delta F=1$ 
box diagrams with $\nu\bar\nu$ and $\mu\bar\mu$, respectively.
Moreover, explicit calculations indicate that
in the case of $\Delta F=1$ box diagrams with $u\bar u$ and $d\bar d$ in
the final state we have, to an excellent approximation:
\be\label{NON}
B^{u\bar u}(v)=B^{\nu\bar\nu}(v),\quad B^{d\bar d}(v)=B^{\mu\bar\mu}(v),
\ee
so that the functions $X$ and $Y$ in (\ref{XYZ}) are valid for 
non-leptonic decays as well. Similarly, it is found that the NP 
contributions to these $\Delta F=1$ box diagrams and to $D$ are rather 
small, so that we may use for them, in the following analysis, their 
SM values ($m_t=167~{\rm GeV}$), which are given by
\be\label{NONSM}
B^{u\bar u}(v)=B^{\nu\bar\nu}(v)=0.73,\quad 
B^{d\bar d}(v)=B^{\mu\bar\mu}(v)=0.18,\quad D(v)=-0.48.
\ee
As a consequence of these approximations, the dominant NP effects 
come from the $Z^0$-penguin function $C(v)$. In the standard MFV scenarios, 
also the one considered in \cite{BFRS-I}, the function $C(v)$ is 
real-valued. Here it contains a weak phase $\theta_C$. Consequently, the 
functions $X$, $Y$ and $Z$ are now given as follows:
\begin{equation}\label{PARXYZ} 
X(v)=|C(v)|e^{i\theta_C}+0.73,\quad  
Y(v)  =|C(v)|e^{i\theta_C}+0.18, \quad
Z(v)  =|C(v)|e^{i\theta_C}-0.12.
\end{equation}

While our analysis does not rely on a particular model with the properties
specified above, concrete models with enhanced CP-violating $Z^0$-mediated 
FCNC couplings generated either at the one-loop level or even at the 
tree level have been discussed in the literature. They are reviewed
in \cite{Buras:1998ed}--\cite{Buchalla:2000sk}, in particular in the last of
these papers; see also \cite{GNR97}. Also models with $Z^\prime$-mediated 
FCNCs could be put in this class, provided their contributions can 
effectively be absorbed in the function $C(v)$. For a recent analysis, 
see~\cite{BCLL03}.

\boldmath
\subsection{Relation Between Rare Decays and $B\to\pi K$ 
Modes}\label{ssec:EWP-REL}
\unboldmath
As already discussed in \cite{BFRS-I}, the connection between the rare 
decays and the $B\to\pi K$ system is established by relating the function 
$C$ to the EW penguin parameter $q$ by means of a renormalization-group 
analysis. In the case of a complex $C$, the relation given in \cite{BFRS-I} 
is generalized as follows:
\be\label{RG}
|C(v)|e^{i\theta_C} = 2.35\, \bar q e^{i\phi} -0.82,\quad 
\bar q= q \left[\frac{|V_{ub}/V_{cb}|}{0.086}\right], 
\ee
where $q$ and $\phi$ characterize the EW penguin sector and enter the 
parametrization of the $B\to\pi K$ decays given in Section~\ref{sec:BpiK}. 
The numerical coefficients in (\ref{RG}) correspond to $\alpha_s(M_Z)=0.119$ 
and depend very weakly on its value.

In order to define $q$ and $\phi$, we recall formula (43) of 
\cite{BF-neutral1} for the ratio of the EW penguin amplitude $P_{\rm EW}'$ 
to the tree contributions in $B\to\pi K$ decays (see also 
Section~\ref{sec:BpiK}), yielding
\be\label{BF43}
\left|\frac{P_{\rm EW}'}{T'+C'}\right| e^{i(\delta_{\rm EW}'-\delta_{T'+C'})}
=-\frac{3}{2}\frac{1}{\lambda |V_{ub}/V_{cb}|} 
\left[\frac{C_9(\mu_b)+C_{10}(\mu_b)\tilde\xi}
{C_1^\prime(\mu_b)+C_2^\prime(\mu_b)\tilde\xi}\right]\,,
\ee
where $\tilde\xi=1$ in the $SU(3)$ symmetry limit \cite{BF-neutral1,NR}. Here 
$C_9(\mu_b)$ and $C_{10}(\mu_b)$ ($\mu_b=\ord(m_b)$) are the Wilson 
coefficients of the $(V-A)\otimes(V-A)$ EW penguin operators $Q_9$ and 
$Q_{10}$, respectively, which enter the effective Hamiltonian for 
$\Delta B=1$ non-leptonic decays \cite{Buchalla:1995vs}, and 
the coefficients $C_{1,2}^\prime(\mu_b)$ are given as follows
\cite{defan}:
\be
C_1^\prime(\mu_b)=C_1(\mu_b)+\frac{3}{2}C_9(\mu_b),\quad
C_2^\prime(\mu_b)=C_2(\mu_b)+\frac{3}{2}C_{10}(\mu_b),
\ee
with  $C_{1,2}$ being the Wilson coefficients of the current--current 
operators $Q_{1,2}$. 
As $C_{9,10}=\ord(\alpha)$ and $C_{1,2}=\ord(1)$, we will approximate 
$C^\prime_{1,2}$ by $C_{1,2}$. Moreover, we will set $\tilde\xi=1$ 
in front of $C_{10}$. This approximation is  justified by 
the fact that $C_{10}(M_W)=0$ and consequently $C_{10}(\mu_b)$ is 
significantly smaller than $C_{9}(\mu_b)$. 
Next, in the NP scenario outlined above, we may write
\be\label{C910}
C_{9}(\mu_b)+C_{10}(\mu_b)= h+g C(v) e^{i\theta_C}\equiv -A_{\rm EW} 
e^{i\phi},
\ee
where $h$ and $g$ are calculable by means of the two-loop 
renormalization-group formula given in \cite{Buras:1993dy}. 
The minus sign in (\ref{C910}) causes the parameter $A_{\rm EW}$ to be 
positive in the SM. 
Writing then
$\tilde\xi=\xi e^{i\tau}$, $\tau$ being a strong phase, we obtain
\be
C_1(\mu_b)+C_2(\mu_b)\xi e^{i\tau}\equiv B_{\rm CC} e^{-i\omega},
\ee
where it was convenient to introduce a strong phase $\omega$ 
\cite{BF-neutral1}, which vanishes in the $SU(3)$ limit 
\cite{NR,neubert-BpiK}. Consequently, (\ref{BF43}) yields the
following simple expression:
\be\label{FBF43}
\left|\frac{P_{\rm EW}'}{T'+C'}\right| e^{i(\delta_{\rm EW}'-\delta_{T'+C'})}
= q e^{i\phi} e^{i\omega},\quad 
q=\frac{3}{2}\frac{1}{\lambda |V_{ub}/V_{cb}|}\frac{A_{\rm EW}}{B_{\rm CC}},
\ee
which we will use for the parametrization of the $B\to\pi K$ decays
in Section~\ref{sec:BpiK}. 

As far as the rare decays considered in Section~\ref{sec:rare} 
are concerned, they can be directly expressed in terms of the $X$, $Y$ 
and $Z$ functions introduced in (\ref{PARXYZ}). Consequently, in order 
to discuss the correlation between the $B\to\pi K$ decays and the rare 
decays, it is useful to express them in terms of $(\bar q,\phi)$ by 
inserting (\ref{RG}) into (\ref{PARXYZ}). The corresponding expressions 
are given in Section~\ref{sec:rare}.

\subsection{CKM Parameters}\label{ssec:CKM}
Concerning the CKM parameters, we will use the Wolfenstein parametrization
\cite{WO}, generalized to include higher orders in $\lambda\equiv|V_{us}|$ 
\cite{BLO}.
Writing then
\be
V_{td}=A R_t\lambda^3 e^{-i\beta},\quad V_{ts}=-|V_{ts}|e^{-i\beta_s},
\ee
with $\tan\beta_s\approx -\lambda^2 \bar\eta$, we have
\be\label{LAMT}
\lambda_t\equiv V_{ts}^*V_{td}=-\tilde r A^2 \lambda^5 R_t e^{-i\beta} 
e^{i\beta_s} \quad\mbox{with}\quad 
\tilde r=\left|\frac{V_{ts}}{V_{cb}}\right|\approx 0.98,
\ee
and
\be
\tilde\lambda^{(d)}_t\equiv V_{tb}^*V_{td}= A \lambda^3 R_t e^{-i\beta},\quad 
\tilde\lambda^{(s)}_t\equiv V_{tb}^*V_{ts}= 
-\tilde r A \lambda^2 e^{-i\beta_s},
\ee
where $R_t$ is one of the sides in the UT, and $A\equiv |V_{cb}|/\lambda^2$ 
is the usual Wolfenstein parameter. Moreover, we have the following useful 
relations for the two UT sides $R_t$ and $R_b$ and the ratio
$|V_{ub}/V_{cb}|$ \cite{schladming}:
\be\label{RTVUB}
R_t=\frac{\sin\gamma}{\sin(\beta+\gamma)},\quad
R_b\equiv \left(1-\frac{\lambda^2}{2}\right)
\frac{1}{\lambda}\left|\frac{V_{ub}}{V_{cb}}\right|,\quad
\left|\frac{V_{ub}}{V_{cb}}\right|=\left(\frac{\lambda}{1-\lambda^2/2}
\right)\frac{\sin\beta}{\sin(\beta+\gamma)}.
\ee
For our numerical analyses, we will use
\be\label{CKM1}
\lambda=0.2240\pm 0.0036,\quad
A=0.83\pm0.02,\quad \left|\frac{V_{ub}}{V_{cb}}\right|=0.086\pm0.008,\quad
R_b=0.37\pm0.04
\ee
\be\label{CKM2}
\beta=(23.5\pm 2.0)^\circ,\quad \beta_s=-1^\circ.
\ee

\subsection{Summary}
Before turning to the $B\to\pi\pi$ system, let us summarize the main
results of this section:
\begin{itemize}
\item The NP scenario considered here involves two parameters, $C(v)$ and 
$\theta_C$.

\item The relevant EW penguin parameters for the $B\to\pi K$ decays 
are $q$, $\phi$ and $\omega$.

\item The parameter sets $(C,\theta_C)$ and $(q,\phi)$ are related 
through (\ref{RG}), which allows us to investigate the correlations 
between rare decays and non-leptonic $B$ decays.
\end{itemize}

\section{\boldmath The $B\to\pi\pi$ System \unboldmath}\label{sec:Bpipi}
\setcounter{equation}{0}
In the literature, $B\to\pi\pi$ decays are usually considered 
in the context of the determination of the UT angle $\alpha$ 
\cite{grolo,alpha-det}. Here we shall use these modes from
a very different point of view, which is inspired by the analyses
performed in \cite{RF-BsKK}--\cite{FIM}. Very recently, the decay 
$B_d\to\pi^+\pi^-$ was also discussed in \cite{BS} in the context
of bounds on $\gamma$ and the UT. However, these approaches differ 
significantly from the one presented in \cite{BFRS-II} and here. 
In particular, the channels involving neutral pions are not discussed 
at all in \cite{BS}.
\subsection{Basic Formulae}\label{ssec:Bpipi-basic}
In order to make our analysis more transparent, we shall first neglect 
the EW penguin contributions to the $B\to\pi\pi$ decays, which play a 
very minor r\^ole \cite{GHLR-EWP,PAPIII}; using the isospin flavour 
symmetry of strong interactions, we include these contributions 
in Subsection~\ref{ssec:Bpipi-EWP}, following \cite{BF-neutral1,GPY}. 
We may then write the $B\to\pi\pi$ amplitudes as
\begin{eqnarray}
\sqrt{2}A(B^+\to\pi^+\pi^0)&=&-[\tilde T+\tilde C] = 
-[T+C]\label{B+pi+pi0}\\
A(B^0_d\to\pi^+\pi^-)&=&-[\tilde T + P]\label{Bdpi+pi-}\\
\sqrt{2}A(B^0_d\to\pi^0\pi^0)&=&-[\tilde C - P],\label{Bdpi0pi0}
\end{eqnarray}
with
\begin{eqnarray}
P&=&\lambda^3 A({\cal P}_t-{\cal P}_c)\equiv\lambda^3 A {\cal P}_{tc}
\label{P-def}\\
\tilde T &=&\lambda^3 A R_b e^{i\gamma}\left[{\cal T}-\left({\cal 
P}_{tu}-{\cal E}\right)\right]\label{T-tilde}\\
 \tilde C &=&\lambda^3 A R_b e^{i\gamma}\left[{\cal C}+\left({\cal P}_{tu}-
{\cal E}\right)\right].\label{C-tilde}
\end{eqnarray}
The parameters of the CKM matrix entering these formulae have been defined 
in Subsection~\ref{ssec:CKM}, whereas the ${\cal P}_q$ describe the strong 
amplitudes of QCD penguins with internal $q$-quark exchanges 
($q\in\{t,c,u\}$),\footnote{Strictly speaking in the case of ${\cal P}_t$
this standard terminology could be misleading, as the top-quark effects in
amplitudes for $B$ decays are only present in the relevant Wilson
coefficients -- also absorbed in ${\cal P}_q$ -- and not in hadronic matrix
elements. For relations between the language used here to that of
the operator product expansion, see \cite{RF-DIPL}--\cite{BSNONF}.} 
including annihilation and exchange penguins; ${\cal T}$ 
and ${\cal C}$ are the strong amplitudes of colour-allowed and 
colour-suppressed tree-diagram-like topologies, respectively; and ${\cal E}$ 
denotes the strong amplitude of an exchange amplitude. In the usual notation 
employed in the literature (see, for instance, \cite{GHLR-EWP}), the 
colour-allowed and colour-suppressed tree-diagram-like amplitudes
\begin{eqnarray}
T &=&\lambda^3 A R_b e^{i\gamma}{\cal T}\\
C &=&\lambda^3 A R_b e^{i\gamma}{\cal C}
\end{eqnarray}
appear; they differ from the $\tilde T$ and $\tilde C$ amplitudes 
through the $({\cal P}_{tu}-{\cal E})$ pieces, which may actually play
an important r\^ole, as was emphasized in \cite{PAP0}. However, 
we observe that these terms cancel in   
\begin{equation}
\tilde T+\tilde C=T+C,
\end{equation}
and that the amplitudes in (\ref{B+pi+pi0})--(\ref{Bdpi0pi0}) satisfy 
the well-known isospin relation \cite{grolo}
\begin{equation}\label{Bpipi-isospin}
\sqrt{2}A(B^+\to\pi^+\pi^0)=A(B^0_d\to\pi^+\pi^-)+
\sqrt{2}A(B^0_d\to\pi^0\pi^0).
\end{equation}
It is convenient to rewrite these amplitudes as
\begin{eqnarray}
\sqrt{2}A(B^+\to\pi^+\pi^0)&=&-|\tilde T|e^{i\delta_{\tilde T}}
e^{i\gamma}\left[1+x e^{i\Delta}\right]\label{ampl-Bpi+pi0}\\
A(B^0_d\to\pi^+\pi^-)&=&-|\tilde T|e^{i\delta_{\tilde T}}
\left[e^{i\gamma}-d e^{i\theta}\right]\label{ampl-Bdpi+pi-}\\
\sqrt{2}A(B^0_d\to\pi^0\pi^0)&=&|P|e^{i\delta_P} 
\left[1+\frac{x}{d}e^{i\gamma}e^{i(\Delta-\theta)}\right],\label{ampl-Bpi0pi0}
\end{eqnarray}
with
\begin{equation}\label{x-Bpipi}
x e^{i\Delta}\equiv \frac{\tilde C}{\tilde T}=
\left|\frac{\tilde C}{\tilde T}\right|e^{i(\delta_{\tilde C}-
\delta_{\tilde T})}=\frac{{\cal C}+\left({\cal P}_{tu}-
{\cal E}\right)}{{\cal T}-\left({\cal 
P}_{tu}-{\cal E}\right)}
\end{equation}
\begin{equation}\label{d-theta-def}
d e^{i\theta}\equiv-\frac{P}{\tilde T}e^{i\gamma}=
-\left|\frac{P}{\tilde T}\right|e^{i(\delta_P-\delta_{\tilde T})}
=-\frac{1}{R_b}\left[\frac{{\cal P}_{tc}}{{\cal T}-\left({\cal 
P}_{tu}-{\cal E}\right)}\right],
\end{equation}
where the latter parameter was already introduced in \cite{RF-BsKK},
and $\delta_{\tilde T}$ and $\delta_P$ denote the CP-conserving strong 
phases of the amplitudes $\tilde T$ and $P$, respectively. 
If we now consider the corresponding CP-averaged decay-amplitude squares,
which are generically defined through
\begin{equation}
\langle |A|^2\rangle\equiv \frac{1}{2}\left[|A(B\to f)|^2+
|A(\bar B\to\bar f)|^2\right],
\end{equation}
we obtain
\begin{equation}\label{ampl-sqr-1}
\langle |A(B^\pm\to\pi^\pm\pi^0)|^2\rangle=\frac{|\tilde T|^2}{2}
\left[1+2x\cos\Delta+x^2\right]
\end{equation}
\begin{equation}
\langle |A(B_d\to\pi^+\pi^-)|^2\rangle=|\tilde T|^2\left[1-
2d\cos\theta\cos\gamma+d^2\right]
\end{equation}
\begin{equation}\label{ampl-sqr-3}
\langle |A(B_d\to\pi^0\pi^0)|^2\rangle=\frac{|\tilde P|^2}{2}\left[1+
2\left(\frac{x}{d}\right)\cos(\Delta-\theta)\cos\gamma+
\left(\frac{x}{d}\right)^2\right].
\end{equation}
These quantities provide two independent ratios of CP-averaged branching
ratios, which we may choose as those introduced in (\ref{Rpm-def}) 
and (\ref{R00-def}). Using (\ref{ampl-sqr-1})--(\ref{ampl-sqr-3}),
we obtain
\begin{equation}\label{Rpm-expr}
R_{+-}^{\pi\pi}=\frac{1+2x\cos\Delta+x^2}{1-2d\cos\theta\cos\gamma+d^2}
\end{equation}
\begin{equation}\label{R00-expr}
R_{00}^{\pi\pi}=\frac{d^2+2dx\cos(\Delta-\theta)\cos\gamma+
x^2}{1-2d\cos\theta\cos\gamma+d^2}.
\end{equation}

In addition to $R_{+-}^{\pi\pi}$ and $R_{00}^{\pi\pi}$, also the 
time-dependent CP asymmetries of the decays $B_d\to\pi^+\pi^-$ and 
$B_d\to\pi^0\pi^0$ provide valuable information. Thanks to the 
efforts at the $B$ factories, experimental results for the former channel
are already available, which is a decay into a CP-even final state, 
exhibiting the following rate asymmetry \cite{RF-Phys-Rep}:
\begin{eqnarray}
\lefteqn{\frac{\Gamma(B^0_d(t)\to \pi^+\pi^-)-\Gamma(\bar B^0_d(t)\to 
\pi^+\pi^-)}{\Gamma(B^0_d(t)\to \pi^+\pi^-)+\Gamma(\bar B^0_d(t)\to 
\pi^+\pi^-)}}\nonumber\\
&&={\cal A}_{\rm CP}^{\rm dir}(B_d\to \pi^+\pi^-)\cos(\Delta M_d t)+
{\cal A}_{\rm CP}^{\rm mix}(B_d\to \pi^+\pi^-)
\sin(\Delta M_d t).\label{rate-asym}
\end{eqnarray}
Here, $\Delta M_d>0$ is the mass difference of the mass eigenstates
of the $B_d$-meson system, while the CP-violating observables 
\begin{equation}\label{obs}
{\cal A}_{\rm CP}^{\rm dir}(B_d\to \pi^+\pi^-)
\equiv\frac{1-\bigl|\xi_{\pi^+\pi^-}^{(d)}\bigr|^2}{1+
\bigl|\xi_{\pi^+\pi^-}^{(d)}\bigr|^2} \quad \mbox{and} \quad
{\cal A}_{\rm CP}^{\rm mix}(B_d\to \pi^+\pi^-)\equiv
\frac{2\,\mbox{Im}\,\xi^{(d)}_{\pi^+\pi^-}}{1+\bigl|\xi^{(d)}_{\pi^+\pi^-}
\bigr|^2}
\end{equation}
describe ``direct'' and ``mixing-induced'' CP violation, respectively, 
and are governed by 
\begin{equation}\label{xi-Bdpi+pi-}
\xi^{(d)}_{\pi^+\pi^-}=-e^{-i\phi_d}
\left[\frac{e^{-i\gamma}-d e^{i\theta}}{e^{+i\gamma}-d e^{i\theta}}\right].
\end{equation}
The quantity $\phi_d$, which equals $2\beta$ in the SM, is the CP-violating
weak $B^0_d$--$\bar B^0_d$ mixing phase. Consequently, we arrive at the 
following expressions:
\begin{equation}\label{Adir-Bpipi}
{\cal A}_{\rm CP}^{\rm dir}(B_d\to \pi^+\pi^-)
=-\left[\frac{2d\sin\theta\sin\gamma}{1-
2d\cos\theta\cos\gamma+d^2}\right]
\end{equation}
\begin{equation}\label{Amix-Bpipi}
{\cal A}_{\rm CP}^{\rm mix}(B_d\to \pi^+\pi^-)
=\frac{\sin(\phi_d+2\gamma)-2d\cos\theta
\sin(\phi_d+\gamma)+d^2\sin\phi_d}{1-2d\cos\theta\cos\gamma+d^2}.
\end{equation}
The current experimental status of these observables is given by
\begin{equation}\label{Adir-exp}
{\cal A}_{\rm CP}^{\rm dir}(B_d\to\pi^+\pi^-)
=\left\{
\begin{array}{ll}
-0.19\pm0.19\pm0.05 & \mbox{(BaBar \cite{BaBar-Bpipi})}\\
-0.77\pm0.27\pm0.08 & \mbox{(Belle \cite{Belle-Bpipi})}
\end{array}
\right.
\end{equation}
\begin{equation}\label{Amix-exp}
{\cal A}_{\rm CP}^{\rm mix}(B_d\to\pi^+\pi^-)
=\left\{
\begin{array}{ll}
+0.40\pm0.22\pm0.03& \mbox{(BaBar \cite{BaBar-Bpipi})}\\
+1.23\pm0.41 ^{+0.07}_{-0.08} & \mbox{(Belle \cite{Belle-Bpipi}).}
\end{array}
\right.
\end{equation}
Unfortunately, the BaBar and Belle results are not fully consistent with 
each other, although both experiments point towards the same signs, and 
the last BaBar update of ${\cal A}_{\rm CP}^{\rm mix}(B_d\to\pi^+\pi^-)$ 
has moved towards Belle. In \cite{HFAG}, the following averages are
given:
\begin{eqnarray}
{\cal A}_{\rm CP}^{\rm dir}(B_d\to\pi^+\pi^-)&=&
-0.38\pm0.16\label{Bpipi-CP-averages-dir}\\ 
{\cal A}_{\rm CP}^{\rm mix}(B_d\to\pi^+\pi^-)&=&
+0.58\pm0.20.\label{Bpipi-CP-averages-mix}
\end{eqnarray}
As was pointed out in \cite{Fl-Ma,FIM}, the CP asymmetries in 
(\ref{Bpipi-CP-averages-dir}) and (\ref{Bpipi-CP-averages-mix}) point 
towards $\gamma\sim 60^\circ$ for $\phi_d\sim47^\circ$, in accordance 
with the SM. Before discussing the determination of $\gamma$ in more 
detail in Subsection~\ref{ssec:Bpipi-gam-det},\footnote{See also 
Appendix~\ref{app:belle-new}, where a very recent update by the
Belle collaboration \cite{Belle-new} is discussed.} it is useful 
to first have a closer look at the hadronic $B\to\pi\pi$ parameters.

\subsection{Determination of the Hadronic Parameters}\label{ssec:Bpipi-hadr}
\begin{figure}
\vspace*{0.3truecm}
\begin{center}
\psfrag{theta}{$\theta$}\psfrag{d}{$d$}
\psfrag{Adir}{${\cal A}_{\rm CP}^{\rm dir}$} 
\psfrag{Amix}{${\cal A}_{\rm CP}^{\rm mix}$} 
\psfrag{H}{$H$}
\includegraphics[width=10cm]{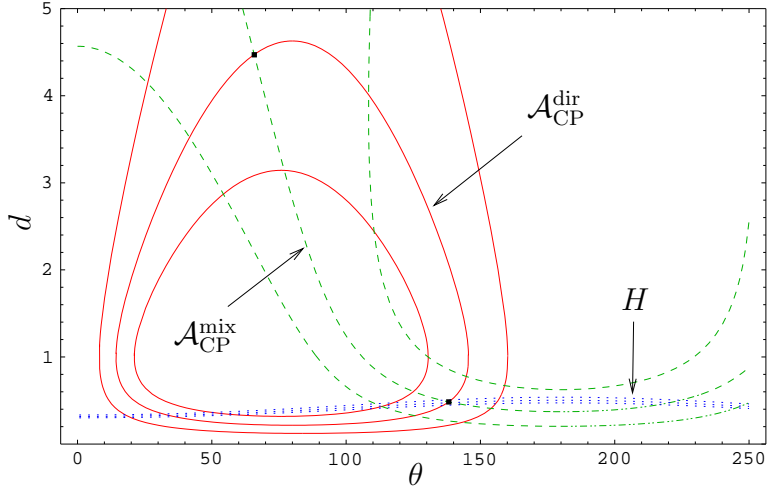}
\end{center}
\caption{The contours in the $\theta$--$d$ plane for $\gamma=65^\circ$ and
$\phi_d = 47^\circ$. The solid lines correspond to the central value 
and 1$\sigma$ upper and lower ranges of 
${\cal A}_{\rm CP}^{\rm dir}(B_d\to\pi^+\pi^-)= -0.38 \pm 0.16$, 
the dashed lines represent ${\cal A}_{\rm CP}^{\rm mix}(B_d\to\pi^+\pi^-) = 
+0.58 \pm 0.20$, and the dotted lines refer to $H = 7.17 \pm 0.75$, as 
discussed in Subsection~\ref{ssec:Bpipi-gam-det}.}\label{fig:theta-d}
\end{figure}

In order to explore the hadronic $B\to\pi\pi$ parameters, we assume that 
\begin{equation}\label{UT-angles}
\gamma=(65\pm 7)^\circ,\quad \phi_d=2\beta=(47\pm4)^\circ, 
\end{equation}
as in the SM \cite{CKM-Book}. If we then look at (\ref{Adir-Bpipi}) and 
(\ref{Amix-Bpipi}), we observe that each of these observables allows us to 
determine $d$ as a function of the strong phase $\theta$. In the case of the 
direct CP asymmetry ${\cal A}_{\rm CP}^{\rm dir}(B_d\to\pi^+\pi^-)$, we obtain
\begin{eqnarray}
\lefteqn{d=\frac{1}{{\cal A}_{\rm CP}^{\rm dir}(B_d\to\pi^+\pi^-)}\Biggl[
{\cal A}_{\rm CP}^{\rm dir}(B_d\to\pi^+\pi^-)\cos\theta\cos\gamma-
\sin\theta\sin\gamma}\nonumber\\
&&\pm\sqrt{\left[{\cal A}_{\rm CP}^{\rm dir}(B_d\to\pi^+\pi^-)
\cos\theta\cos\gamma-\sin\theta\sin\gamma\right]^2-
{\cal A}_{\rm CP}^{\rm dir}(B_d\to\pi^+\pi^-)^2}\Biggr],\label{Adir-cont}
\end{eqnarray}
whereas its mixing-induced counterpart
${\cal A}_{\rm CP}^{\rm mix}(B_d\to\pi^+\pi^-)$ implies
\begin{equation}\label{Amix-cont}
d=k\pm\sqrt{k^2-l},
\end{equation}
with
\begin{equation}
k=\left[\frac{\sin(\phi_d+\gamma)-{\cal A}_{\rm CP}^{\rm mix}(B_d\to\pi^+\pi^-)
\cos\gamma}{\sin\phi_d
-{\cal A}_{\rm CP}^{\rm mix}(B_d\to\pi^+\pi^-)}\right]\cos\theta
\end{equation}
\begin{equation}
l=\frac{\sin(\phi_d+2\gamma)-{\cal A}_{\rm CP}^{\rm mix}
(B_d\to\pi^+\pi^-)}{\sin\phi_d-{\cal A}_{\rm CP}^{\rm mix}(B_d\to\pi^+\pi^-)}.
\end{equation}
In Fig.~\ref{fig:theta-d}, we show the corresponding contours, and 
observe that we obtain a twofold solution for $(d,\theta)$. It should
be emphasized that these contours and the corresponding determination of
$(d,\theta)$ for a given value of $\gamma$ are {\it theoretically clean}.

\begin{figure}
\vspace*{0.3truecm}
\begin{center}
\psfrag{Delta}{$\Delta$}\psfrag{x}{$x$}
\psfrag{R+-}{$R_{+-}^{\pi\pi}$}\psfrag{R00}{$R_{00}^{\pi\pi}$}
\psfrag{physical}{physical}\psfrag{solution}{solution}
\includegraphics[width=10cm]{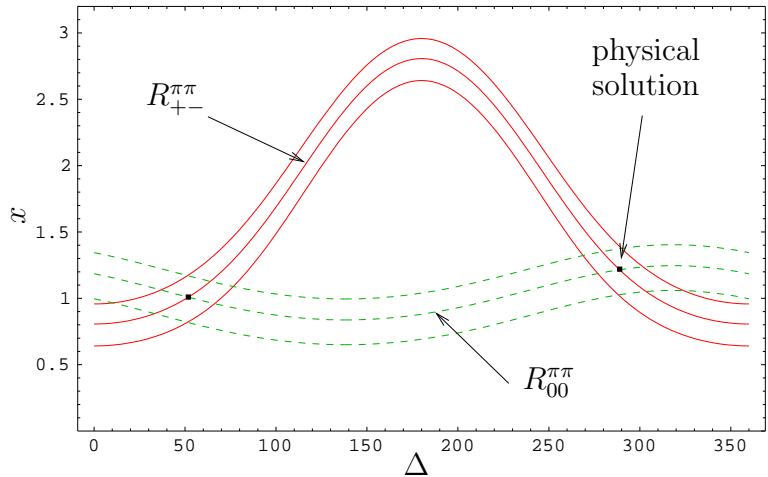}
\end{center}
\caption{The contours in the $\Delta$--$x$ plane. Using $d = 0.48$ and
$\theta = 138^\circ$ following from the central values of 
Fig.~\ref{fig:theta-d}, we obtain the solid set of contours for 
$R_{+-}^{\pi\pi} = 2.12 \pm 0.37$ and the dashed set of contours for 
$R_{00}^{\pi\pi} = 0.83 \pm 0.23$.}\label{fig:Delta-x}
\end{figure}

Let us now also consider the observables $R_{+-}^{\pi\pi}$ and 
$R_{00}^{\pi\pi}$. If we complement them with the CP-violating 
observables ${\cal A}_{\rm CP}^{\rm dir}(B_d\to\pi^+\pi^-)$
and ${\cal A}_{\rm CP}^{\rm mix}(B_d\to\pi^+\pi^-)$, which allow us
to determine $d$ and $\theta$ as we have just seen, we may extract 
the parameters $x$ and $\Delta$ as well. To this end, it is convenient 
to introduce
\begin{equation}
\tilde R_{+-}^{\pi\pi}\equiv D R_{+-}^{\pi\pi}
\end{equation}  
\begin{equation}
\tilde R_{00}^{\pi\pi}\equiv D R_{00}^{\pi\pi},
\end{equation}
with
\begin{equation}
D\equiv 1-2d\cos\theta\cos\gamma+d^2.
\end{equation}
Using then (\ref{Rpm-expr}) and (\ref{R00-expr}), we obtain 
\begin{equation}\label{x-cont1}
x=-\cos\Delta\pm\sqrt{\tilde R_{+-}^{\pi\pi}-\sin^2\Delta}
\end{equation}
and
\begin{equation}\label{x-cont2}
x=-d\cos\gamma\cos(\Delta-\theta)\pm\sqrt{\tilde R_{00}^{\pi\pi}-\left[1-
\cos^2\gamma\cos^2(\Delta-\theta)\right]d^2},
\end{equation}
respectively, allowing us to calculate two contours in the $\Delta$--$x$ 
plane. The intersections of these curves then allow us to extract $x$ 
and $\Delta$. Interestingly, the $(d,\theta)$ solution with $d\sim5$
gives only complex solutions and can therefore by excluded. On the
other hand, in Fig.~\ref{fig:Delta-x}, we show the contours in the 
$\Delta$--$x$ plane corresponding to the physical $d\sim0.5$ case, 
yielding a twofold solution for $(x,\Delta)$.

Using (\ref{UT-angles}) as an input, the expressions in (\ref{Rpm-expr}), 
(\ref{R00-expr}), (\ref{Adir-Bpipi}) and (\ref{Amix-Bpipi}) allow us
to convert the data in (\ref{Rpm-def}), (\ref{R00-def}), 
(\ref{Bpipi-CP-averages-dir}) and (\ref{Bpipi-CP-averages-mix})
into the hadronic parameters $(d,\theta)$ and $(x,\Delta)$. Following
these lines, we obtain
\begin{equation}\label{d-det}
d=0.48^{+0.35}_{-0.22},\quad \theta=+\left(138^{+19}_{-23}\right)^\circ,
\end{equation}
as well as the twofold solution
\begin{equation}\label{x-det}
x=1.22^{+0.26}_{-0.21},\quad \Delta=-\left(71^{+19}_{-26}\right)^\circ,
\end{equation}
\begin{equation}\label{x-det2}
x=1.01^{+0.25}_{-0.19},\quad \Delta=+\left(52^{+24}_{-34}\right)^\circ,
\end{equation}
where our treatment of errors is as described in 
Appendix~\ref{app:error-treatment}.
At this stage, we cannot distinguish between the two solutions for
$xe^{i\Delta}$. However, as we will show in Subsection~\ref{ssec:x-elim},
the solution in (\ref{x-det2}) can be {\it excluded} through the $B\to\pi K$ 
data, as it would correspond to large direct CP violation in 
$B^\pm\to \pi^0K^\pm$, which is ruled out by experiment. For the time being, 
we will hence focus on (\ref{x-det}). Let us finally note that the 
determination of the hadronic parameters given in 
(\ref{d-det})--(\ref{x-det2}) is essentially theoretically clean, and 
that the experimental picture will improve significantly in the future.

\subsection{Interpretation of the Hadronic 
Parameters}\label{ssec:Bpipi-hadr-int}
The result of $x={\cal O}(1)$ obtained above, which implies that 
$|\tilde C|\sim |\tilde T|$, is in stark contrast to the na\"\i ve 
expectation of $|\tilde C|\sim 0.25\times|\tilde T|$. At first sight, 
this feature seems to imply a complete breakdown of the concept of 
colour suppression in $B\to\pi\pi$ decays. However, one should realize 
that the usual arguments leading to $|\tilde C|\sim 0.25\times|\tilde T|$ 
neglect the contributions $({\cal P}_{tu}-{\cal E})$ in (\ref{T-tilde}) 
and (\ref{C-tilde}), which may be important \cite{PAP0}. In 
fact, our analysis suggests that these contributions cannot be neglected 
if we want to understand the pattern of the $B\to\pi\pi$ observables
in a plausible manner. Because of the different signs in (\ref{T-tilde}) 
and (\ref{C-tilde}), we may explain the surprisingly small 
$B_d\to\pi^+\pi^-$ branching ratio naturally through {\it destructive} 
interference between the ${\cal T}$ and $({\cal P}_{tu}-{\cal E})$ 
amplitudes, whereas the puzzling large $B_d\to\pi^0\pi^0$ branching 
ratio originates from {\it constructive} interference between the 
${\cal C}$ and $({\cal P}_{tu}-{\cal E})$ amplitudes. 

On the other hand, we have $\left.\theta\right|_{\rm fact}=180^\circ$. 
Consequently, as can be seen in (\ref{ampl-Bdpi+pi-}), $B_d\to\pi^+\pi^-$ 
would favour -- in contrast to the SM expectation -- $\gamma>90^\circ$
within the factorization approach, since $\mbox{BR}(B_d\to\pi^+\pi^-)$ 
would then be reduced through destructive interference between trees and 
penguins. In contrast, we arrive at a picture with large non-factorizable
contributions, exhibiting certain interference effects at the hadronic 
level that allow us to accommodate straightforwardly {\it any} currently 
observed feature of the $B\to\pi\pi$ system in the SM.

\boldmath
\subsubsection{Insights into the Substructure of $xe^{i\Delta}$}
\begin{figure}
\vspace*{0.3truecm}
\begin{center}
\psfrag{z}{$\zeta$}\psfrag{Dz}{$\Delta_\zeta$}
\psfrag{a=.2}{$a_2^{\pi\pi}=0.2$}\psfrag{a=.5}{$a_2^{\pi\pi}=0.5$}
\psfrag{d2=0}{$\Delta_2^{\pi\pi}\!=\!0^\circ$}\psfrag{d2=90}{$90^\circ$}
\psfrag{d2=180}{$180^\circ$}\psfrag{d2=270}{$270^\circ$}
\includegraphics[width=10cm]{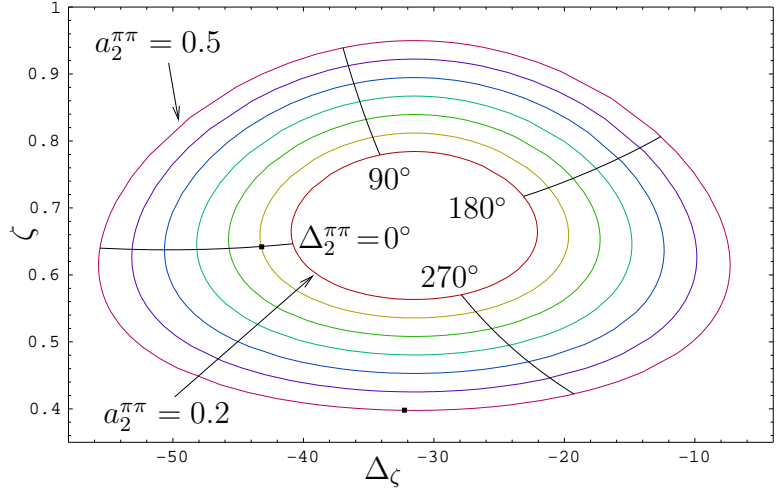}
\end{center}
\caption{The contours in the $\Delta_\zeta$--$\zeta$ plane 
corresponding to the central values of $xe^{i\Delta}$ in (\ref{x-det})
for various values of $a_2^{\pi\pi}$ between 0.2 and 0.5 and 
$\Delta_2^{\pi\pi}\in[0^\circ,360^\circ]$. The two dots refer to
cases discussed in the text.\label{fig:zeta}}
\end{figure}

\unboldmath\label{ssec:x-Del-struct}
The calculation of the hadronic parameter $xe^{i\Delta}$ from 
first principles is extremely challenging and cannot be done 
in a reliable manner because of our limited knowledge of 
non-perturbative strong-interaction physics. However, in order 
to obtain deeper insights into the substructure of $xe^{i\Delta}$, 
it is instructive to write
\begin{equation}
x e^{i\Delta}=\frac{a_2^{\pi\pi} e^{i\Delta_2^{\pi\pi}}+
\zeta e^{i\Delta_\zeta}}{1-
\zeta e^{i\Delta_\zeta}},
\end{equation}
with
\begin{equation}\label{a2-def}
a_2^{\pi\pi} e^{i\Delta_2^{\pi\pi}}\equiv\frac{{\cal C}}{{\cal T}}
\end{equation}
\begin{equation}\label{zeta-def}
\zeta e^{i\Delta_\zeta}\equiv \frac{{\cal P}_{tu}-{\cal E}}{{\cal T}}.
\end{equation}
Consequently, we may convert $xe^{i\Delta}$ into the complex quantity
$\zeta e^{i\Delta_\zeta}$ with the help of 
\begin{equation}
\zeta e^{i\Delta_\zeta}
=\frac{x^2+x\left[e^{i\Delta}-a_2^{\pi\pi}e^{i(\Delta_2^{\pi\pi}-
\Delta)}\right]-a_2^{\pi\pi} e^{i\Delta_2^{\pi\pi}}}{1+2x\cos\Delta+x^2}.
\end{equation}
In the special case of $a_2^{\pi\pi} e^{i\Delta_2^{\pi\pi}}=xe^{i\Delta}$, 
this expression reduces to $\zeta e^{i\Delta_\zeta}=0$. However, a value of 
$a_2^{\pi\pi} e^{i\Delta_2^{\pi\pi}}\sim 1.22 \times e^{-i71^\circ}$ would 
appear as completely unrealistic. On the other hand, in view of the large 
non-factorizable effects exhibited by (\ref{d-det}) and (\ref{x-det}), 
we think that $a_2^{\pi\pi}$ may well take values as large as 0.5, with a 
large strong phase $\Delta_2^{\pi\pi}$. In Fig.~\ref{fig:zeta}, we 
consider the central values of $x$ and $\Delta$ in (\ref{x-det}), and 
show the contours in the $\Delta_\zeta$--$\zeta$ plane corresponding to 
different values of $a_2^{\pi\pi}$, where each point is parametrized by 
the value of $\Delta_2^{\pi\pi}\in[0^\circ,360^\circ]$. We observe that the
``na\"\i ve'' value of $a_2^{\pi\pi}e^{i\Delta_2^{\pi\pi}}\sim0.25$ would 
result in a rather large value of $\zeta\sim 0.65$. On the other 
hand, $a_2^{\pi\pi}e^{i\Delta_2^{\pi\pi}}\sim0.5\times e^{i290^\circ}$ gives
a significantly smaller $\zeta\sim0.4$, which may well originate from 
{\it constructive} interference between the ${\cal P}_t$ and ${\cal P}_u$ 
amplitudes, satisfying, for instance, 
$|{\cal P}_t/{\cal T}|\sim|{\cal P}_u/{\cal T}|\sim0.25$. Moreover, for
such values of $a_2^{\pi\pi}e^{i\Delta_2^{\pi\pi}}$, we have
$|1+a_2^{\pi\pi}e^{i\Delta_2^{\pi\pi}}|\sim1.25$, in accordance with
the ``na\"\i ve'' expectation. Consequently, in the $B^\pm\to\pi^\pm\pi^0$ 
channel, which is measured in agreement with the conventional theoretical 
expectations, not only the hadronic interference effects originating from 
the $({\cal P}_{tu}-{\cal E})$ amplitude cancel, but also the large 
non-factorizable effects in $a_2^{\pi\pi}e^{i\Delta_2^{\pi\pi}}$
would not manifest themselves.

\begin{figure}
\vspace*{0.3truecm}
\begin{center}
\psfrag{zt}{$\tilde\zeta$}\psfrag{Dzt}{$\Delta_{\tilde \zeta}$}
\psfrag{a=.2}{$a_2^{\pi\pi}=0.2$}\psfrag{a=.5}{$a_2^{\pi\pi}=0.5$}
\psfrag{d2=0}{$\Delta_2^{\pi\pi}\!=\!0^\circ$}\psfrag{d2=90}{$90^\circ$}
\psfrag{d2=180}{$180^\circ$}\psfrag{d2=270}{$270^\circ$}
\includegraphics[width=10cm]{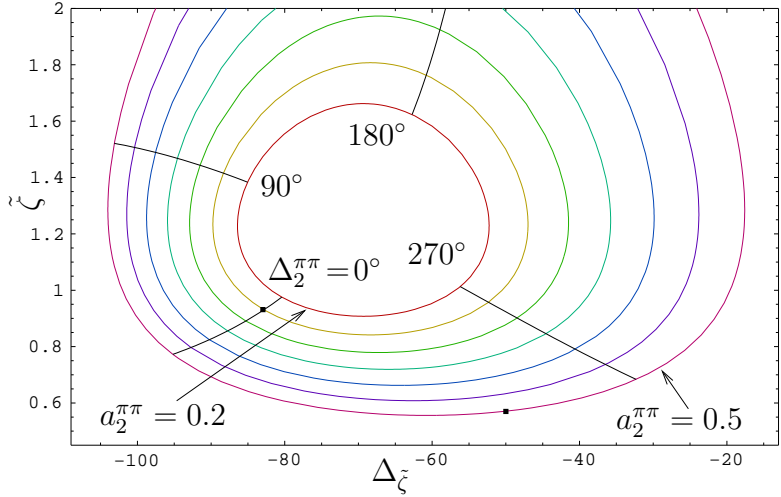}
\end{center}
\caption{The contours in the $\Delta_{\tilde\zeta}$--$\tilde\zeta$ plane 
corresponding to the central values of $xe^{i\Delta}$ in (\ref{x-det})
for various values of $a_2^{\pi\pi}$ between 0.2 and 0.5 and 
$\Delta_2^{\pi\pi}\in[0^\circ,360^\circ]$. \label{fig:zetatilde}}
\end{figure}

We shall come back to these considerations in Subsection~\ref{ssec:rho_c-th},
where the quantity 
\begin{equation}\label{zetatilde-def}
\tilde\zeta e^{i\Delta_{\tilde\zeta}}\equiv\frac{\zeta e^{i\Delta_\zeta}}{1-
\zeta e^{i\Delta_\zeta}}=\frac{{\cal P}_{tu}-{\cal E}}{\tilde {\cal T}}
\end{equation}
will play an important r\^ole in the context of a hadronic parameter
$\rho_{\rm c}$ entering the charged $B\to\pi K$ decays. In 
Fig.~\ref{fig:zetatilde}, we show -- in analogy to Fig.~\ref{fig:zeta} --
the contour plots in the $\Delta_{\tilde\zeta}$--$\tilde\zeta$ 
plane for given values of $a_2^{\pi\pi}$ and 
$\Delta_2^{\pi\pi}\in[0^\circ,360^\circ]$. Interestingly, the analysis
of $\rho_{\rm c}$ will also point towards
$a_2^{\pi\pi}e^{i\Delta_2^{\pi\pi}}\sim0.5\times e^{i290^\circ}$, thereby
complementing the picture described above.

Let us finally note that the hadronic parameters in (\ref{d-det}) and 
(\ref{x-det}) also imply
\begin{equation}
\left[\frac{P}{T+C}\right]e^{i\gamma}=
-\left[\frac{de^{i\theta}}{1+xe^{i\Delta}}\right]=\frac{1}{R_b}
\left[\frac{{\cal P}_{tc}}{{\cal T}+{\cal C}}\right]=
\left(0.27^{+0.13}_{-0.10}\right)\times e^{i(-2^{+18}_{-23})^\circ},
\end{equation}
yielding
\begin{equation}\label{Ptc-det}
\frac{{\cal P}_{t}-{\cal P}_{c}}{{\cal T}+{\cal C}}=
\left(0.10^{+0.05}_{-0.04}\right)\times e^{i(-2^{+18}_{-23})^\circ},
\end{equation}
where we have used the value of $R_b$ in (\ref{CKM1}). Consequently, 
the small numerical value in (\ref{Ptc-det}) would then require 
{\it destructive} interference between the ${\cal P}_t$ and ${\cal P}_c$ 
amplitudes, where ${\cal P}_c$ contains ``charming'' penguins 
\cite{c-pen}, in contrast to the constructive interference between 
${\cal P}_t$ and ${\cal P}_u$. Interesting insights into these subtle 
dynamical issues can be obtained with the help of the penguin-induced decay 
$B_d\to K^0\bar K^0$ \cite{RF-BdK0K0}, which can nicely be 
complemented with the $B_s\to K^0\bar K^0$ channel \cite{RF-Phys-Rep}.

\boldmath
\subsubsection{Probing Penguin Annihilation and Exchange Topologies through
the $B_d\to K^+K^-$, $B_s\to\pi^+\pi^-$ System}\label{ssec:PA-E}
\unboldmath
The origin of the value of $xe^{i\Delta}$ in (\ref{x-det}) could also be 
due to anomalously enhanced penguin annihilation and exchange topologies 
(which contribute to the $B\to\pi\pi$ modes but not to their $B\to\pi K$ 
counterparts discussed in Section~\ref{sec:BpiK}). However,
the importance of these topologies can be probed through the 
decay $B_d\to K^+K^-$, where the current experimental upper bound 
is given as follows \cite{HFAG}:
\begin{equation}
\mbox{BR}(B_d\to K^+K^-)< 0.6\times 10^{-6} \, \mbox{(90\% C.L.)}.
\end{equation}
The corresponding decay amplitude has the same structure as 
(\ref{ampl-Bdpi+pi-}), with the important difference that now {\it only} 
penguin annihilation and exchange 
topologies contribute. Employing the $SU(3)$ flavour symmetry of strong
interactions, we may hence write
\begin{equation}
A(B^0_d\to K^+K^-)=-\lambda^3AR_b\left[{\cal E}-({\cal PA})_{tu}\right]
\left[e^{i\gamma}+\varrho_{\cal PA}e^{i\vartheta_{\cal PA}}\right],
\end{equation}
with
\begin{equation}
\varrho_{\cal PA}e^{i\vartheta_{\cal PA}}\equiv
\frac{1}{R_b}\left[\frac{({\cal PA})_{tc}}{{\cal E}-({\cal PA})_{tu}}\right],
\end{equation}
where the penguin annihilation amplitudes $({\cal PA})_{tu}$ 
and $({\cal PA})_{tc}$ are implicitly included in the $B\to\pi\pi$
amplitudes ${\cal P}_{tu}$ and ${\cal P}_{tc}$, respectively. 
Consequently, we obtain 
\begin{equation}\label{BdKK-bound}
\sqrt{\frac{1}{2}\left[\frac{\mbox{BR}(B_d\to 
K^+K^-)}{\mbox{BR}(B^\pm\to\pi^\pm\pi^0)}\right]
\frac{\tau_{B^+}}{\tau_{B^0_d}}}\approx
\left|\frac{{\cal E}-({\cal PA})_{tu}}{{\cal T}+{\cal C}}\right|
\sqrt{1+2\varrho_{\cal PA}\cos\vartheta_{\cal PA}\cos\gamma+
\varrho_{\cal PA}^2}\lsim 0.2,
\end{equation}
which does not indicate any anomalous behaviour. 
In the future, this bound can be improved significantly. The decay 
$B_d\to K^+K^-$ is very accessible at LHCb, where it may be possible 
to reach the $10^{-8}$ level for its CP-averaged branching ratio 
\cite{wilkinson}. At this experiment, we may also exploit the physics
potential of the $B_s^0\to\pi^+\pi^-$ channel, which is the $U$-spin
partner of the $B^0_d\to K^+K^-$ mode,\footnote{The $U$-spin flavour symmetry
of strong interactions connects the down and strange quarks in the same 
manner as the ordinary $SU(2)$ isospin flavour symmetry connects the down 
and up quarks.} and has a decay amplitude
of the same structure as the $B^0_s\to K^+K^-$ mode \cite{RF-BsKK} 
(or the $B^0_d\to\pi^-K^+$ transition discussed in 
Section~\ref{sec:BpiK}).
If we then use the $U$-spin flavour symmetry and consider the 
CP-averaged branching ratios, we obtain
\begin{equation}\label{HPA-def}
H_{\cal PA}\equiv\frac{1}{\epsilon}\left[
\frac{\mbox{BR}(B_d\to K^+K^-)}{\mbox{BR}(B_s\to\pi^+\pi^-)}\right]
\frac{\tau_{B^0_s}}{\tau_{B^0_d}}
=\frac{1+2\varrho_{\cal PA}\cos\vartheta_{\cal PA}\cos\gamma+
\varrho_{\cal PA}^2}{\epsilon^2-2\epsilon 
\varrho_{\cal PA}\cos\vartheta_{\cal PA}\cos\gamma+
\varrho_{\cal PA}^2},
\end{equation}
where we neglect tiny phase-space effects, and
\begin{equation}\label{epsi}
\epsilon\equiv\frac{\lambda^2}{1-\lambda^2}=0.053.
\end{equation}
The quantity $H_{\cal PA}$ allows us to obtain constraints for the parameter
$\varrho_{\cal PA}$. Since the amplitude structure of the
$B_d\to K^+K^-$, $B_s\to \pi^+\pi^-$ system is analogous to that of
the $B_d\to \pi^+\pi^-$, $B_s\to K^+K^-$ ($B_d\to\pi^\mp K^\pm$) system,
we may apply the bounds derived in \cite{RF-Bpipi}, yielding
\begin{equation}\label{PA-E-bounds}
\frac{1-\epsilon\sqrt{H_{\cal PA}}}{1+\sqrt{H_{\cal PA}}} \leq 
\varrho_{\cal PA}\leq \frac{1+\epsilon\sqrt{H_{\cal PA}}}{|1-
\sqrt{H_{\cal PA}}|}.
\end{equation}
In complete analogy to the $B_d\to\pi^+\pi^-$, $B_s\to K^+K^-$ strategy
proposed in \cite{RF-BsKK}, a measurement of the CP-violating
$B_d\to K^+K^-$, $B_s\to \pi^+\pi^-$ observables would allow a determination 
of both $\gamma$ and the hadronic parameters $\varrho_{\cal PA}$ 
and $\vartheta_{\cal PA}$. However, the main interest in an analysis of the 
$B_d\to K^+K^-$, $B_s\to \pi^+\pi^-$ system is obviously to obtain insights 
into the importance of penguin annihilation and exchange topologies. Using
the corresponding information on $\varrho_{\cal PA}$ and 
$\vartheta_{\cal PA}$, the expression in (\ref{BdKK-bound}) allows us to 
determine $|({\cal E}-({\cal PA})_{tu})/({\cal T}+{\cal C})|$, while
\begin{equation}
\sqrt{\frac{\epsilon}{2}
\left[\frac{\mbox{BR}(B_s\to \pi^+\pi^-)}{\mbox{BR}(B^\pm\to\pi^\pm\pi^0)}
\right]\frac{\tau_{B^+}}{\tau_{B^0_s}}}\approx
\frac{1}{R_b}\left|\frac{({\cal PA})_{tc}}{{\cal T}+{\cal C}}\right|.
\end{equation}
It will be very interesting to confront these considerations with experimental
data for the $B_s\to\pi^+\pi^-$ channel. First constraints may soon be 
available from run II of the Tevatron, which should be considerably 
improved at LHCb.

\boldmath
\subsection{Determination of $\gamma$}\label{ssec:Bpipi-gam-det}
\unboldmath
Let us now come back to Fig.~\ref{fig:theta-d}. There we have also included 
a contour, which is related to the following quantity \cite{RF-Bpipi}:
\begin{eqnarray}\label{H-expr}
H&\equiv&\frac{1}{\epsilon}\left|\frac{{\cal C}'}{{\cal C}}\right|^2
\left[\frac{M_{B_d}}{M_{B_s}}
\frac{\Phi(M_K/M_{B_s},M_K/M_{B_s})}{\Phi(M_\pi/M_{B_d},M_\pi/M_{B_d})}
\frac{\tau_{B_s^0}}{\tau_{B_d^0}}\right]
\left[\frac{\mbox{BR}(B_d\to\pi^+\pi^-)}{\mbox{BR}(B_s\to K^+K^-)}
\right]\nonumber\\
&\stackrel{U \, {\rm spin}}{=}&
\frac{1-2d\cos\theta\cos\gamma+d^2}{\epsilon^2+
2\epsilon d\cos\theta\cos\gamma+d^2},
\end{eqnarray}
where $|{\cal C}'/{\cal C}|$ is a $U$-spin-breaking parameter, and   
\begin{equation}
\Phi(x,y)\equiv\sqrt{\left[1-(x+y)^2\right]\left[1-(x-y)^2\right]}
\end{equation}
denotes the usual two-body phase-space function. The introduction and 
the use of $H$, which has the same structure as (\ref{HPA-def}), is 
inspired by a variant of the strategy presented in \cite{RF-BsKK}, 
allowing us to determine $\gamma$ and the hadronic parameters 
$(d,\theta)$ by relating the CP-violating $B_d\to\pi^+\pi^-$ observables 
to their $B_s\to K^+K^-$ counterparts through the $U$-spin flavour 
symmetry of strong interactions (see also 
Subsection~\ref{ssec:BsKK}). This strategy is very promising for $B$-decay 
experiments at hadron colliders. First important steps are already expected 
at run II of the Tevatron \cite{TEV-BOOK}, whereas it can be fully exploited in
the era of the LHC \cite{LHC-BOOK}, in particular at LHCb, where experimental
uncertainties of $\gamma$ at the few-degree level may be achieved 
\cite{LHCb}. Unfortunately, the physics potential of $B_s\to K^+K^-$
cannot yet be exploited. However, if we assume that the penguin annihilation 
and exchange topologies discussed in Subsection~\ref{ssec:PA-E} play a 
minor r\^ole and employ $SU(3)$ flavour-symmetry arguments, we may replace 
$B_s\to K^+K^-$ through $B_d\to\pi^\mp K^\pm$, and arrive 
straightforwardly at 
\begin{equation}\label{H-simple}
H=\frac{1}{\epsilon}\left(\frac{f_K}{f_\pi}\right)^2\left[
\frac{\mbox{BR}(B_d\to\pi^+\pi^-)}{\mbox{BR}(B_d\to\pi^\mp K^\pm)}\right]
=7.17\pm0.75,
\end{equation}
where the ratio $f_K/f_{\pi}=160/131$ describes factorizable $SU(3)$-breaking 
corrections \cite{RF-Bpipi}.  In fact, 
the utility of $B_d\to\pi^\mp K^\pm$ transitions to address the ``penguin 
problem'' in $B_d\to\pi^+\pi^-$ was already pointed out about ten years 
ago \cite{si-wo}. If we now use the expression for $H$ in terms of the 
hadronic parameters given in (\ref{H-expr}), we obtain \cite{RF-Bpipi}:
\begin{equation}\label{H-cont}
d=-b\cos\theta\cos\gamma\pm\sqrt{a+(b\cos\theta\cos\gamma)^2},
\end{equation}
with
\begin{equation}
a=\frac{1-\epsilon^2 H}{H-1}, \quad 
b=\frac{1+\epsilon H}{H-1},
\end{equation}
which allows us to calculate $d$ as a function of $\theta$ for $\gamma$ and 
$H$ as given in (\ref{UT-angles}) and (\ref{H-simple}), respectively. 
The resulting contour is shown in Fig.~\ref{fig:theta-d} and fits 
perfectly into the picture of the {\it theoretically clean} $\theta$--$d$ 
contours following from the CP-violating $B_d\to\pi^+\pi^-$ observables. The 
fact that the $H$ contour described by (\ref{H-cont}) goes right through the 
intersection of (\ref{Adir-cont}) and (\ref{Amix-cont}) is very remarkable and 
indicates that the penguin annihilation and exchange amplitudes as well as 
non-factorizable $SU(3)$-breaking effects actually play a minor r\^ole. 
Otherwise, the corresponding hadronic uncertainties would have to conspire 
in a very contrived way, which does not look plausible to us. 

If we do not fix $\gamma$ through the UT fits, but use instead $H$, 
as discussed in \cite{RF-BsKK}--\cite{FIM}, we may {\it determine} this 
angle, yielding the following twofold solution:
\begin{equation}\label{gam-H}
\gamma=(40.5^{+5.4}_{-6.1})^\circ  \, \lor \, 
(64.7^{+6.3}_{-6.9})^\circ.
\end{equation}
The errors are obtained with the help of a standard $\chi^2$ analysis from 
the errors of ${\cal A}_{\rm CP}^{\rm dir}(B_d\to\pi^+\pi^-)$,
${\cal A}_{\rm CP}^{\rm mix}(B_d\to\pi^+\pi^-)$ and $H$ by fitting to the
hadronic parameters and $\gamma$. In addition, there is an error of
$\pm 2.5^\circ$ on $\gamma$ from the uncertainty in $\phi_d$.
A detailed analysis of the whole $B\to\pi K$ system, as discussed in 
Subsection~\ref{ssec:gam-refine}, allows us to lift this degeneracy, 
leaving us essentially with the solution around $65^\circ$, which 
agrees perfectly with the range for $\gamma$ given in (\ref{UT-angles}). 

An interesting alternative to determine $\gamma$ is provided by the
direct CP asymmetry
\begin{equation}\label{ACPdir-exp}
{\cal A}_{\rm CP}^{\rm dir}(B_d\to\pi^\mp K^\pm)\equiv
\frac{\mbox{BR}(B^0_d\to\pi^-K^+)-
\mbox{BR}(\bar B^0_d\to\pi^+K^-)}{\mbox{BR}(B^0_d\to\pi^-K^+)+
\mbox{BR}(\bar B^0_d\to\pi^+K^-)}=+0.095\pm 0.028.
\end{equation}
If we employ again the $SU(3)$ flavour symmetry and the plausible dynamical
arguments specified above, we may express this observable in terms of 
the hadronic parameters introduced in (\ref{d-theta-def}) as follows:
\begin{equation}\label{ACPdir}
{\cal A}_{\rm CP}^{\rm dir}(B_d\to\pi^\mp K^\pm)=\frac{2\epsilon d
\sin\theta\sin\gamma}{\epsilon^2+2\epsilon d\cos\theta\cos\gamma+d^2}.
\end{equation}
In contrast to expression (\ref{H-simple}) for $H$, (\ref{ACPdir}) does
not involve a $(f_K/f_\pi)^2$ term and is hence affected to a smaller 
extent by $SU(3)$-breaking corrections, i.e.\ is more favourable from 
a theoretical point of view. On the other hand, as can be seen in 
(\ref{ACPdir-exp}), a non-vanishing value of
${\cal A}_{\rm CP}^{\rm dir}(B_d\to\pi^\mp K^\pm)$ has not yet been
established by the $B$ factories, although the experimental evidence 
for this CP asymmetry is steadily increasing. 

In this context, it is important to note that we have the following 
relation \cite{RF-BsKK,RF-Bpipi}:
\begin{equation}\label{CP-rel}
\underbrace{-\left[\frac{{\cal A}_{\rm CP}^{\rm dir}(B_d\to\pi^\mp 
K^\pm)}{{\cal A}_{\rm CP}^{\rm dir}(B_d\to\pi^+\pi^-)}\right]}_{0.25\pm0.13}
=\underbrace{\left(\frac{f_K}{f_\pi}\right)^2\left[\frac{\mbox{BR}
(B_d\to\pi^+\pi^-)}{\mbox{BR}(B_d\to\pi^\mp K^\pm)}\right]}_{0.38\pm0.04}=
\epsilon H,
\end{equation}
where we have also indicated the current experimental results. Within
the uncertainties, this relation is satisfied by the data. In particular,
also the pattern of the signs of the direct CP asymmetries is in 
accordance with (\ref{CP-rel}). Should the central values be correct,
they would correspond to non-factorizable $SU(3)$-breaking corrections 
at the $B\to\pi\pi,\pi K$ amplitude level of ${\cal O}(20\%)$. On the 
other hand, the current central value for 
${\cal A}_{\rm CP}^{\rm dir}(B_d\to\pi^\mp K^\pm)$ may also be on the
lower side, which would be our preferred scenario. It will be exciting 
to observe how the data evolve in the future. 

In our analysis, we have assumed that $\phi_d\sim47^\circ$, in agreement 
with the SM. However, as discussed in \cite{Fl-Ma,FIM}, it is interesting 
to consider also the second, unconventional solution of $\phi_d\sim133^\circ$. 
If $\phi_d$, $\gamma$, $d$, $\theta$, $x$, $\Delta$ are solutions of 
(\ref{Rpm-expr}), (\ref{R00-expr}), (\ref{Adir-Bpipi}), (\ref{Amix-Bpipi}), 
(\ref{H-expr}), (\ref{ACPdir}), then  
\begin{equation}\label{phid-rel}
\pi-\phi_d,\quad \pi-\gamma,\quad d,\quad \pi-\theta,\quad x,\quad
-\Delta
\end{equation}
are solutions as well. Consequently, (\ref{phid-rel}) allows us to go 
easily from the $\phi_d\sim47^\circ$ to the $\phi_d\sim133^\circ$ case. 
Interestingly, for the value of $\theta$ in (\ref{d-det}), we obtain 
$\cos\theta\sim-0.7<0$, exhibiting the same sign as within factorization, 
where $\left.\theta\right|_{\rm fact}=180^\circ$. On the other hand, the
value of $\theta$ corresponding to $\phi_d\sim133^\circ$ yields
$\cos\theta\sim+0.7>0$, i.e.\ the opposite sign, thereby disfavouring
the $\phi_d\sim133^\circ$ solution. Moreover, also the experimental sign 
information about CP violation in $B_d\to D^{(\ast)\pm}\pi^\mp$ modes, 
obtained recently by the BaBar collaboration, points towards 
$\phi_d\sim47^\circ$ \cite{RF-BdDpi}. Unfortunately, the Belle collaboration
favours the opposite sign \cite{HFAG}. 

We shall return to ${\cal A}_{\rm CP}^{\rm dir}(B_d\to\pi^\mp K^\pm)$
in Subsection~\ref{ssec:BpiK-mixed}, and to the determination of $\gamma$ 
in Subsection~\ref{ssec:gam-refine}, where we shall also address the
other two angles, $\alpha$ and $\beta$, of the UT. For the following 
numerical analyses, we will continue to use $\gamma$ as given in 
(\ref{UT-angles}).

\begin{figure}
\vspace*{0.3truecm}
\begin{center}
\psfrag{gamma}{$\gamma$}
\psfrag{Adir}{${\cal A}_{\rm CP}^{\rm dir}(B_d\to\pi^0\pi^0)$}
\includegraphics[width=9cm]{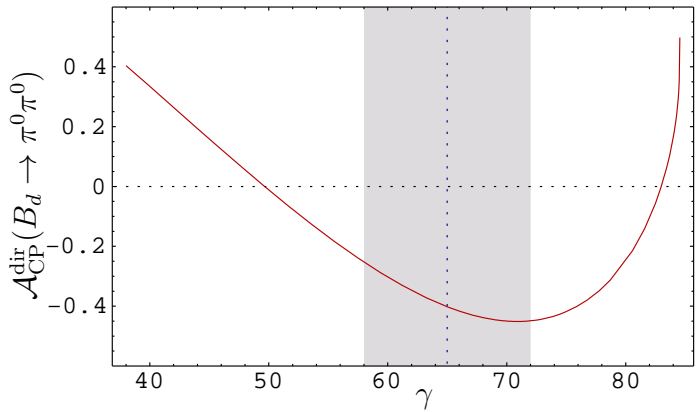}
\end{center}
\caption{The dependence of ${\cal A}_{\rm CP}^{\rm dir}(B_d\to\pi^0\pi^0)$
on $\gamma$, as described in the text.}\label{fig:Bdpi0pi0-1}
\end{figure}

\boldmath
\subsection{Prediction of CP Violation in $B_d\to\pi^0\pi^0$}
\label{ssec:Bpi0pi0-CPV}
\unboldmath
Let us consider, as the last element of our $B\to\pi\pi$ analysis, the 
CP-violating observables of the decay $B_d\to\pi^0\pi^0$, which is -- 
in analogy to $B_d\to\pi^+\pi^-$ -- another transition into a CP-even 
final state. If we apply the standard formalism as discussed for
$B_d\to\pi^+\pi^-$ in (\ref{rate-asym})--(\ref{xi-Bdpi+pi-}), we obtain
\begin{equation}\label{Adir-Bpipi0}
{\cal A}_{\rm CP}^{\rm dir}(B_d\to\pi^0\pi^0)=\frac{2dx\sin(\theta-\Delta)
\sin\gamma}{d^2+2dx\cos(\theta-\Delta)\cos\gamma+x^2}
\end{equation}
\begin{equation}\label{Amix-Bpipi0}
{\cal A}_{\rm CP}^{\rm mix}(B_d\to\pi^0\pi^0)=\frac{d^2\sin\phi_d+
2dx\cos(\theta-\Delta)\sin(\phi_d+\gamma)+x^2\sin(\phi_d+2\gamma)}{d^2+
2dx\cos(\theta-\Delta)\cos\gamma+x^2}.
\end{equation}
Complementing these formulae with the analysis outlined above, we may
{\it predict} these CP-violating observables, with the following SM
result:
\begin{eqnarray}
\left.{\cal A}_{\rm CP}^{\rm dir}(B_d\to\pi^0\pi^0)\right|_{\rm SM}&=&
-0.41^{+0.35}_{-0.17}\label{detpipi-dir}\\
\left.{\cal A}_{\rm CP}^{\rm mix}(B_d\to\pi^0\pi^0)\right|_{\rm SM}&=&
-0.55^{+0.43}_{-0.45},\label{detpipi-mix}
\end{eqnarray}
where the errors come from the procedure described in 
Appendix~\ref{app:error-treatment}.
In the future, when the errors of our input quantities will decrease, 
more accurate predictions for these quantities will be possible. 

\begin{figure}
\vspace*{0.3truecm}
\begin{center}
\psfrag{gamma}{$\gamma$}
\psfrag{Amix}{${\cal A}_{\rm CP}^{\rm mix}(B_d\to\pi^0\pi^0)$}
\includegraphics[width=9cm]{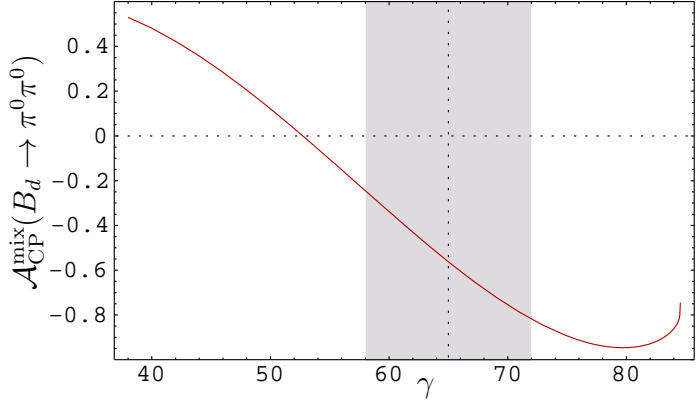}
\end{center}
\caption{The dependence of ${\cal A}_{\rm CP}^{\rm mix}(B_d\to\pi^0\pi^0)$
on $\gamma$, as described in the text.}\label{fig:Bdpi0pi0-2}
\end{figure}

It is interesting to note
that a measurement of one of the CP asymmetries of the $B_d\to\pi^0\pi^0$
mode would allow an essentially clean determination of $\gamma$. In
Figs.~\ref{fig:Bdpi0pi0-1} and \ref{fig:Bdpi0pi0-2}, we illustrate
the dependences of ${\cal A}_{\rm CP}^{\rm dir}(B_d\to\pi^0\pi^0)$
and ${\cal A}_{\rm CP}^{\rm mix}(B_d\to\pi^0\pi^0)$ on this angle,
respectively, where we consider the central experimental values 
of $R_{+-}^{\pi\pi}$, $R_{00}^{\pi\pi}$, 
${\cal A}_{\rm CP}^{\rm dir}(B_d\to\pi^+\pi^-)$ and
${\cal A}_{\rm CP}^{\rm mix}(B_d\to\pi^+\pi^-)$ for 
simplicity.\footnote{We suppress also certain unphysical branches in these
figures, which are related to unphysical solutions for the hadronic
parameters $d$, $\theta$, $x$ and $\Delta$.}
We observe that in particular 
${\cal A}_{\rm CP}^{\rm mix}(B_d\to\pi^0\pi^0)$ -- being more 
sensitive to $\gamma$ than ${\cal A}_{\rm CP}^{\rm dir}(B_d\to\pi^0\pi^0)$ --
would allow a useful determination of this angle, which would also be
unambiguous in the range of $\gamma$ considered.

\boldmath
\subsection{EW Penguin Contributions}
\label{ssec:Bpipi-EWP}
\unboldmath
Let us finally come back to the EW penguin contributions to the
$B\to\pi\pi$ modes, which we have neglected so far. Using the results 
of \cite{BF-neutral1,GPY}, we may take them into account with the help 
of the isospin symmetry of strong interactions. If we consider the SM
and absorb the colour-suppressed EW penguins in the amplitude $P$, 
(\ref{B+pi+pi0})--(\ref{Bdpi0pi0}) are modified as follows:
\begin{eqnarray}
\sqrt{2}A(B^+\to\pi^+\pi^0)&=&-|\tilde T|e^{i\delta_{\tilde T}}
\left[1+x e^{i\Delta}\right]\left[e^{i\gamma}+\tilde qe^{-i\beta}\right]\\
A(B^0_d\to\pi^+\pi^-)&=&-|\tilde T|e^{i\delta_{\tilde T}}
\left[e^{i\gamma}-d e^{i\theta}\right]\label{Bdpi+pi-EWP}\\
\sqrt{2}A(B^0_d\to\pi^0\pi^0)&=&|P|e^{i\delta_P} 
\left[1+\frac{x}{d}e^{i\gamma}e^{i(\Delta-\theta)}+\tilde q\left(
\frac{1+x e^{i\Delta}}{d}\right)e^{-i\theta}e^{-i\beta}
\right],\label{Bdpi0pi0EWP}
\end{eqnarray}
where
\begin{equation}\label{qtilde}
\tilde q\equiv\left|\frac{P_{\rm EW}}{T+C}\right|\approx1.3\times10^{-2}\times
\left|\frac{V_{td}}{V_{ub}}\right|=
1.3\times10^{-2}\times\left(1-\frac{\lambda^2}{2}\right)
\left|\frac{\sin\gamma}{\sin\beta}\right|\approx 3\times10^{-2}.
\end{equation}
The impact of the EW penguin contributions on the determination of the
hadronic parameters $(d,\theta)$ and $(x,\Delta)$ discussed above 
is negligible. For $\tilde q$ as in (\ref{qtilde}), $x$ changes by
about 1\% and $\Delta$ by $1^\circ$, whereas $d$ and $\theta$ are not affected
by $\tilde q$. Even if an arbitrary (NP) weak phase for $\tilde q$ is admitted,
$x$ changes by at most $\pm 4\%$ and $\Delta$ by $\pm 2^\circ$. 

In contrast to the $B\to\pi\pi$ decays, there are modes in the
$B\to\pi K$ system that are significantly affected by EW penguins
\cite{GHLR-EWP,PAPIII,DeHe}, since the CKM structure of these channels 
is very different from the $B\to\pi\pi$ case. Despite complications for
the exploration of the $B\to\pi K$ system, this feature offers a 
nice avenue for NP to manifest itself in the $B$-factory data.

\boldmath
\subsection{Summary}\label{ssec:Bpipi-sum}
\unboldmath
Before having a closer look at the $B\to\pi K$ system in the next section, 
let us summarize the main results of our $B\to\pi\pi$ analysis:
\begin{itemize}
\item The starting point is a general parametrization of the relevant decay 
amplitudes. Assuming that $\gamma$ and $\phi_d=2\beta$ agree with the SM 
expectations, we may extract the hadronic parameters from the $B\to\pi\pi$ 
data in an essentially {\it clean} manner. 
\item This analysis results in values of the hadronic parameters
that show large non-factorizable contributions. In particular, we have
seen that the amazingly large $B_d\to\pi^0\pi^0$ branching ratio 
and the surprisingly small $B_d\to\pi^+\pi^-$ branching ratio can be
conveniently accommodated in the SM through constructive and destructive 
hadronic interference effects, respectively. The $B^\pm\to\pi^\pm\pi^0$ 
channel, which is measured in accordance with the conventional theoretical 
estimates, is not affected by this mechanism. Moreover, we expect that
also the $B\to\pi\pi$ colour-suppression factor is sizeably affected 
by non-factorizable contributions, where, for example, 
$a_2^{\pi\pi}e^{i\Delta_2^{\pi\pi}}\sim0.5\times e^{-i70^\circ}$
would fit nicely into our picture. 
\item The $B_d\to K^+K^-$, $B_s\to\pi^+\pi^-$ system allows us to probe
the importance of penguin annihilation and exchange topologies, which 
are expected to play a minor r\^ole. This strategy is particularly 
promising for the era of the LHC. The current $B$-factory bounds on 
$B_d\to K^+K^-$ do not indicate any anomalous behaviour. 
\item Assuming that penguin annihilation and exchange topologies are 
negligible, we may complement the $B\to\pi\pi$ data in a variety 
of ways with the experimental information provided by $B_d\to\pi^\mp K^\pm$ 
modes, which are only marginally affected by EW penguins. Following these 
lines, we may obtain insights into $SU(3)$-breaking effects and may also 
determine $\gamma$.  Interestingly, we obtain results in good agreement 
with the well-known UT fits, and arrive at a remarkably consistent overall 
picture.
\item Finally, having all relevant parameters at hand, we may {\it predict} 
the CP-violating observables of the decay $B_d\to\pi^0\pi^0$, with the 
promising perspective of having large direct and mixing-induced CP 
violation in this channel. A measurement of one of the corresponding
CP asymmetries would allow a clean determination of $\gamma$.
\end{itemize}

\section{\boldmath The $B\to\pi K$ System \unboldmath}\label{sec:BpiK}
\setcounter{equation}{0}
Decays of the kind $B\to\pi K$ have received a lot of attention in the
literature over the last couple of years (for reviews, see
\cite{RF-Phys-Rep,BpiK-Revs}), mainly in the context of the determination
of the UT angle $\gamma$. However, as was emphasized in 
\cite{BF-neutral1,PAPIII,PAPI}, these modes offer also valuable insights into 
the world of EW penguins. Interestingly, puzzling features of the current 
$B$-factory data for the $B\to\pi K$ modes may point towards NP effects 
in the EW penguin sector, as we will show in this section. A crucial 
ingredient of this analysis is given by the hadronic $B\to\pi\pi$ parameters 
determined in Section~\ref{sec:Bpipi}, which emerged from our resolution 
of the ``$B\to\pi\pi$ puzzle'' within the SM. 
\subsection{Basic Formulae}\label{ssec:BpiK-basic}
If we employ the isospin flavour symmetry of strong interactions, 
we may decompose the $B\to\pi K$ amplitudes in the following manner:
\begin{eqnarray}
A(B^+\to\pi^+K^0)&=&-P'\left[1+\rho_{\rm c}e^{i\theta_{\rm c}}e^{i\gamma}
\right]\label{B+pi+K0}\\
\sqrt{2}A(B^+\to\pi^0K^+)&=&P'\left[1+\rho_{\rm c}e^{i\theta_{\rm c}}
e^{i\gamma}-\left(e^{i\gamma}-qe^{i\phi}e^{i\omega}\right)
r_{\rm c}e^{i\delta_{\rm c}}\right]\label{B+pi0K+}\\
A(B^0_d\to\pi^-K^+)&=&P'\left[1-re^{i\delta}e^{i\gamma}\right]
\label{B0pi-K+}\\
\sqrt{2}A(B^0_d\to\pi^0K^0)&=&-P'\left[1+\rho_{\rm n}e^{i\theta_{\rm n}}
e^{i\gamma}-qe^{i\phi}e^{i\omega}r_{\rm c}e^{i\delta_{\rm c}}\right].
\label{B0pi0K0}
\end{eqnarray}
Here the CP-conserving strong amplitude
\begin{equation}
P'\equiv\left(1-\frac{\lambda^2}{2}\right)A\lambda^2({\cal P}_t'-{\cal P}_c')
\end{equation}
is the $B\to\pi K$ counterpart of (\ref{P-def}),\footnote{The primes
remind us that we are dealing with $\bar b\to\bar s$ transitions.} 
describing the difference of the QCD penguins with internal top- and 
charm-quark exchanges,
\begin{equation}\label{rho-c-def}
\rho_{\rm c}e^{i\theta_{\rm c}}\equiv\left(\frac{\lambda^2R_b}{1-\lambda^2}
\right)\left[\frac{{\cal P}_t'-
\tilde {\cal P}_u'-{\cal A}'}{{\cal P}_t'-{\cal P}_c'}\right],
\end{equation}
where $\tilde {\cal P}_u'$ is the strong amplitude of QCD penguins with
internal up-quark exchanges contributing to the {\it charged} $B\to\pi K$ 
decays and ${\cal A}'$ denotes an annihilation amplitude, 
\begin{equation}\label{rc-def}
r_{\rm c}e^{i\delta_{\rm c}}\equiv\left(\frac{\lambda^2R_b}{1-\lambda^2}
\right)\left[\frac{{\cal T}'+{\cal C}'}{{\cal P}_t'-{\cal P}_c'}\right],
\end{equation}
where ${\cal T}'$ and ${\cal C}'$ are the colour-allowed and colour-suppressed
tree-diagram-like topologies corresponding to their $B\to\pi\pi$
counterparts ${\cal T}$ and ${\cal C}$ in (\ref{T-tilde}) and 
(\ref{C-tilde}), respectively,
\begin{equation}\label{r-def}
re^{i\delta}\equiv\left(\frac{\lambda^2R_b}{1-\lambda^2}
\right)\left[\frac{{\cal T}'-({\cal P}_t'-{\cal P}_u')}{{\cal P}_t'-
{\cal P}_c'}\right],
\end{equation}
where ${\cal P}_u'$ is the strong amplitude of QCD penguins with
internal up-quark exchanges contributing to the {\it neutral} 
$B\to\pi K$ decays, 
\begin{equation}\label{rho-n-def}
\rho_{\rm n}e^{i\theta_{\rm n}}\equiv\left(\frac{\lambda^2R_b}{1-\lambda^2}
\right)\left[\frac{{\cal C}'+({\cal P}_t'-{\cal P}_u')}{{\cal P}_t'-
{\cal P}_c'}\right],
\end{equation}
and the EW penguin parameter $qe^{i\phi}e^{i\omega}$ was introduced in 
(\ref{FBF43}). In (\ref{B+pi+K0})--(\ref{B0pi0K0}), we have neglected
certain contributions from colour-suppressed EW penguins, which are 
expected to have a minor impact on our analysis. Since a detailed 
discussion of these topologies is rather technical, we have left it
to Appendix~\ref{APP:BpiK-CSEWP}, where also the corresponding
generalizations of (\ref{B+pi+K0})--(\ref{B0pi0K0}) can be found.
As we will see in Subsection~\ref{ssec:BpiK-analysis}, the current
$B$-factory data for those $B\to\pi K$ modes where EW penguins may 
{\it only} contribute in colour-suppressed form do not indicate any anomalous 
behaviour, i.e.\ do not point towards an unexpected enhancement of the
colour-suppressed EW penguins. The amplitudes in 
(\ref{B+pi+K0})--(\ref{B0pi0K0}) satisfy the following well-known 
isospin relation \cite{LNQS}:
\begin{displaymath}
A(B^+\to\pi^+K^0)+\sqrt{2}A(B^+\to\pi^0K^+)=A(B^0_d\to\pi^-K^+)+
\sqrt{2}A(B^0_d\to\pi^0K^0)
\end{displaymath}
\begin{equation}\label{BpiK-iso-rel}
=-\left[e^{i\gamma}-qe^{i\phi}e^{i\omega}\right]|T'+C'|e^{i\delta_{T'+C'}},
\end{equation}
as well as \cite{PAPIII}
\begin{displaymath}
A(B^+\to\pi^+K^0)+A(B^0_d\to\pi^-K^+)=-e^{i\gamma}\left[re^{i\delta}
+\rho_{\rm c}e^{i\theta_{\rm c}}\right]|P'|e^{i\delta_{P'}}.
\end{displaymath}
Let us also note that the hadronic parameters 
$\rho_{\rm c}e^{i\theta_{\rm c}}$, $r_{\rm c}e^{i\delta_{\rm c}}$, 
$re^{i\delta}$ and $\rho_{\rm n}e^{i\theta_{\rm n}}$ introduced above do 
{\it not} involve any EW penguin contributions, in contrast to the 
quantities appearing in the parametrization proposed in \cite{BF-neutral1}. 
This feature is important for the following considerations.

\subsection{Determination of the Hadronic Parameters}\label{ssec:BpiK-hadr}
\boldmath
\subsubsection{$\rho_{\rm c}e^{i\theta_{\rm c}}$}
\unboldmath
As can be seen in (\ref{rho-c-def}), the parameter 
$\rho_{\rm c}e^{i\theta_{\rm c}}$ entering the charged $B\to\pi K$ decays
is expected to be tiny because of $\lambda^2 R_b\sim0.02$. On the other 
hand, a sizeable value of $\rho_{\rm c}$ would be indicated by a significant 
direct CP asymmetry 
\begin{eqnarray}
{\cal A}_{\rm CP}^{\rm dir}(B^\pm\to\pi^\pm K)&\equiv&
\frac{\mbox{BR}(B^+\to\pi^+K^0)-
\mbox{BR}(B^-\to\pi^-\bar K^0)}{\mbox{BR}(B^+\to\pi^+K^0)+
\mbox{BR}(B^-\to\pi^-\bar K^0)}\nonumber\\
&&=-\left[\frac{2\rho_{\rm c}\sin\theta_{\rm c}\sin\gamma}{1+2\rho_{\rm c}
\cos\theta_{\rm c}\cos\gamma+\rho_{\rm c}^2}\right],\label{ACP-B+pi+K}
\end{eqnarray}
and an enhancement of the CP-averaged $B^\pm\to K^\pm K$ branching 
ratio \cite{BFM,FKNP}. However, 
the current $B$-factory results for these quantities \cite{HFAG},
\begin{equation}\label{ACP-B+pi+K-HFAG}
{\cal A}_{\rm CP}^{\rm dir}(B^\pm\to\pi^\pm K)=-0.02\pm 0.06
\end{equation}
\begin{equation}\label{BR-BKK}
\mbox{BR}(B^\pm\to K^\pm K)< 2.4\times 10^{-6} \, \mbox{(90\% C.L.)},
\end{equation}
do not indicate any anomalous behaviour. In particular, if we employ
the $U$-spin flavour symmetry of strong interactions and introduce 
(for a detailed discussion, see \cite{RF-Phys-Rep})
\begin{equation}
K\equiv\left[\frac{1}{\epsilon R_{SU(3)}^2}\right]
\left[\frac{\mbox{BR}(B^\pm\to \pi^\pm K)}{\mbox{BR}(B^\pm\to K^\pm K)}
\right]=\frac{1+2\rho_{\rm c}\cos\theta_{\rm c}\cos\gamma+
\rho_{\rm c}^2}{\epsilon^2-2\epsilon\rho_{\rm c}\cos\theta_{\rm c}\cos\gamma+
\rho_{\rm c}^2},
\end{equation}
we obtain the following allowed range for $\rho_{\rm c}$, 
which has the same structure as
(\ref{PA-E-bounds}):
\begin{equation}
\frac{1-\epsilon\sqrt{K}}{1+\sqrt{K}}\leq\rho_{\rm c}\leq
\frac{1+\epsilon\sqrt{K}}{|1-\sqrt{K}|}.
\end{equation}
Using now $R_{SU(3)}=0.7$, which describes factorizable $U$-spin-breaking
corrections \cite{RF-Phys-Rep}, and
$\mbox{BR}(B^\pm\to K^\pm K)/\mbox{BR}(B^\pm\to \pi^\pm K)<0.1$,
which follows from (\ref{BR-BKK}) and the experimental result 
$\mbox{BR}(B^\pm\to \pi^\pm K)=(21.8\pm1.4)\times 10^{-6}$ \cite{HFAG}, 
we arrive at 
\begin{equation}
\rho_{\rm c}<0.1.
\end{equation}
In the future, this bound can be improved significantly. We shall 
neglect the parameter $\rho_{\rm c}$ in the following discussion. Should 
$B^\pm\to K^\pm K$ decays soon be observed at the $B$ factories, thereby 
indicating a value of $\rho_{\rm c}$ at the 0.1 level, this parameter 
could be taken into account by following the strategies discussed in
\cite{BF-neutral1,defan,BFM,FKNP}. In this context, it should be emphasized
that only the analysis of the charged $B\to\pi K$ system may be affected
by $\rho_{\rm c}$, whereas this quantity does not affect the neutral 
$B\to\pi K$ decays. 

\boldmath
\subsubsection{$r_{\rm c}e^{i\delta_{\rm c}}$, $re^{i\delta}$ and 
$\rho_{\rm n}e^{i\theta_{\rm n}}$}
\unboldmath
Let us now turn to the other hadronic parameters 
$r_{\rm c}e^{i\delta_{\rm c}}$, $re^{i\delta}$ and 
$\rho_{\rm n}e^{i\theta_{\rm n}}$ appearing in 
(\ref{B+pi+K0})--(\ref{B0pi0K0}). If we look at (\ref{rc-def}), 
(\ref{r-def}) and (\ref{rho-n-def}), we may derive the following
relations:
\begin{equation}\label{r-rho-rel}
r_{\rm c}e^{i\delta_{\rm c}}=re^{i\delta}+\rho_{\rm n}e^{i\theta_{\rm n}}
\end{equation}
\begin{equation}\label{rho-n-det}
\rho_{\rm n}e^{i\theta_{\rm n}}=re^{i\delta}x'e^{i\Delta'},
\end{equation}
with 
\begin{equation}\label{x-BpiK}
x'e^{i\Delta'}\equiv \frac{{\cal C}'+{\cal P}_{tu}'}{{\cal T}'-{\cal P}_{tu}'}.
\end{equation}
Consequently, $r_{\rm c}e^{i\delta_{\rm c}}$ and $re^{i\delta}$ 
differ through the quantity $\rho_{\rm n}e^{i\theta_{\rm n}}$, which is 
proportional to $x'e^{i\Delta'}$. Let us next assume that the penguin 
annihilation and exchange topologies discussed in 
Subsection~\ref{ssec:PA-E} play a minor r\^ole. Using then the $SU(3)$ 
flavour symmetry of strong interactions, we may relate the hadronic 
$B\to\pi K$ parameter $x'e^{i\Delta'}$ to its $B\to\pi\pi$ counterpart 
$xe^{i\Delta}$ introduced in (\ref{x-Bpipi}) through the simple relation 
\begin{equation}\label{x-rel}
x'e^{i\Delta'}=xe^{i\Delta}.
\end{equation}
As far as the parameter $re^{i\delta}$ is concerned, we have
\begin{equation}\label{r-det}
re^{i\delta}=\frac{\epsilon}{d}e^{i(\pi-\theta)}
\end{equation}
with $\epsilon$ defined in (\ref{epsi}). 
Consequently, (\ref{r-rho-rel}), (\ref{rho-n-det}), (\ref{x-rel}) and
(\ref{r-det}) allow us to determine $re^{i\delta}$,
$\rho_{\rm n}e^{i\theta_{\rm n}}$ and $r_{\rm c}e^{i\delta_{\rm c}}$ from 
the $B\to\pi\pi$ analysis performed in Section~\ref{sec:Bpipi}. Following
these lines, the numerical values given in (\ref{d-det}) and (\ref{x-det})
imply
\begin{equation}\label{r-det-Bpipi}
r=0.11^{+0.07}_{-0.05},\quad \delta=+(42^{+23}_{-19})^\circ
\end{equation}
\vspace*{-0.6truecm}
\begin{equation}\label{rho-n-det-Bpipi}
\rho_{\rm n}=0.13^{+0.07}_{-0.05},\quad 
\theta_{\rm n}=-(29^{+21}_{-26})^\circ
\end{equation}
\vspace*{-0.4truecm}
\begin{equation}\label{r-c-det-Bpipi}
r_{\rm c}=0.20^{+0.09}_{-0.07},\quad \delta_{\rm c}=+(2^{+23}_{-18})^\circ,
\end{equation}
where our treatment of errors is described in 
Appendix~\ref{app:error-treatment}. 
Interestingly, the values in (\ref{r-c-det-Bpipi}) mimic the picture
of QCD factorization \cite{Be-Ne,BBNS2}, whereas (\ref{r-det-Bpipi}) and 
(\ref{rho-n-det-Bpipi}) differ strongly from the corresponding predictions.

The parameter $r_{\rm c}$ in (\ref{r-c-det-Bpipi}), which follows 
directly from the simple expression
\begin{equation}\label{rc-simple-expr}
r_{\rm c}=\frac{\epsilon}{d}\sqrt{\left[1-
2d\cos\theta\cos\gamma+d^2\right]R_{+-}^{\pi\pi}},
\end{equation}
can be determined alternatively with the
help of the following well-known relation \cite{GRL}:
\begin{equation}\label{rc-alt-det}
r_{\rm c}=\sqrt{2}\left|\frac{V_{us}}{V_{ud}}\right|\frac{f_K}{f_\pi}
\sqrt{\frac{{\rm BR}(B^\pm\to \pi^\pm\pi^0)}{{\rm BR}(B^\pm\to \pi^\pm K^0)}}
=0.196\pm0.016,
\end{equation}
which relies on the $SU(3)$ flavour symmetry and the neglect of the
$\rho_{\rm c}$ term in (\ref{B+pi+K0}). We consider the agreement between
the numerical values in (\ref{r-c-det-Bpipi}) and (\ref{rc-alt-det}) 
as very remarkable. In particular, (\ref{r-c-det-Bpipi}) does {\it not}
rely on any assumption about the $\rho_{\rm c}$ parameter, thereby supporting
its neglect, in addition to the arguments given above.
Whereas the impact of $\rho_{\rm c}$ on (\ref{rc-alt-det}) is maximal
for $\theta_{\rm c}\sim 0^\circ \lor 180^\circ$, as this strong phase 
enters there through $\cos\theta_{\rm c}$, (\ref{ACP-B+pi+K}) is governed
by $\sin\theta_{\rm c}$, and is hence affected most for 
$\theta_{\rm c}\sim \pm90^\circ$. Consequently, we obtain complementary 
information on $\rho_{\rm c}$, suggesting that this parameter is indeed 
negligible. Interestingly, also enhanced colour-suppressed EW penguins
(see Appendix~\ref{APP:BpiK-CSEWP})
could affect the determination of $r_{\rm c}$ through (\ref{rc-alt-det}), 
thereby leading to a possible 
discrepancy with the value following from (\ref{rc-simple-expr}). The 
agreement between these two determinations of $r_{\rm c}$ does therefore
also not point towards an anomalous behaviour of the 
colour-suppressed EW penguins. Moreover, it suggests moderate 
non-factorizable $SU(3)$-breaking corrections.\footnote{Otherwise, 
these effects have to conspire in a very contrived way with the
$\rho_{\rm c}e^{i\theta_{\rm c}}$ parameter and the
colour-suppressed EW penguins, which does not seem plausible to us.} 
A comprehensive study of the $SU(3)$-breaking corrections to the relation 
in (\ref{rc-alt-det}) was recently performed with the help of QCD sum rules 
\cite{KMM}, with the result that the dominant effects are actually described 
by the ``factorizable'' $f_K/f_\pi$ factor, which is in accordance with
the picture following from our phenomenological analysis. 
In this context, it is also interesting to note that (\ref{r-det}) is not 
affected by any $SU(3)$-breaking corrections within the factorization 
approximation, as the corresponding decay constants and form factors 
cancel \cite{RF-BsKK}. 
On the other hand, (\ref{x-rel}) is affected by factorizable $SU(3)$-breaking 
corrections in the following manner:
\begin{equation}\label{x-rel-SU3}
x'e^{i\Delta'}=\left[\frac{f_\pi F_{BK}(M_\pi^2;0^+)}{f_K F_{B\pi}
(M_K^2;0^+)}\right]xe^{i\Delta},
\end{equation}
where the Bauer--Stech--Wirbel model \cite{BSW} points to a small 
deviation from 1 of the correction factor, which we shall neglect in 
the following discussion. These $SU(3)$-breaking effects are neglected in
(\ref{rc-alt-det}) as well. 

\boldmath
\subsubsection{Further Theoretical Aspects of 
$\rho_{\rm c}e^{i\theta_{\rm c}}$}\label{ssec:rho_c-th}
\unboldmath
 From the theoretical point of view, it is instructive to complement
the considerations of Subsection~\ref{ssec:x-Del-struct} with 
our $B\to\pi K$ results, and to explore the implications for 
$\rho_{\rm c}e^{i\theta_{\rm c}}$. If we employ (\ref{zeta-def}) and 
the relations given above, we obtain
\begin{eqnarray}
\rho_{\rm c}e^{i\theta_{\rm c}}&=&\left[\left(\frac{{\cal P}_t'-
\tilde {\cal P}_u'}{{\cal P}_t'-{\cal P}_u'}\right)
-\left(\frac{{\cal A}'}{{\cal P}_t'-{\cal P}_u'}\right)\right]
\left[\frac{\zeta e^{i\Delta_{\zeta}}}{1-\zeta e^{i\Delta_{\zeta}}}\right]
r e^{i\delta}
\nonumber\\
&\sim& \left[\left(\frac{{\cal P}_t'-
\tilde {\cal P}_u'}{{\cal P}_t'-{\cal P}_u'}\right)
-\left(\frac{{\cal A}'}{{\cal P}_t'-{\cal P}_u'}\right)\right]
\tilde \zeta e^{i\Delta_{\tilde\zeta}}
\times 0.1\times e^{i42^\circ},\label{rho-c-estimate}
\end{eqnarray}
where we have used the parameter $\tilde\zeta e^{i\Delta_{\tilde\zeta}}$
introduced in (\ref{zetatilde-def}) and have taken the numerical values in 
(\ref{r-det-Bpipi}) into account. As can be seen in Fig.~\ref{fig:zetatilde}, 
$\tilde \zeta$ may -- in principle -- be as large as ${\cal O}(2)$. 
However, for values of $a_2^{\pi\pi}e^{i\Delta_2^{\pi\pi}}\sim0.5\times 
e^{-i70^\circ}$, which fit nicely into the picture developed in
Subsection~\ref{ssec:x-Del-struct}, we obtain 
$\tilde\zeta e^{i\Delta_{\tilde\zeta}}\sim 0.57\times e^{-i50^\circ}$,
thereby suppressing $\rho_{\rm c}$ to a sufficient extent.
Interestingly,
$\Delta_{\tilde\zeta}$ would then largely cancel $\delta$ in 
(\ref{rho-c-estimate}). Moreover, also the first term in square 
brackets on the right-hand side of this equation may suppress 
$\rho_{\rm c}$ even further.

\boldmath
\subsection{Refined Determination of $\gamma$}\label{ssec:gam-refine}
\begin{figure}
\vspace*{0.3truecm}
\begin{center}
\psfrag{g}{$\gamma$}\psfrag{c2}{$\chi^2$}
\psfrag{incl}{incl.\ constraint from $r_{\rm c}$}
\psfrag{excl}{excl.\ constraint from $r_{\rm c}$}
\includegraphics[width=10cm]{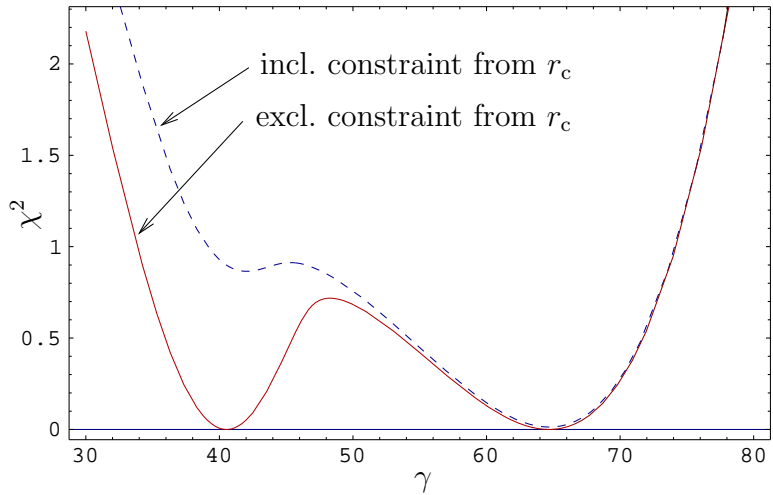}
\end{center}
\caption{$\chi^2$ of a fit of the hadronic parameters to
${\cal A}_{\rm CP}^{\rm dir}(B_d\to\pi^+\pi^-)$,
${\cal A}_{\rm CP}^{\rm mix}(B_d\to\pi^+\pi^-)$ and $H$ with (dashed) and
without (solid) a simultaneous fit to $r_{\rm c}$ as a function of 
$\gamma$. By including the constraint from $r_{\rm c}$, the degeneracy is 
lifted.}\label{fig:chi2-gamma}
\end{figure}

\unboldmath
Imposing the constraint that the values of $r_{\rm c}$ following from 
(\ref{rc-simple-expr}) and (\ref{rc-alt-det}) agree with each other allows
us also to refine the determination of $\gamma$ discussed in 
Subsection~\ref{ssec:Bpipi-gam-det}. In particular, it allows us to
lift the degeneracy of the two solutions for $\gamma$ given in 
(\ref{gam-H}). In order to illustrate this feature, we show in 
Fig.~\ref{fig:chi2-gamma} how the $\chi^2$ of a fit of the hadronic 
parameters to ${\cal A}_{\rm CP}^{\rm dir}(B_d\to\pi^+\pi^-)$,
${\cal A}_{\rm CP}^{\rm mix}(B_d\to\pi^+\pi^-)$ and $H$ with and
without a simultaneous fit to $r_{\rm c}$ varies as a function of $\gamma$.
These fits are performed by fixing $\gamma$ and then fitting the
hadronic parameters for each value of $\gamma$.
We observe that the inclusion of the constraint from $r_{\rm c}$ does
actually lift the degeneracy, leaving us with
\begin{equation}\label{gam-fin}
\gamma=(64.7^{+6.3}_{-6.9})^\circ.
\end{equation}
The fact that the curve corresponding to 
the fit including $r_{\rm c}$ has its minimum in one of the minima of 
the other curve and practically touches the $x$-axis is non-trivial, but 
yet another sign of the consistency of our approach.

\begin{figure}
\vspace*{0.3truecm}
\begin{center}
\psfrag{eta}{$\bar\eta$}\psfrag{rho}{$\bar\rho$}
\includegraphics[width=10cm]{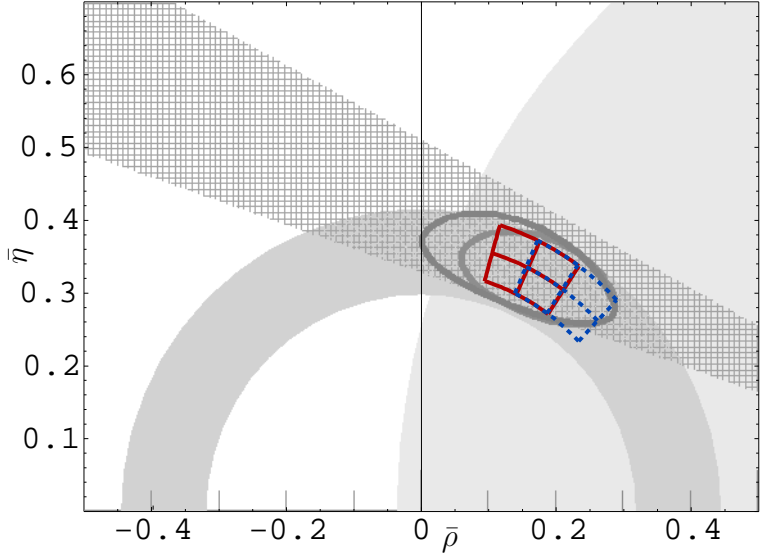}
\end{center}
\caption{Comparison of our determination of $\gamma$ from the 
$B\to\pi\pi,\pi K$ data with the standard UT fit, where our 
1$\sigma$ results are indicated by the solid and dashed 
windows, as described in the text.}\label{fig:ut-compare}
\end{figure}

Using $R_b$ as given in (\ref{CKM1}) and the result for $\gamma$
in (\ref{gam-fin}), the simple relations
\begin{equation}
\bar\rho=R_b\cos\gamma,\quad 
\bar\eta=R_b\sin\gamma,
\end{equation}
which follow from the unitarity of the CKM matrix, allow us to calculate
straightforwardly the other two angles $\alpha$ and $\beta$ of the
UT, where we obtain 
\begin{equation}\label{alpha-beta}
\alpha=(93.6^{+10.3}_{-9.1})^\circ,  \quad  
\beta=(21.7^{+2.5}_{-2.6})^\circ.
\end{equation}
In Fig.~\ref{fig:ut-compare}, we follow \cite{FIM}, and compare these 
results with the allowed region for the apex of the UT following from 
the ``standard analysis'' 
\cite{BurasParodiStocchi}.\footnote{The small and large ellipses 
in Fig.~\ref{fig:ut-compare} correspond
to the SM and MFV, respectively, as obtained in an update \cite{schladming} of 
\cite{BurasParodiStocchi}.} 
The solid window corresponds to the range for $\gamma$
in (\ref{gam-fin}), whereas the dashed window indicates how the results 
change when the new Belle data (see Appendix \ref{app:belle-new}) are 
used. It should be noted that we show 1$\sigma$ regions, while the 
elliptic areas in the original UT plot show 95\% C.L.\ contours.
Needless to note, the consistency of the overall picture is very
remarkable.

\boldmath
\subsection{Analysis of the $B\to\pi K$ Observables}\label{ssec:BpiK-analysis}
\unboldmath
Having all relevant hadronic parameters at hand, we may now analyse 
the observables provided by the $B\to\pi K$ modes within the SM, and
may search for discrepancies that may shed light on NP. 
\boldmath
\subsubsection{The $B_d\to\pi^\mp K^\pm$, $B^\pm\to\pi^\pm K$ 
System}\label{ssec:BpiK-mixed}
\unboldmath
Let us first turn to the decays $B_d\to\pi^\mp K^\pm$ and $B^\pm\to\pi^\pm K$.
The characteristic feature of these modes is that EW penguins may {\it only}
contribute in colour-suppressed form, as discussed in detail in 
Appendix~\ref{APP:BpiK-CSEWP}. Consequently, EW penguins are expected
to have a marginal impact on the $B_d\to\pi^\mp K^\pm$, $B^\pm\to\pi^\pm K$ 
system. The relevant observables are the ratio $R$ introduced in 
(\ref{R-def}), which involves the CP-averaged branching ratios, and 
the corresponding direct 
CP asymmetries. If we assume again that $\rho_{\rm c}=0$, (\ref{B+pi+K0}) 
and (\ref{B0pi-K+}) imply the following well-known expression \cite{FM}:
\begin{equation}\label{R-expr}
R=1-2r\cos\delta\cos\gamma+r^2. 
\end{equation}
Using the SM value of $\gamma$ in (\ref{UT-angles}) and the hadronic 
parameters given in (\ref{r-det-Bpipi}), we obtain
\begin{equation}\label{R-pred}
\left.R\right|_{\rm SM}=0.943^{+0.033}_{-0.026},
\end{equation}
which agrees well with the experimental result in (\ref{R-def}). 

Additional information is provided by the $B_d\to\pi^\mp K^\pm$,
$B^\pm\to\pi^\pm K$ system through direct CP violation. As we have 
seen in (\ref{ACP-B+pi+K}), the CP asymmetry 
${\cal A}_{\rm CP}^{\rm dir}(B^\pm\to\pi^\pm K)$ vanishes for 
$\rho_{\rm c}=0$, in accordance with the average of the current 
$B$-factory data given in (\ref{ACP-B+pi+K-HFAG}). Consequently,
we are left with the direct CP asymmetry of the $B_d\to\pi^\mp K^\pm$
modes, which we addressed already in (\ref{ACPdir-exp})
and (\ref{ACPdir}). In terms
of $r$ and $\delta$, this observable is given by
\begin{equation}\label{ACPdir-r}
{\cal A}_{\rm CP}^{\rm dir}(B_d\to\pi^\mp K^\pm)=
\frac{2r\sin\delta\sin\gamma}{1-2r\cos\delta\cos\gamma+r^2}.
\end{equation}
If we employ again -- as in (\ref{R-pred}) -- the SM expectation for $\gamma$ 
in (\ref{UT-angles}) and the values of $r$ and $\delta$ in 
(\ref{r-det-Bpipi}), this expression yields the prediction
\begin{equation}
\left.{\cal A}_{\rm CP}^{\rm dir}(B_d\to\pi^\mp K^\pm)\right|_{\rm SM}
=0.140^{+0.139}_{-0.087},
\end{equation}
which is in accordance with the current $B$-factory average given in
(\ref{ACPdir-exp}). Although we find that $re^{i\delta}$ is strongly
affected by non-factorizable effects, which is in particular reflected
by the sizeable strong phase $\delta$ in (\ref{r-det-Bpipi}), the rather
small value of $r$ suppresses the direct $B_d\to\pi^\mp K^\pm$ CP
asymmetry, in agreement with the data. 

To conclude our analysis of the $B_d\to\pi^\mp K^\pm$, $B^\pm\to\pi^\pm K$ 
system, we emphasize that the corresponding $B$-factory measurements give 
a current picture that does not show any anomalous behaviour and is nicely
consistent with the SM description. From this feature, we may in particular 
conclude that no anomalous enhancement of colour-suppressed EW penguins 
is indicated by the data. Further strategies to address these topologies 
are discussed in Appendix~\ref{APP:BpiK-CSEWP}.

\boldmath
\subsubsection{The Charged and Neutral $B\to\pi K$ 
Systems}\label{ssec:BpiK-c-n}
\unboldmath
Let us now analyse the data provided by the charged and neutral
$B\to\pi K$ decays, where the quantities $R_{\rm c}$ and $R_{\rm n}$
introduced in (\ref{Rc-def}) and (\ref{Rn-def}), respectively, 
play a key r\^ole. In analogy to $R$, they involve CP-averaged
branching ratios. Using (\ref{B+pi+K0})--(\ref{B0pi0K0}) with
$\rho_{\rm c}=0$, we obtain
\begin{eqnarray}\label{Rc-expr}
R_{\rm c}&=&1-2r_{\rm c}\cos\delta_{\rm c}\cos\gamma+r_{\rm c}^2\nonumber\\
&&+q r_{\rm c}\left[2\left\{\cos(\delta_{\rm c}+\omega)\cos\phi
-r_{\rm c}\cos\omega\cos(\gamma-\phi)\right\}+qr_{\rm c}\right],
\end{eqnarray}
whereas
\begin{equation}\label{Rn-expr}
R_{\rm n}=\frac{1}{b}\left[1-2r\cos\delta\cos\gamma+r^2\right],
\end{equation}
with
\begin{eqnarray}
b&=&1-2qr_{\rm c}\cos(\delta_{\rm c}+\omega)\cos\phi
+q^2r_{\rm c}^2\nonumber\\
&&+2\rho_{\rm n}\left[\cos\theta_{\rm n}\cos\gamma-
qr_{\rm c}\cos(\theta_{\rm n}-\delta_{\rm c}-\omega)\cos(\gamma-\phi)
\right]+\rho_{\rm n}^2.\label{b-expr}
\end{eqnarray}
The quantity $b$ was introduced in \cite{BFRS-I} through
\begin{equation}
b\equiv\frac{R}{R_{\rm n}}=2\left[\frac{{\mbox{BR}(B^0_d\to\pi^0K^0)+
\mbox{BR}(\bar B^0_d\to\pi^0\bar K^0)}}{\mbox{BR}(B^+\to\pi^+K^0)+
\mbox{BR}(B^-\to\pi^-\bar K^0)}\right]\frac{\tau_{B^+}}{\tau_{B^0_d}}
=1.19\pm 0.16,
\end{equation}
where the experimental value follows from \cite{HFAG}. This variable 
coincides with $R_{00}$ in \cite{Be-Ne}. 
In contrast to (\ref{R-expr}), (\ref{Rc-expr}) and (\ref{Rn-expr}) are 
significantly affected by the EW penguin parameter $qe^{i\phi}e^{i\omega}$.
Using the SM result
\begin{equation}\label{q-SM}
\left.qe^{i\phi}e^{i\omega}\right|_{\rm SM}=0.69 \times
\left[\frac{0.086}{|V_{ub}/V_{cb}|}\right],
\end{equation}
with $|V_{ub}/V_{cb}|$ given in (\ref{CKM1}) and the SM value of
$\gamma$ in (\ref{UT-angles}), the hadronic parameters in 
(\ref{r-det-Bpipi})--(\ref{r-c-det-Bpipi}) yield
\begin{equation}\label{Rc-SM}
\left.R_{\rm c}\right|_{\rm SM}=1.14^{+0.08}_{-0.07}
\end{equation}
and
\begin{equation}\label{Rn-SM}
\left.R_{\rm n}\right|_{\rm SM}=1.11^{+0.06}_{-0.07},
\end{equation}
exhibiting a pattern that is {\it not} in accordance with the current 
experimental picture given in (\ref{Rc-def}) and (\ref{Rn-def}), respectively. 

In this context, it is also interesting to consider the quantity
\begin{equation}\label{L-def}
L\equiv\frac{(R_{\rm c}-1)+b(1-R_{\rm n})}{2 r_{\rm c}^2}=5.9\pm2.8,
\end{equation}
which was introduced in \cite{BFRS-I} and measures the violation of the 
Lipkin sum rule \cite{Lipkin}; the numerical value in this expression
corresponds to (\ref{rc-alt-det}) and the averages of the $B$-factory 
data compiled in \cite{HFAG}. Using (\ref{Rc-expr}) and (\ref{Rn-expr}) 
with  (\ref{b-expr}), we arrive at
\begin{eqnarray}
r_{\rm c}^2 L &=&qr_{\rm c}\left[qr_{\rm c}-\left\{r_{\rm c}\cos\omega
+\rho_{\rm n}\cos(\theta_{\rm n}-\delta_{\rm c}-\omega)\right\}
\cos(\gamma-\phi)\right]\nonumber\\
&&+\left[r\cos\delta-r_{\rm c}\cos\delta_{\rm c}\right]
\cos\gamma+\rho_{\rm n}\cos\theta_{\rm n}\cos\gamma+
\frac{1}{2}\left(r_{\rm c}^2-r^2+\rho_{\rm n}^2\right).
\end{eqnarray}
Taking, moreover, (\ref{r-rho-rel}) into account, this expression
can be simplified as follows:
\begin{eqnarray}\label{Lform}
L&=&q\left[q-\cos\omega\cos(\gamma-\phi)\right]\nonumber\\
&&+\frac{\rho_{\rm n}}{r_{\rm c}}\left[\left\{\frac{\rho_{\rm n}+r\cos(\delta
-\theta_{\rm n})}{r_{\rm c}}\right\}-q\cos(\theta_{\rm n}-\delta_{\rm c}
-\omega)\cos(\gamma-\phi)\right].
\end{eqnarray}
In analogy to (\ref{Rc-SM}) and (\ref{Rn-SM}), we may then calculate
\begin{equation}
\left.L\right|_{\rm SM}=0.59^{+0.14}_{-0.13},
\end{equation}
which is in conflict with the experimental number in (\ref{L-def}). 

Already back in 2000, when the observation of the $B^0_d\to\pi^0K^0$ channel
was announced by the CLEO collaboration, two of us pointed out puzzling
features that were indicated by the corresponding results for $R_{\rm n}$
and $R_{\rm c}$, emphasizing the possibility of having NP in the EW penguin
sector \cite{BF-neutral2}. Now we have a much better experimental picture, 
where the BaBar, Belle and CLEO data show the same pattern for the observables 
$R_{\rm n}$ and $R_{\rm c}$. In a recent paper \cite{BFRS-I}, we showed 
that enhanced EW penguins may in fact provide a solution to this puzzle. 
Here we perform a considerably more refined analysis and go beyond our 
previous study in the following respects:
\begin{itemize}
\item We fix the relevant hadronic parameters $re^{i\delta}$, 
$r_{\rm c}e^{i\delta_{\rm c}}$ and $\rho_{\rm n}e^{i\theta_{\rm n}}$
through the $B\to\pi\pi$ analysis, as discussed above. In particular,
we take also $\rho_{\rm n}e^{i\theta_{\rm n}}$ into account, which is found to 
be sizeable because of the very recent data for the 
$B_d\to\pi^0\pi^0$ channel. 
\item In addition to an enhancement of $q$, we also allow for a CP-violating 
NP phase $\phi$ in the EW penguin sector, and for a CP-conserving strong 
phase $\omega$, which could be induced by large non-factorizable 
$SU(3)$-breaking corrections. 
\end{itemize}

\begin{figure}
\vspace*{0.3truecm}
\begin{center}
\psfrag{Rn}{$R_{\rm n}$}\psfrag{Rc}{$R_{\rm c}$}
\psfrag{f}{\small $\!\!\phi$}\psfrag{expRegion}{exp.\ region}\psfrag{SM}{SM}
\psfrag{q069}{$q=0.69$}\psfrag{q122}{$q=1.22$}\psfrag{q175}{$q=1.75$}
\includegraphics[width=10.2cm]{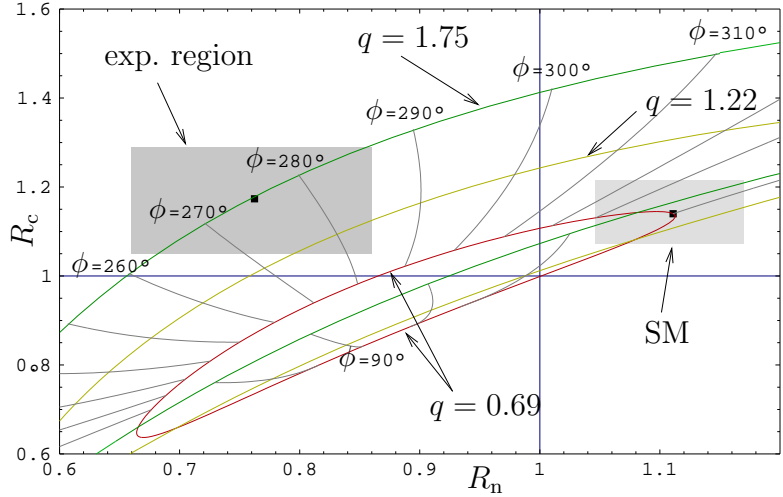}
\end{center}
\caption{The situation in the $R_{\rm n}$--$R_{\rm c}$ plane. We show contours
for values of $q=0.69$, $q=1.22$ and $q=1.75$, with
$\phi \in [0^\circ,360^\circ]$. 
The ranges from (\ref{Rc-def}) and (\ref{Rn-def}) (experiment) as well as 
(\ref{Rc-SM}) and (\ref{Rn-SM}) (SM) are indicated in grey.}\label{fig:Rn-Rc}
\end{figure}

In order to make our analysis more transparent, let us first assume that
$\omega=0^\circ$, as implied by the $SU(3)$ flavour symmetry 
\cite{BF-neutral1,NR,neubert-BpiK}. We are then left with the two 
EW penguin parameters $q$ and $\phi$. In Fig.~\ref{fig:Rn-Rc}, we
follow \cite{BFRS-I} and consider the $R_{\rm n}$--$R_{\rm c}$ plane
for the central values of the hadronic parameters in 
(\ref{r-det-Bpipi})--(\ref{r-c-det-Bpipi}) and $\gamma=65^\circ$, showing
contours for different values of $q$ and $\phi\in[0^\circ,360^\circ]$.
We observe that we may in fact move to the experimental region for an
enhanced value of $q\sim 1.8$ and $\phi\sim -90^\circ$, where in particular
the large CP-violating phase is in stark contrast
to the SM picture characterized by (\ref{q-SM}). In order to put
these observations on a more quantitative level, we use the value
of $\gamma$ in (\ref{UT-angles}) and the hadronic parameters in 
(\ref{r-det-Bpipi})--(\ref{r-c-det-Bpipi}), which allow us to convert 
the experimental results for $R_{\rm c}$ and $R_{\rm n}$ in (\ref{Rc-def}) 
and (\ref{Rn-def}), respectively, into values of $q$ and $\phi$. Following 
these lines, we obtain
\begin{equation}\label{q-det}
q=1.75^{+1.27}_{-0.99},\quad  \phi=-(85^{+11}_{-14})^\circ,
\end{equation}
where our treatment of errors is discussed in 
Appendix~\ref{app:error-treatment}. 
We may now also calculate the quantity $L$ in (\ref{Lform}) within our
NP scenario, yielding
\begin{equation}\label{L-det}
L=6.02^{+7.37}_{-4.67},
\end{equation}
in accordance with the experimental result given in (\ref{L-def}).

In addition to the CP-violating asymmetries that we considered already in 
our analysis, there is yet another one that is strongly constrained 
by the $B$-factory data \cite{HFAG}:
\begin{equation}\label{ACP-Bpi0K-def}
{\cal A}_{\rm CP}^{\rm dir}(B^\pm\to\pi^0 K^\pm)\equiv
\frac{\mbox{BR}(B^+\to\pi^0K^+)-
\mbox{BR}(B^-\to\pi^0 K^-)}{\mbox{BR}(B^+\to\pi^0K^+)+
\mbox{BR}(B^-\to\pi^0K^-)}=0.00\pm0.07.
\end{equation}
Using (\ref{B+pi0K+}) with $\rho_{\rm c}=0$, we obtain
\begin{equation}\label{ACP-Bpi0K-defa}
{\cal A}_{\rm CP}^{\rm dir}(B^\pm\to\pi^0 K^\pm)=
\frac{2}{R_{\rm c}}\left[r_{\rm c}\sin\delta_{\rm c}\sin\gamma
-qr_{\rm c}\left\{\sin(\delta_{\rm c}+\omega)\sin\phi+r_{\rm c}
\sin\omega\sin(\gamma-\phi)\right\}\right],
\end{equation}
where the expression for $R_{\rm c}$ is given in (\ref{Rc-expr}). We are
now in a position to calculate this CP asymmetry, where we obtain 
\begin{equation}
\left.{\cal A}_{\rm CP}^{\rm dir}(B^\pm\to\pi^0 K^\pm)\right|_{\rm SM}
=0.01^{+0.14}_{-0.10}
\end{equation}
within the SM, and
\begin{equation}
{\cal A}_{\rm CP}^{\rm dir}(B^\pm\to\pi^0 K^\pm)=0.04^{+0.37}_{-0.28}
\end{equation}
for our NP scenario, in accordance with the experimental value given in 
(\ref{ACP-Bpi0K-def}). On the other hand, the analysis performed in
\cite{BFRS-I} favoured generically larger CP asymmetries. Consequently, the
determination of the hadronic parameters through the $B\to\pi\pi$ analysis
and the introduction of the CP-violating NP phase $\phi$ allow us now to 
achieve a much better agreement with the experimental picture. 

Interestingly, we may also employ the experimental information on 
$R_{\rm c}$, $R_{\rm n}$ and the CP asymmetry
${\cal A}_{\rm CP}^{\rm dir}(B^\pm\to\pi^0 K^\pm)$ to determine 
the three EW penguin parameters $q$, $\phi$ and $\omega$ simultaneously. 
This analysis gives
\begin{equation}\label{q-omega-det}
q=1.74^{+1.28}_{-0.95},\quad \phi=-(85^{+11}_{-15})^\circ,\quad 
\omega=-(4^{+44}_{-32})^\circ,
\end{equation}
where a central value of $\omega$ being nicely consistent with 
$0^\circ$ -- as expected in the strict $SU(3)$ limit -- is very 
remarkable. While this feature supports also our working 
assumption of using the $SU(3)$ flavour symmetry, the reduction of the
error on $\omega$ would be very desirable.

Let us finally note, for completeness, that there is also a
mathematical solution with $\phi\sim0^\circ$ and $\omega\sim90^\circ$,
\begin{equation}\label{omega-crazy}
q=2.32^{+1.38}_{-1.04},\quad \phi=-(10^{+11}_{-12})^\circ,\quad 
\omega=(82^{+20}_{-28})^\circ.
\end{equation}
However, since such a value of $\omega$ looks completely unrealistic,
we will not consider this solution further.

\boldmath
\subsubsection{Elimination of the Second Solution for 
$xe^{i\Delta}$ in (\ref{x-det2})}\label{ssec:x-elim}
\unboldmath
Let us now come back to the second solution for $xe^{i\Delta}$
represented by (\ref{x-det2}). Using (\ref{r-det-Bpipi}) and the relations 
in (\ref{r-rho-rel})--(\ref{x-rel}), we obtain
\begin{equation}\label{rho-n-det-Bpipi2}
\rho_{\rm n}=0.11^{+0.05}_{-0.04},\quad 
\theta_{\rm n}=+(94^{+36}_{-40})^\circ
\end{equation}
\vspace*{-0.3truecm}
\begin{equation}\label{r-c-det-Bpipi2}
r_{\rm c}=0.20^{+0.09}_{-0.07},\quad \delta_{\rm c}=+(68^{+26}_{-25})^\circ,
\end{equation}
where the agreement between the values of $r_{\rm c}$ in 
(\ref{r-c-det-Bpipi}) and (\ref{r-c-det-Bpipi2}) is obvious from 
(\ref{rc-simple-expr}). If we assume that $\omega$ vanishes, 
(\ref{Rc-def}) and (\ref{Rn-def}) give
\begin{equation}
q=1.73^{+1.32}_{-0.69},\quad \phi=-(102^{+130}_{-29})^\circ,\end{equation}
yielding
\begin{equation}
{\cal A}_{\rm CP}^{\rm dir}(B^\pm\to\pi^0 K^\pm)=0.81^{+0.07}_{-0.82},
\end{equation}
which is very much disfavoured by the experimental result given in 
(\ref{ACP-Bpi0K-def}). Even the inclusion of a non-zero strong phase 
$\omega$ does not provide any physically attractive solution for the 
EW penguin parameters, implying, for instance, values for $q$ as high 
as $2.4$ that are totally exluded by the rare-decay constraints 
discussed in Section~\ref{sec:rare}. Consequently, we will not consider 
the second solution for $xe^{i\Delta}$ in (\ref{x-det2}) further.

\boldmath
\subsection{Prediction of CP Violation in $B_d\to\pi^0K_{\rm S}$}
\label{ssec:Bpi0KS-CPV}
\unboldmath
The decay $B_d\to\pi^0K_{\rm S}$ is a transition into a final state
with CP eigenvalue $-1$. Using the standard formalism for the
calculation of the observables provided by the corresponding 
time-dependent CP asymmetry \cite{RF-Phys-Rep}, as discussed for 
$B_d\to\pi^+\pi^-$ in Subsection~\ref{ssec:Bpipi-basic}, 
(\ref{B0pi0K0}) yields
\begin{eqnarray}
\lefteqn{{\cal A}_{\rm CP}^{\rm dir}(B_d\to\pi^0 K_{\rm S})=
\frac{2}{b}\Bigl[qr_{\rm c}\sin(\delta_{\rm c}+\omega)\sin\phi}\nonumber\\
&&-\rho_{\rm n}\left\{\sin\theta_{\rm n}\sin\gamma-qr_{\rm c}
\sin(\theta_{\rm n}-\delta_{\rm c}-\omega)
\sin(\gamma-\phi)\right\}\Bigr]\label{ACPdir-Bpi0K0}
\end{eqnarray}
\begin{eqnarray}\label{ACPmix-Bpi0K0}
\lefteqn{{\cal A}_{\rm CP}^{\rm mix}(B_d\to\pi^0 K_{\rm S})=
-\frac{1}{b}\Bigl[\sin\phi_d-2qr_{\rm c}\cos(\delta_{\rm c}+\omega)
\sin(\phi_d+\phi)+q^2r_{\rm c}^2\sin(\phi_d+2\phi)}\quad\\
&&+2\rho_{\rm n}\left\{\cos\theta_{\rm n}\sin(\phi_d+\gamma)-
qr_{\rm c}\cos(\theta_{\rm n}-\delta_{\rm c}-\omega)
\sin(\phi_d+\gamma+\phi)\right\}+\rho_{\rm n}^2\sin(\phi_d+2\gamma)
\Bigr],
\nonumber
\end{eqnarray}
where the expression for $b$ is given in (\ref{b-expr}). In the
special case of $\rho_{\rm n}=0$ and $\phi=0^\circ$, we arrive at the
following, well-known results \cite{PAPIII}:
\begin{eqnarray}
{\cal A}_{\rm CP}^{\rm dir}(B_d\to\pi^0 K_{\rm S})&=&0\\
{\cal A}_{\rm CP}^{\rm mix}(B_d\to\pi^0 K_{\rm S})&=&-\sin\phi_d
={\cal A}_{\rm CP}^{\rm mix}(B_d\to J/\psi K_{\rm S}),
\end{eqnarray}
where the average of the current $B$-factory data for the mixing-induced 
$B_d\to J/\psi K_{\rm S}$ CP asymmetry is given in (\ref{ACPmix-gol}). 
Recently, the BaBar collaboration reported the first results for these
observables \cite{browder-talk}, which are given by
\begin{eqnarray}
{\cal A}_{\rm CP}^{\rm dir}(B_d\to\pi^0 K_{\rm S})&=&
+0.40^{+0.27}_{-0.28}\pm0.10\label{Adir-Bdpi0KS}\\
{\cal A}_{\rm CP}^{\rm mix}(B_d\to\pi^0 K_{\rm S})&=&
-0.48_{-0.38}^{+0.47}\pm0.11.
\end{eqnarray}
Moreover, there is also the following measurement of the direct CP
asymmetry of the $B_d^0\to \pi^0K^0$ channel available \cite{HFAG}:
\begin{equation}
{\cal A}_{\rm CP}^{\rm dir}(B_d^0\to\pi^0 K^0)=-0.03\pm0.36\pm0.09,
\end{equation}
which is supposed to agree with (\ref{Adir-Bdpi0KS}). Consequently,
these experimental numbers are expected to change significantly in the
future. It will also be very exciting to see the corrsponding first Belle 
results.

\begin{figure}
\vspace*{0.3truecm}
\begin{center}
\psfrag{Amix}{${\cal A}_{\rm CP}^{\rm mix}(B_d\to\pi^0 K_{\rm S})$}
\psfrag{Adir}{${\cal A}_{\rm CP}^{\rm dir}(B_d\to\pi^0 K_{\rm S})$}
\psfrag{f}{\small $\!\!\phi$}\psfrag{expRegion}{exp.\ region}\psfrag{SM}{SM}
\psfrag{q069}{$q=0.69$}\psfrag{q175}{$q=1.75$}\psfrag{NP}{NP}
\includegraphics[width=10cm]{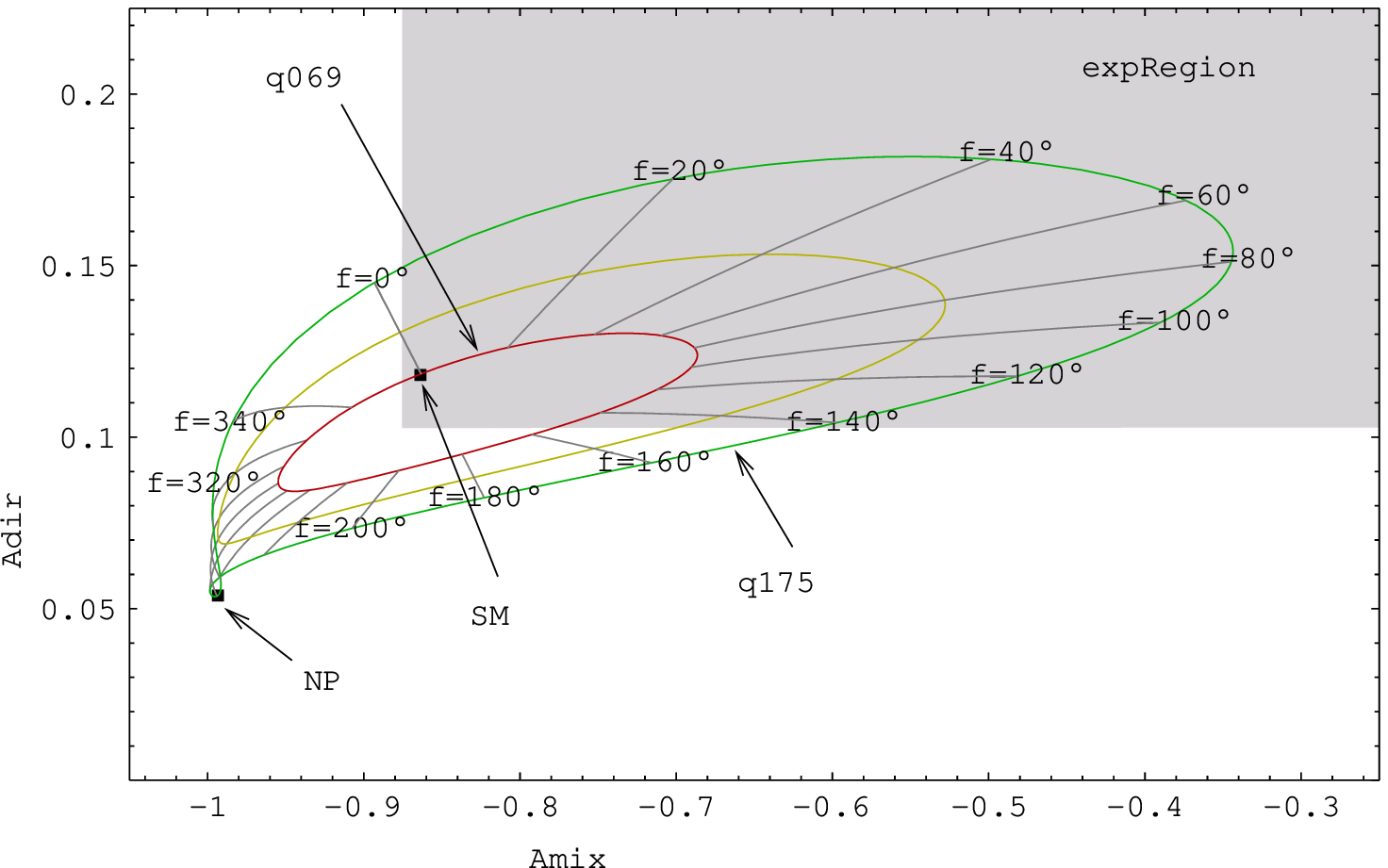}
\end{center}
\caption{The situation in the 
${\cal A}_{\rm CP}^{\rm mix}(B_d\to\pi^0 
K_{\rm S})$--${\cal A}_{\rm CP}^{\rm dir}(B_d\to\pi^0 K_{\rm S})$ plane. 
We show contours for values of 
$q=0.69$ to $q=1.75$ and $\phi \in
[0^\circ,360^\circ]$. The grey area shows the lower left-hand corner
of the BaBar 1$\sigma$ range, whilst the central value is outside 
the plotted area. The central values of 
(\ref{CP-dir-Bdpi0KS-pred-SM}), (\ref{CP-mix-Bdpi0KS-pred-SM})
and (\ref{CP-dir-Bdpi0KS-pred}), (\ref{CP-mix-Bdpi0KS-pred})
are indicated by the labels ``SM'' and ``NP'', respectively.
\label{fig:Adirpi0KS-Amixpi0K}}
\end{figure}

Following the analysis performed in Subsection~\ref{ssec:BpiK-c-n},
we may predict the CP-violating $B_d\to\pi^0 K_{\rm S}$ observables. 
In order to illustrate the dependence on the EW penguin parameters,
we employ -- in analogy to Fig.~\ref{fig:Rn-Rc} -- the central values of 
the hadronic parameters in (\ref{r-det-Bpipi})--(\ref{r-c-det-Bpipi}),
$\omega=0^\circ$, $\gamma=65^\circ$, and show in 
Fig.~\ref{fig:Adirpi0KS-Amixpi0K} the contours in the 
${\cal A}_{\rm CP}^{\rm mix}(B_d\to\pi^0 
K_{\rm S})$--${\cal A}_{\rm CP}^{\rm dir}(B_d\to\pi^0 K_{\rm S})$ plane 
corresponding to various values of $q$ with $\phi\in[0^\circ,360^\circ]$. 
If we take the uncertainties both of the parameters in 
(\ref{r-det-Bpipi})--(\ref{r-c-det-Bpipi}) and of $\gamma$ in 
(\ref{UT-angles}) into account, and assume again that $\omega=0^\circ$, 
the SM expression (\ref{q-SM}) yields
\begin{eqnarray}
\left.{\cal A}_{\rm CP}^{\rm dir}(B_d\to\pi^0 K_{\rm S})\right|_{\rm SM}&=&
+0.12^{+0.11}_{-0.13}\label{CP-dir-Bdpi0KS-pred-SM}\\
\left.{\cal A}_{\rm CP}^{\rm mix}(B_d\to\pi^0 K_{\rm S})\right|_{\rm SM}&=&
-0.86^{+0.05}_{-0.07},\label{CP-mix-Bdpi0KS-pred-SM}
\end{eqnarray}
whereas our NP scenario, which is characterized by the values
of $q$ and $\phi$ in (\ref{q-det}), corresponds to the following 
prediction:
\begin{eqnarray}
{\cal A}_{\rm CP}^{\rm dir}(B_d\to\pi^0 K_{\rm S})&=&
+0.05^{+0.24}_{-0.29}\label{CP-dir-Bdpi0KS-pred}\\
{\cal A}_{\rm CP}^{\rm mix}(B_d\to\pi^0 K_{\rm S})&=&
-0.99^{+0.04}_{-0.01}.\label{CP-mix-Bdpi0KS-pred}
\end{eqnarray}
The measurement of these CP asymmetries will allow a crucial test of
our NP scenario. By the time solid experimental numbers are available, 
the uncertainties of the parameters entering (\ref{CP-dir-Bdpi0KS-pred}) 
and (\ref{CP-mix-Bdpi0KS-pred}) are expected to be significantly smaller,
thereby leading to much more stringent predictions of CP violation in 
$B_d\to\pi^0K_{\rm S}$.\footnote{It is evident from
Fig.\ \ref{fig:Adirpi0KS-Amixpi0K} that the error in 
(\ref{CP-mix-Bdpi0KS-pred}) is accidentally small.}  Constraints 
for these observables were also
recently derived, within the framework of the SM, in \cite{GGR}.

\boldmath
\subsection{Future Avenues Offered by $B_s$ Decays}
\unboldmath
The physics potential of $B_s$-meson decays, which can be exploited at
hadronic $B$-decay experiments, i.e.\ at run II of the Tevatron 
\cite{TEV-BOOK} and later on at the LHC \cite{LHC-BOOK}, provides
interesting strategies to explore CP violation and to obtain insights into
hadronic physics. These results will nicely complement the 
$B_{u,d}\to\pi\pi,\pi K$ methods proposed above. 
\boldmath
\subsubsection{$B_s\to K^+K^-$}\label{ssec:BsKK}
\unboldmath
As pointed out in \cite{RF-BsKK}, the decay $B_s\to K^+K^-$ can be
related to $B_d\to\pi^+\pi^-$ through the $U$-spin flavour symmetry 
of strong interactions, thereby providing attractive simultaneous 
determinations of $\gamma$, $d$ and $\theta$, as well as insights 
into $U$-spin-breaking effects. It will be very exciting to
see whether we will arrive at a picture that is consistent with
the one developed in Section~\ref{sec:Bpipi}. Interestingly, we may 
use the hadronic parameters determined there to make {\it predictions} 
for the CP-violating $B_s\to K^+K^-$ observables with the help of
the $U$-spin flavour symmetry, which implies the following
expressions:
\begin{equation}\label{ACPdir-BsKK}
{\cal A}_{\rm CP}^{\rm dir}(B_s\to K^+K^-)=\frac{2\epsilon d
\sin\theta\sin\gamma}{\epsilon^2+2\epsilon d\cos\theta\cos\gamma+d^2}
\end{equation}
\begin{equation}\label{ACPmix-BsKK}
{\cal A}_{\rm CP}^{\rm mix}(B_s\to K^+K^-)=
\frac{\epsilon^2\sin(\phi_s+2\gamma)+2\epsilon d\cos\theta\sin(\phi_s+\gamma)
+d^2\sin\phi_s}{\epsilon^2+2\epsilon d\cos\theta\cos\gamma+d^2},
\end{equation}
where the $B^0_s$--$\bar B^0_s$ mixing phase $\phi_s$ is given by
\begin{equation}
\phi_s=-2\lambda^2\bar\eta=2\beta_s
\end{equation}
in the SM, with the numerical value of $\beta_s$ in (\ref{CKM2}). 
Note that the expression for 
${\cal A}_{\rm CP}^{\rm dir}(B_s\to K^+K^-)$ agrees with the one
for ${\cal A}_{\rm CP}^{\rm dir}(B_d\to \pi^\mp K^\pm)$ in 
(\ref{ACPdir}). Using now the range for $\gamma$ in (\ref{UT-angles}) 
and the hadronic parameters in (\ref{d-det}), we obtain the following 
SM predictions:
\begin{eqnarray}
\left.{\cal A}_{\rm CP}^{\rm dir}(B_s\to K^+K^-)\right|_{\rm SM}&=&
0.14^{+0.14}_{-0.09}\label{ACP-dir-BsKK}\\
\left.{\cal A}_{\rm CP}^{\rm mix}(B_s\to K^+K^-)\right|_{\rm SM}&=&
-0.18^{+0.08}_{-0.07},\label{ACP-mix-BsKK}
\end{eqnarray}
where the latter observable may be affected by NP contributions 
to $B^0_s$--$\bar B^0_s$ mixing. By the time the CP-violating $B_s\to K^+K^-$
observables can be measured, more precise SM predictions will be
available. 

Moreover, we may also explore the branching ratio of this channel.
To this end, we use the quantity $H$ introduced in (\ref{H-expr}).
If we use again (\ref{UT-angles}) and (\ref{d-det}), we obtain
\begin{equation}\label{H-SM}
\left. H \right|_{\rm SM}=7.0^{+7.4}_{-4.7},
\end{equation}
which is in nice accordance with the numerical value in 
(\ref{H-simple}).\footnote{Because of the discussion in 
Subsection~\ref{ssec:Bpipi-gam-det}, this feature is of course 
not surprising.}
In order to be able to predict $\mbox{BR}(B_s\to K^+K^-)$, we have
to know the $U$-spin-breaking factor $|{\cal C}'/{\cal C}|$, which is
given -- within the factorization approximation -- as follows \cite{RF-BsKK}:
\begin{equation}\label{U-fact}
\left|\frac{{\cal C}'}{{\cal C}}\right|_{\rm fact}=
\frac{f_K}{f_\pi}\frac{F_{B_sK}(M_K^2;0^+)}{F_{B_d\pi}(M_\pi^2;0^+)}
\left(\frac{M_{B_s}^2-M_K^2}{M_{B_d}^2-M_\pi^2}\right).
\end{equation}
In a recent analysis \cite{KMM}, this parameter has been calculated
through QCD sum rules, with the following result:
\begin{equation}\label{U-fact-est}
\left|\frac{{\cal C}'}{{\cal C}}\right|_{\rm fact}=
1.76^{+0.15}_{-0.17}.
\end{equation}
If we now complement (\ref{H-expr}) with (\ref{H-SM}) and
(\ref{U-fact-est}), and use the experimental result
$\mbox{BR}(B_d\to\pi^+\pi^-)=(4.6\pm0.4)\times 10^{-6}$ 
\cite{HFAG}, we obtain
\begin{equation}\label{BR-BsKK}
\mbox{BR}(B_s\to K^+K^-)=(35^{+73}_{-20})\times 10^{-6}.
\end{equation}
The very large uncertainty reflects the fact that our analysis
is rather insensitive to $H$ (this can also be seen in Fig.~\ref{fig:theta-d}),
and therefore the predicted value (\ref{H-SM}) has a large
uncertainty that propagates to (\ref{BR-BsKK}).

Alternatively, we may -- in the spirit of (\ref{H-simple}) --
assume that the penguin annihilation and exchange topologies 
discussed in Subsection~\ref{ssec:PA-E} play a minor r\^ole.
Taking, moreover, factorizable $SU(3)$-breaking corrections into
account, we obtain
\begin{eqnarray}
\frac{\mbox{BR}(B_s\to K^+K^-)}{\mbox{BR}(B_d\to \pi^\mp K^\pm)}
&=&\left[\frac{M_{B_d}}{M_{B_s}}
\frac{\Phi(M_K/M_{B_s},M_K/M_{B_s})}{\Phi(M_\pi/M_{B_d},M_K/M_{B_d})}
\frac{\tau_{B_s^0}}{\tau_{B_d^0}}\right]\nonumber\\
&&\times\left[\frac{F_{B_sK}(M_K^2;0^+)}{F_{B_d\pi}(M_\pi^2;0^+)}
\left(\frac{M_{B_s}^2-M_K^2}{M_{B_d}^2-M_\pi^2}\right)\right]^2,
\end{eqnarray}
where
\begin{equation}
\frac{F_{B_sK}(M_K^2;0^+)}{F_{B_d\pi}(M_\pi^2;0^+)}
\left(\frac{M_{B_s}^2-M_K^2}{M_{B_d}^2-M_\pi^2}\right)=1.45^{+0.13}_{-0.14}
\end{equation}
corresponds to (\ref{U-fact-est}) \cite{KMM}. If we then use the experimental
result $\mbox{BR}(B_d\to\pi^\mp K^\pm)=(18.2\pm0.8)\times 10^{-6}$
\cite{HFAG}, we arrive at
\begin{equation}
\mbox{BR}(B_s\to K^+K^-)=(35\pm7)\times 10^{-6}.
\end{equation}
A measurement of this branching ratio, which should soon be
available from run II of the Tevatron, will be very interesting, 
allowing in particular valuable insights into (\ref{U-fact-est}). It is
interesting to note that -- in contrast to (\ref{BR-BsKK}) -- 
the predictions of the CP-violating observables in (\ref{ACP-dir-BsKK}) and 
(\ref{ACP-mix-BsKK}) are not affected by factorizable $U$-spin-breaking 
corrections \cite{RF-BsKK}, i.e.\ do not involve a ratio of form 
factors as in (\ref{U-fact}).

\boldmath
\subsubsection{$B_s\to\pi^\pm K^\mp$}
\unboldmath
The decay $B^0_s\to \pi^+K^-$ is related to the $B^0_d\to \pi^-K^+$ mode
through the $U$-spin flavour symmetry of strong interactions \cite{GR-U-spin}, 
which allows us to write 
\begin{equation}
A(B^0_s\to \pi^+K^-)=\lambda^3 A ({\cal P}_{t}'-{\cal P}_{c}')\left[1+
\frac{1}{\epsilon}re^{i\delta}e^{i\gamma}\right],
\end{equation}
complementing the $B^0_d\to \pi^-K^+$ amplitude in (\ref{B0pi-K+}).
If we then combine the direct CP asymmetry
\begin{equation}\label{ACPdir-BspiK}
{\cal A}_{\rm CP}^{\rm dir}(B_s\to \pi^\pm K^\mp)=
-\left[\frac{2\epsilon r\sin\delta\sin\gamma}{\epsilon^2+2\epsilon r
\cos\delta\cos\gamma+r^2}\right]
\end{equation}
with its $B_d\to\pi^\mp K^\pm$ counterpart in (\ref{ACPdir-r}), 
we may determine $r$ and $\delta$ for given values of 
$\gamma$.\footnote{If we consider, in addition, the ratio of the 
CP-averaged $B_s\to \pi^\pm K^\mp$ and $B^\pm\to\pi^\pm K$ branching 
ratios, we may determine $\gamma$ as well \cite{GR-U-spin}. This
extraction involves, however, the ratio 
$F_{B_sK}(M_\pi^2;0^+)/F_{B_d\pi}(M_K^2;0^+)$ of $SU(3)$-breaking form
factors.}
Fixing then $r_{\rm c}$ through (\ref{rc-alt-det}), we may use 
(\ref{r-rho-rel}) to eliminate $\rho_{\rm n}e^{i\theta_{\rm n}}$ in the 
$B^0_d\to\pi^0K^0$ amplitude, so that $R_{\rm n}$ and the two 
$B_d\to\pi^0K_{\rm S}$ CP asymmetries depend on the strong phase 
$\delta_{\rm c}$ and the EW penguin parameters $q$, $\phi$ (and $\omega$). 
If we complement these observables with $R_{\rm c}$ and the direct 
$B^\pm\to\pi^0 K^\pm$ CP asymmetry, we may extract these parameters 
and may perform internal consistency checks. 

The advantage of this avenue is that it is {\it not} affected by the 
penguin annihilation and exchange topologies discussed in 
Subsection~\ref{ssec:PA-E}. Consequently, it will be very interesting 
to see whether we will eventually arrive at a consistent overall picture.
If we neglect the penguin annihilation and exchange topologies and use the 
$SU(3)$ flavour symmetry, we obtain the simple relation
\begin{equation}
{\cal A}_{\rm CP}^{\rm dir}(B_s\to \pi^\pm K^\mp)=
{\cal A}_{\rm CP}^{\rm dir}(B_d\to \pi^+\pi^-),
\end{equation}
which should already provide important insights into this issue. 

Let us finally note that EW penguins enter the $B_s$ modes considered
above only in colour-suppressed form. In the case of $\phi=0^\circ$,
these topologies would {\it not} affect our $B_s$ strategies at all,
as becomes obvious from the discussion in Appendix~\ref{APP:BpiK-CSEWP}. 
On the other hand, for $\phi\not=0^\circ$, anomalously enhanced 
colour-suppressed EW penguins would manifest themselves in the 
corresponding data. Consequently, the $B_s$ studies complement the 
strategies to address the colour-suppressed EW penguins discussed 
in Appendix~\ref{APP:BpiK-CSEWP}.

\subsection{Summary}
Before turning to rare $K$ and $B$ decays as well as $\epe$ in the
next section, let us summarize the main results of our analysis
of the $B\to\pi K$ system:
\begin{itemize}
\item Employing the $SU(2)$ isospin flavour symmetry of strong 
interactions, we have given a parametrization of the $B\to\pi K$ 
amplitudes, which allows us to deal with CP-violating NP effects in 
the EW penguin sector. Moreover, the relevant hadronic parameters are 
introduced in such a manner that we may determine them with the help 
of the $B\to\pi\pi$ analysis performed in Section~\ref{sec:Bpipi}. 
To this end, we have to neglect penguin annihilation and exchange 
topologies, and have to employ the $SU(3)$ flavour symmetry of
strong interactions. 
\item We find a remarkable agreement between the
corresponding determination of $r_{\rm c}$ and the value following from 
an alternative strategy, i.e.\ we arrive at a consistent picture, which
would be spoiled if the working assumptions specified in the previous
item were not satisfied. On the other hand, if we impose the 
constraint that the two values of $r_{\rm c}$ agree with each other, 
we may refine the extraction of $\gamma$ discussed in 
Subsection~\ref{ssec:Bpipi-gam-det}. In particular, this additional
input allows us to resolve the twofold ambiguity, leaving us with
$\gamma=(64.7^{+6.3}_{-6.9})^\circ$, which is in excellent 
agreement with the SM expectation given in (\ref{UT-angles}). 
\item Having all relevant parameters at hand, we may analyse the $B\to\pi K$
observables within the SM. As far as the 
$B_d\to\pi^\mp K^\pm$, $B^\pm\to \pi^\pm K$ system is concerned, where
EW penguins may only enter in colour-suppressed form and are hence
expected to play a minor r\^ole, we arrive at a picture 
that is in nice agreement with the $B$-factory data and does not indicate 
any anomalous behaviour. On the other hand, the SM analysis of the 
observables $R_{\rm c}$ and $R_{\rm n}$ of the charged and neutral 
$B\to\pi K$ systems, respectively, which are significantly affected by 
EW penguins, is in conflict with the pattern of the $B$-factory data. 
\item Interestingly, we can resolve this ``puzzle'' with the help of
NP in the EW penguin sector. In particular, we arrive at the values
of $q=1.75^{+1.27}_{-0.99}$ and $\phi=-(85^{+11}_{-14})^\circ$,
i.e.\ at enhanced EW
penguins with a large CP-violating NP phase. Moreover, we find the
strong phase $\omega=-(4^{+44}_{-32})^\circ$, in  agreement with
the implication of the $SU(3)$ flavour symmetry that this phase
vanishes. Let us emphasize that this picture corresponds to small
direct CP violation in $B^\pm\to\pi^0K^\pm$, as indicated by the
$B$-factory data. Finally, we may {\it predict} the CP-violating observables
of the $B_d\to\pi^0K_{\rm S}$ channel, which will provide a crucial 
test of our NP scenario. 
\item The impact of colour-suppressed 
EW penguins on our analysis has been addressed in 
Appendix~\ref{APP:BpiK-CSEWP}. 
The current $B$-factory data do not indicate 
any enhancement of these topologies. 
\item The decays $B_s\to K^+K^-$ and $B_s\to\pi^\pm K^\mp$, which are
very accessible at hadronic $B$-decay experiments, complement the
analysis of the $B_{u,d}\to\pi\pi,\pi K$ modes proposed above in a variety
of ways. In particular the latter channel allows us to avoid assumptions
about penguin annihilation and exchange topologies. 
\end{itemize}

\section{\boldmath Rare $K$ and $B$ Decays 
and $\epe$ \unboldmath}\label{sec:rare}
\setcounter{equation}{0}
\subsection{Preliminaries}
As discussed in Section~\ref{sec:Basic}, the rare $K$ and $B$ decays are 
governed by the functions $X$, $Y$ and $Z$, for which the parametrization 
in terms of $|C|$ and $\theta_C$ was given in (\ref{PARXYZ}). In order 
to obtain elegant expressions for rare decays, it is useful to go one step 
further and to rewrite (\ref{PARXYZ}) as follows:
\be\label{NXYZ}
X=|X|e^{i\theta_X},\quad Y=|Y|e^{i\theta_Y},\quad Z=|Z|e^{i\theta_Z},
\ee
with 
\be\label{Xv}
|X|=\sqrt{5.52\,\bar q^2-0.42\,\bar q \cos\phi+0.01},\quad  
\tan\theta_X=\frac{2.35\,\bar q\sin\phi}{2.35\,\bar q\cos\phi-0.09}
\ee

\be\label{Yv}
|Y|=\sqrt{5.52\,\bar q^2-3.00\,\bar q \cos\phi+0.41},\quad  
\tan\theta_Y=\frac{2.35\,\bar q\sin\phi}{2.35\,\bar q\cos\phi-0.64}
\ee

\be\label{Zv}
|Z|=\sqrt{5.52\,\bar q^2-4.42\,\bar q \cos\phi+0.88},\quad  
\tan\theta_Z=\frac{2.35\,\bar q\sin\phi}{2.35\,\bar q\cos\phi-0.94}.
\ee
It should be emphasized that only two independent parameters, $\bar q$ and 
$\phi$, appear in these expressions.

Next we define the following weak phases:
\be\label{betas}
\beta_X\equiv \beta-\beta_s-\theta_X,\quad 
\beta_Y\equiv \beta-\beta_s-\theta_Y,\quad
\beta_Z\equiv \beta-\beta_s-\theta_Z.
\ee
In \cite{BFRS-II}, we have suppressed $\beta_s$ in all formulae for rare
decays but we have included its effect in the numerical analysis.

Finally, the numerical constants in the formulae below correspond to 
\cite{PDG} and 
\be\label{PDG0}
\sin^2\theta_{\rm w}=0.231,\quad \alpha=\frac{1}{128},\quad \lambda=0.224,
\ee
with the first two given in the $\overline{\mbox{MS}}$ scheme. 
They are the same as in \cite{Cracow}. 
The values of the remaining CKM parameters have been given in 
Section~\ref{sec:Basic}.

With the result for $(q,\phi)$ in (\ref{q-det}) at hand, we can calculate 
the functions
$X$, $Y$ and $Z$ and the weak phases $\beta_X$, $\beta_Y$ and $\beta_Z$. 
Setting $|V_{ub}/V_{cb}|=0.086$, that is $\bar q=q$, we find 
\be
|X|\approx |Y|\approx |Z| \approx 4.3^{+3.0}_{-2.4}~.
\ee
While the central value of $|X|$ is still compatible with the data on 
$K\to\pi \nu\bar\nu$ and $B\to X_{s,d}\nu\bar\nu$, the central value 
of $|Y|$  violates the 
upper bound $|Y|\le 2.2$ following from the BaBar and Belle data on 
$B\to X_s\mu^+\mu^-$ \cite{Kaneko:2002mr}, and the upper bound on
$\mbox{BR}(K_{\rm L}\to \pi^0 e^+e^-)$ in (\ref{ktev1}) from KTeV 
\cite{KTEVKL}. In addition, $|Z|$ is too large to be consistent with the 
data on $\epe$, even if the hadronic uncertainties in this ratio are large. 

On the other hand, the $B\to \pi K$ data seem to signal the possibility of 
enhanced values of $|X|$, $|Y|$ and $|Z|$ and of large weak phases
$\theta_i$. Consequently, we may still encounter significant deviations from
the SM predictions for rare decays, while being consistent with all
experimental data.
In order to illustrate this exciting feature, we consider
only the subset of those values of $(q,\phi)$ in (\ref{q-det}) that 
satisfy the constraint of $|Y|= 2.2$.
Using (\ref{Yv}) and expressing $\bar q$ in terms of $|Y|$ and $\phi$, and 
subsequently varying $\phi$ in the full range given in (\ref{q-det}), 
we obtain
\be\label{qnew}
\bar q= 0.92^{+0.07}_{-0.05} ,\quad \phi=-(85^{+11}_{-14})^\circ.
\ee
These values are compatible with all data on rare decays and also with the
$B\to\pi K$ data. In particular, we find the values for  
$R_{\rm c}$, $R_{\rm n}$, 
${\cal A}_{\rm CP}^{\rm dir}(B^\pm\to\pi^0 K^\pm)$,
${\cal A}_{\rm CP}^{\rm dir}(B_d\to\pi^0 K_{\rm S})$,  
${\cal A}_{\rm CP}^{\rm mix}(B_d\to\pi^0 K_{\rm S})$,
$L$ and $b$ given in the third column of 
Table~\ref{tab:with-withoutconstraints}.
To this end we have varied 
$\phi$ in the range given in (\ref{qnew}), keeping $|Y|=2.2$.
We compare the result of this exercise with the values for the $B\to\pi K$ 
observables obtained without the $|Y|=2.2$ constraint and with the data.

\begin{table}[hbt]
\vspace{0.4cm}
\begin{center}
\begin{tabular}{|c||c|c|c|}
\hline
Quantity & Without RD constraint & With RD constraint & Experiment
 \\ \hline
$R_{\rm c}$ & $1.17\pm0.12$ (input) & 
$1.00^{+0.12}_{-0.08}$ \rule{0em}{1.05em} & $1.17\pm0.12$
\\ \hline
$R_{\rm n}$ & $0.76\pm0.10$ (input) & 
$0.82^{+0.12}_{-0.11}$ \rule{0em}{1.05em} & $0.76\pm0.10$
\\ \hline
${\cal A}_{\rm CP}^{\rm dir}(B^\pm\!\to\!\pi^0 K^\pm) $ 
 & $0.04^{+0.37}_{-0.28}$ & $0.03^{+0.32}_{-0.24}$ 
\rule{0em}{1.05em} & $0.00\pm0.07$
\\ \hline
${\cal A}_{\rm CP}^{\rm dir}(B_d\!\to\!\pi^0 K_{\rm S})$ 
 & $0.05^{+0.24}_{-0.29}$ & $0.08^{+0.18}_{-0.22}$ \rule{0em}{1.05em} 
& $0.40^{+0.29}_{-0.30}$
\\ \hline 
${\cal A}_{\rm CP}^{\rm mix}(B_d\!\to\!\pi^0 K_{\rm S})$ 
 & $-0.99^{+0.04}_{-0.01}$ & $-0.98^{+0.05}_{-0.02}$ \rule{0em}{1.05em} 
& $-0.48_{-0.40}^{+0.48}$
\\ \hline
$L$ & $6.02^{+7.37}_{-4.67}$ & $2.67^{+0.34}_{-0.30}$ \rule{0em}{1.05em} 
& $5.9_{-2.7}^{+3.0}$
\\ \hline
$b$ & $1.24^{+0.19}_{-0.15}$ & $1.15^{+0.16}_{-0.13}$ \rule{0em}{1.05em} 
& $1.19\pm 0.16$
\\ \hline
\end{tabular}
\end{center}
\caption[]{\small Theoretical predictions with and without the $|Y|=2.2$ 
rare decays (RD) constraint and
experimental data for the most important observables.
\label{tab:with-withoutconstraints}}
\end{table}

Proceeding in the same manner, we also  find
\be\label{rC}
|C|=2.24\pm0.04 , \qquad  \theta_C= -(105\pm12)^\circ
\ee
\be\label{rX}
|X|=2.17\pm 0.12, \qquad  \theta_X= -(86\pm 12)^\circ, \qquad
\beta_X= (111\pm 12)^\circ, 
\ee
\be\label{rY}
|Y|=2.2~{\rm (input)}, \qquad  \theta_Y= -(100\pm 12)^\circ, \qquad 
\beta_Y=(124\pm 12)^\circ,
\ee
\be\label{rZ}
|Z|=2.27\pm0.06   , \qquad  \theta_Z=-(108\pm 12)^\circ, \qquad
 \beta_Z= (132\pm 12)^\circ,
\ee
to be compared with $C=0.79$, $X=1.53$, $Y=0.98$ and $Z=0.68$ in the SM for 
$m_t=167~{\rm GeV}$. We will now turn to the implications of these results 
for rare $K$ and $B$ decays.

\boldmath
\subsection{$K\to\pi\nu\bar\nu$}
\unboldmath
\subsubsection{Basic Formulae}
The rare decays $\kpn$ and $\klpn$  
proceed through $Z^0$-penguin
 and box diagrams. As the required hadronic matrix elements can be extracted 
from the leading semileptonic decays and other long-distance contributions 
turn out to be negligible \cite{Gino03}, 
the relevant branching ratios can be 
computed
to an exceptionally high degree of precision \cite{BB}--\cite{MU98}. 
In \cite{Buras:1998ed}--\cite{Buras:1999da}, these decays have already 
been discussed 
in the NP scenario considered here. Below we update the formulae of 
these papers, adapt them to our notation, derive a few new ones,
and include $\ord(\lambda^2)$ terms that were neglected there.
The branching ratios are then given as follows:
\begin{equation}\label{bkpn}
\mbox{BR}(\kpn)=4.78\times 10^{-11}\times\left[\tilde r^2 A^4 R_t^2 |X|^2
+2 \tilde r \bar P_c(X) A^2 R_t |X|\cos\beta_X
+ \bar P_c(X)^2  \right]
\end{equation}
\begin{equation}\label{bklpn}
\mbox{BR}(K_{\rm L}\to\pi^0\nu\bar\nu)=2.09\times 10^{-10}\times
\tilde r^2 A^4 R_t^2 |X|^2\sin^2\beta_X,
\end{equation}
with
\be
\bar P_c(X)=\left(1-\frac{\lambda^2}{2}\right) P_c(X),
\ee
where
$P_c(X)=0.39\pm 0.06$ results from the internal charm 
contribution \cite{BB,BB98}, $\beta_X$ is defined in (\ref{betas}) and 
$\tilde r$ in (\ref{LAMT}). 

Once $\mbox{BR}(\kpn)$ and $\mbox{BR}(\klpn)$ have been measured, the
parameters $|X|$ and $\beta_X$ can be determined, subject to ambiguities that
can be resolved by considering other processes, such as the non-leptonic 
$B$ decays discussed before and the rare decays discussed below. 
Combining (\ref{bkpn}) and (\ref{bklpn}), we find
\be\label{rs-def}
r_s=\frac{\varepsilon_1\sqrt{B_1-B_2}-\bar P_c(X)}
{\varepsilon_2\sqrt{B_2}}=\cot\beta_X,
\ee
where $\varepsilon_i=\pm 1$, and consequently
\be\label{sin2bnunu}
\sin 2\beta_X= \frac{2 r_s}{1+r_s^2}. 
\ee
Moreover,
\be\label{modX}
|X|=\frac{\varepsilon_2\sqrt{B_2}}{\tilde r A^2R_t\sin\beta_X},\quad
\varepsilon_2\sin\beta_X >0.
\ee
 The ``reduced'' branching ratios
$B_i$ are given by
\begin{equation}\label{b1b2}
B_1={\mbox{BR}(\kpn)\over 4.78\times 10^{-11}},\quad
B_2={\mbox{BR}(\klpn)\over 2.09\times 10^{-10}}.
\end{equation}

The formulae (\ref{sin2bnunu}) and (\ref{modX})
were already presented in \cite{BRS}. They are valid for arbitrary 
$\beta_X\not=0^\circ$ and generalize the ones given for the SM and MFV 
models in 
\cite{BBSIN} and \cite{BF01}, respectively. 
For $\theta_X=0^\circ$ and $\varepsilon_1=\varepsilon_2=1$, one obtains from 
(\ref{sin2bnunu}) the SM result that differs at 
first sight from the one given in \cite{BBSIN}. However, in that 
paper and subsequent studies in the literature, a formula for $\sin 2\beta$ 
and not $\sin 2(\beta-\beta_s)$ was given. Using the fact that 
$\beta_s=\ord(\lambda^2)$, one can verify that (\ref{sin2bnunu}), while being 
slightly more accurate, is numerically very close to the formula of 
\cite{BBSIN}. In the scenario considered here, we have
 $99^\circ\le \beta_X\le 125^\circ$ and, consequently, 
$\varepsilon_1=-1$ and $\varepsilon_2=1$.

It should be stressed that $\sin 2\beta_X$ determined 
this way depends only on two measurable branching ratios and on 
$\bar P_c(X)$, which is completely calculable in perturbation theory.
Consequently, this determination is free from any hadronic
uncertainties and its accuracy can be estimated with a high degree
of confidence. 
As in our scenario $\beta$ and $\beta_s$ are already known from the usual 
analysis of the UT, the measurement of $r_s$ in $K\to\pi\nu\bar\nu$ decays 
will provide  a theoretically clean determination of $\theta_X$. Similarly, 
a clean determination of $|X|$ is possible by means of (\ref{modX}), 
so that (\ref{Xv}) will allow us to determine $\bar q$ and $\phi$. 
Assuming that 
the measurements of $\mbox{BR}(\kpn)$ and $\mbox{BR}(\klpn)$
with $10\%$ accuracy will be performed one day, the decays in question will 
most probably provide the cleanest measurements of $\bar q$ and $\phi$. 
This determination could then be compared with the one from other decays, 
in particular from
${\cal A}_{\rm CP}^{\rm dir}(B_d\to\pi^0K_{\rm S})$ and
${\cal A}_{\rm CP}^{\rm mix}(B_d\to\pi^0K_{\rm S})$ that we proposed 
in Section~\ref{sec:BpiK}. It could also be used to calculate other 
$B\to \pi K$ observables.

\subsubsection{Numerical Results}
Using the results for $\bar q$ and $\phi$ in (\ref{qnew}) 
and the parameters in (\ref{CKM1}), (\ref{CKM2}) and 
(\ref{PDG0}), we find
\begin{equation}\label{NPkpnr}
\mbox{BR}(\kpn)=
(7.5 \pm 2.1)\times 10^{-11},\quad \mbox{BR}(\klpn)=
(3.1 \pm 1.0)\times 10^{-10}.  
\end{equation}
This should be compared with the SM prediction, for which we find 
\begin{equation}\label{SMkpnr}
\mbox{BR}(\kpn)_{\rm SM}=
(8.0 \pm 1.1)\times 10^{-11},\quad \mbox{BR}(\klpn)_{\rm SM}=
(3.2 \pm 0.6)\times 10^{-11} , 
\end{equation}
in the ballpark of other recent estimates \cite{Gino03,KENDAL}.
On the experimental side, the results of the   
AGS E787 \cite{Adler97} and KTeV \cite{KTeV00X} collaborations are 
\be\label{kp01}
\mbox{BR}(K^+ \rightarrow \pi^+ \nu \bar{\nu})=
(15.7^{+17.5}_{-8.2})\times 10^{-11} \quad\mbox{and}\quad
\mbox{BR}(\klpn)<5.9 \times 10^{-7},
\end{equation}
respectively. 

We observe that in our scenario $\mbox{BR}(\kpn)$ does not significantly
differ from the SM estimate because the enhancement of the first term in 
(\ref{bkpn}) is to a large extent compensated by the suppression of the
second term ($\cos \beta_X\ll\cos(\beta-\beta_s)$) and its reversed sign.
Consequently, $\mbox{BR}(K^+ \rightarrow \pi^+ \nu \bar{\nu})$ is very 
strongly
dominated by the ``top" contribution given by the function $X$.

On the other hand, we observe a spectacular enhancement of $\mbox{BR}(\klpn)$
by one order of magnitude.
Consequently,  while $\mbox{BR}(\klpn)\approx (1/3)\mbox{BR}(\kpn)$ 
in the SM, it is substantially larger than 
$\mbox{BR}(K^+ \rightarrow \pi^+ \nu \bar{\nu})$ in our scenario.
The huge enhancement of $\mbox{BR}(\klpn)$ found here
is mainly due to 
the large weak phase $\beta_X\approx 111^\circ$, as 
\be
\frac{\mbox{BR}(\klpn)}{\mbox{BR}(\klpn)_{\rm SM}}=
\left|\frac{X}{X_{\rm SM}}\right|^2
\left[\frac{\sin\beta_X}{\sin(\beta-\beta_s)}\right]^2. 
\ee

Inspecting (\ref{bkpn}) and (\ref{bklpn}), we observe that the very strong 
dominance of the ``top" contribution in these expressions implies a simple 
approximate expression:
\be
\frac{\mbox{BR}(\klpn)}{\mbox{BR}(\kpn)}\approx 4.4\times (\sin\beta_X)^2
\approx 4.2\pm 0.2.  
\ee
We note that $\mbox{BR}(\klpn)$ is then rather close to its model-independent
upper bound~\cite{GRNR}
\be\label{BOUND}
\mbox{BR}(\klpn)\le 4.4\, \mbox{BR}(\kpn).
\ee
It is evident from (\ref{rs-def}) that this bound
is reached when the reduced branching ratios $B_1$ and $B_2$ in
(\ref{b1b2}) are equal to each other.

A spectacular 
implication of these findings is a strong violation of the relation 
\cite{BBSIN}
\be 
(\sin 2 \beta)_{\pi \nu\bar\nu}=(\sin 2 \beta)_{\psi K_{\rm S}},
\ee 
which is valid in the SM and any model with MFV. Indeed, we find
\be
(\sin 2 \beta)_{\pi \nu\bar\nu}=\sin 2\beta_X =-(0.69^{+0.23}_{-0.41}),
\ee
in striking disagreement with $(\sin 2 \beta)_{\psi K_{\rm S}}= 
0.736\pm0.049$ following from (\ref{ACPmix-gol}).

\begin{figure}
\vspace*{0.3truecm}
\begin{center}
\psfrag{K+}{BR($K^+\rightarrow\pi^+\nu\bar\nu$)}
\psfrag{KL}{BR($K_{\rm L}\rightarrow\pi^0\nu\bar\nu$)}
\psfrag{SM}{SM}\psfrag{expRange}{exp.\ range}
\psfrag{b25}{$\beta_X=25^\circ$}\psfrag{b50}{$50^\circ$}\psfrag{b70}{$70^\circ$}
\psfrag{b111}{$111^\circ$}\psfrag{b130}{$130^\circ$}\psfrag{b150}{$150^\circ$}
\psfrag{GN}{GN bound}
\includegraphics[width=10cm]{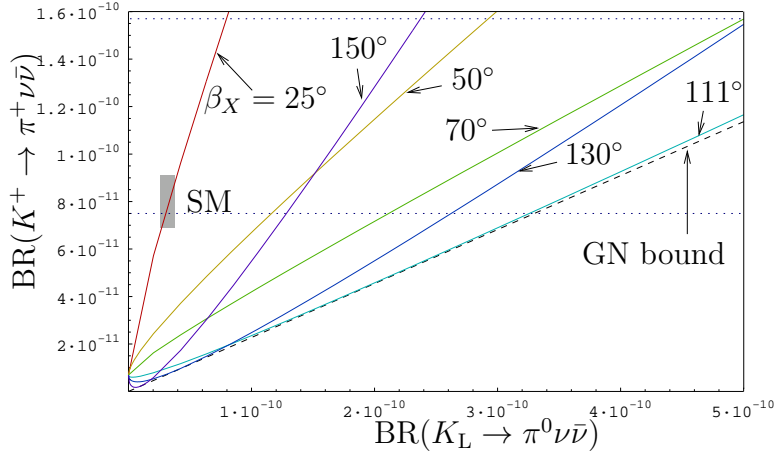}
\end{center}
\caption{$\mbox{BR}(\kpn)$ as a function of $\mbox{BR}(\klpn)$
for various values of $\beta_X$. The dotted horizontal lines indicate 
the lower part of the experimental range
(\ref{kp01}) and the grey area the SM prediction. We also show the 
bound in (\ref{BOUND}).  \label{fig:KpKl}}
\end{figure}

In Fig.~\ref{fig:KpKl}, we plot -- in the spirit of \cite{BF01} -- 
$\mbox{BR}(\kpn)$ as a function of $\mbox{BR}(\klpn)$ for fixed values of
$\beta_X$. As this plot is independent of $|X|$, it offers a direct
measurement of the phase $\beta_X$. 
The first line on the left represents the MFV models with
$\beta_X=\beta-\beta_s$, whereas the first line on the right corresponds 
to the model-independent Grossman--Nir bound \cite{GRNR} given in 
(\ref{BOUND}). The central 
value $\beta_X=111^\circ$ found here is very close to this bound. Note 
that the value of $\beta_X$ corresponding to this bound, where
$B_1=B_2$, depends on the actual value of these reduced branching ratios. 
As can be easily seen from (\ref{rs-def}), we have
\be\label{BX-bound}
(\cot\beta_X)_{\rm Bound}=-\frac{\bar P_c(X)}
{\varepsilon_2\sqrt{B_2}}.
\ee
For the central values of $\bar P_c(X)$ and $B_2$ found here 
the bound corresponds 
to $\beta_X=107.3^\circ$. As only $\cot\beta_X$ and not $\beta_X$ is 
directly determined by the values of the branching ratios in question, 
the angle $\beta_X$ is determined only up to discrete ambiguities,  
seen already in Fig.~\ref{fig:KpKl}. These ambiguities can be resolved by
considering simultaneously other quantities discussed in our paper.
The corresponding plot for different values of $\beta_X$ that are close to 
$\beta$ can be found in \cite{BF01}.

\boldmath
\subsection{$B\to X_{s,d}\nu\bar\nu$ and $B_{s,d}\to \mu^+\mu^-$}
\unboldmath
The inclusive decays $B\to X_{s,d}\nu\bar\nu$ are also theoretically 
clean \cite{BB,BI98}. Generalizing the known MFV formula to our NP
scenario, we obtain
\begin{equation}\label{bbxnn}
\mbox{BR}(B\to X_q\nu\bar\nu)= 1.58\times 10^{-5}
\left[\frac{\mbox{BR}(B\to X_c e\bar\nu)}{0.104}\right]
\left[\frac{0.54}{f(z)}\right]
\frac{|V_{tq}|^2}{|V_{cb}|^2}|X(v)|^2,
\end{equation}
where $q=s,d$, and $f(z)$ is a phase-space factor for $B\to X_c e\bar\nu$.
The SM expectation 
\be
\mbox{BR}(B\to X_s\nu\bar\nu)_{\rm SM}=(3.5\pm 0.5)\times 10^{-5}
\ee
is to be compared with the $90\%$ C.L. ALEPH upper bound of
$6.4\times 10^{-4}$. The exclusive channels are less clean but 
experimentally more easily accessible, with the $90\%$ C.L. BaBar upper bound
of $7.0\times 10^{-5}$. 

Next, the branching ratios for the rare decays $B_q\to\mu^+\mu^-$ are
given by
\begin{equation}\label{bblls}
\mbox{BR}(B_s\to \mu^+\mu^-)=2.42\times 10^{-6}\times 
\left[\frac{\tau_{B_s}}{1.46~{\rm ps}}\right]
\left[\frac{F_{B_s}}{238~ {\rm MeV}}\right]^2
|V^\ast_{tb}V_{ts}|^2 |Y(v)|^2
\end{equation}
\begin{equation}\label{bblld}
\mbox{BR}(B_d\to \mu^+\mu^-)=1.82\times 10^{-6} \times
\left[\frac{\tau_{B_d}}{1.54~{\rm ps}}\right]
\left[\frac{F_{B_d}}{203~ {\rm MeV}}\right]^2
|V^\ast_{tb}V_{td}|^2 |Y(v)|^2,
\end{equation}
with $\tau_{B_{d,s}}$ and $F_{B_{d,s}}$ being the lifetime and decay
constants, respectively.

In the SM, we find \cite{AJB03} 
\be\label{Results1}
 \mbox{BR}(B_{s}\to\mu^+\mu^-)_{\rm SM}=
(3.42\pm 0.53) \left[\frac{\Delta M_s}{18.0/{\rm ps}}\right]\times 10^{-9},
\ee
\be\label{Results2}
\mbox{BR}(B_{d}\to\mu^+\mu^-)_{\rm SM}=(1.00\pm 0.14)\times 10^{-10}.
\ee
This should be compared with the $90\%$ C.L.
bounds 
\be
\mbox{BR}(B_{s}\to\mu^+\mu^-)< 9.5~(16) \times 10^{-7} \quad\mbox{and}\quad
\mbox{BR}(B_{d}\to\mu^+\mu^-)< 1.6 \times 10^{-7}
\ee
from the CDF (D0) and Belle collaborations, respectively \cite{Nakao,BelleBd}.

Inspecting formulae (\ref{Xv}) and (\ref{Yv}), we observe that for a fixed
value of $\bar q$, the 
effect of a non-vanishing phase $\phi\not=0^\circ$ is to suppress
$|X|$ and $|Y|$. Thus, in spite of an
enhancement of $\bar q$, it is possible that for certain values of $\phi$ a 
suppression of the branching ratios given in (\ref{bbxnn}), 
(\ref{bblls}) and (\ref{bblld}) relative to the SM expectations 
could be found. However,  this does not happen in our 
case, and we find
\be\label{Rnunu+mumu}
\frac{\mbox{BR}(B\to X_s\nu\bar\nu)}
{\mbox{BR}(B\to X_s\nu\bar\nu)_{\rm SM}}
=\left|\frac{X}{X_{\rm SM}}\right|^2 \approx 2.0
\quad\mbox{and}\quad
\frac{\mbox{BR}(B_{s}\to\mu^+\mu^-)}{\mbox{BR}(B_{s}\to\mu^+\mu^-)_{\rm SM}}
=\left|\frac{Y}{Y_{\rm SM}}\right|^2 \approx 5.0,
\ee
with the same enhancements for $B\to X_d\nu\bar\nu$ and $B_{d}\to\mu^+\mu^-$,
respectively. We also find
\be
\mbox{BR}(B\to X_s\nu\bar\nu) \approx 7 \times 10^{-5}, 
 \;
\mbox{BR}(B_{s}\to\mu^+\mu^-) \approx 17 \times 10^{-9}, 
 \; 
\mbox{BR}(B_{d}\to\mu^+\mu^-) \approx 5 \times 10^{-10},
\ee
which are still well below the experimental bounds.
\boldmath
\subsection{$K_{\rm L}\to \mu^+\mu^-$}
\unboldmath
For the short-distance contribution to the dispersive part of $\kmm$, 
we obtain in our NP scenario 
\begin{equation}\label{bklm}
\mbox{BR}(\kmm)_{\rm SD}=1.95\times 10^{-9}\times
\left[\bar P_c(Y)+
A^2R_t|Y(v)| \cos\beta_Y\right]^2,
\end{equation}
where $\beta_Y$ is defined in (\ref{betas}), and
\be
\bar P_c(Y)=\left(1-\frac{\lambda^2}{2}\right)P_c(Y),
\ee
with $P_c(Y) = 0.121\pm 0.012$ \cite{BB98}.  
Unfortunately, because of long-distance contributions to the dispersive part 
of $K_{\rm L}\to \mu^+\mu^-$, the extraction of 
$\mbox{BR}(\kmm)_{\rm SD}$ from the data is 
subject to considerable uncertainties \cite{ISIDORI,EDUARDO}. 
While the chapter on this extraction is certainly not closed, let us quote 
the estimate of \cite{ISIDORI}, which reads 
 \begin{equation}\label{SD}
\mbox{BR}(\kmm)_{\rm SD}\le 2.5\times 10^{-9},
\end{equation}
to be compared with 
$\mbox{BR}(\kmm)_{\rm SD}^{\rm SM}=(0.8\pm 0.3) \times 10^{-9}$ 
in the SM.

In the scenario with enhanced $Z^0$ penguins but no new weak phases 
considered by us in \cite{BFRS-I}, $\mbox{BR}(\kmm)_{\rm SD}$ was
substantially enhanced; in the present case, however, with 
$\theta_Y\not=0^\circ$, the small value of $\cos\beta_Y$ compensates the 
enhancement of $|Y|$, so that we find
 \begin{equation}\label{NEWSD}
\mbox{BR}(\kmm)_{\rm SD}= (0.9\pm 0.6)\times 10^{-9}.
\end{equation}

\boldmath
\subsection{Forward--Backward Asymmetries 
in $b\to s\mu^+\mu^-$ Modes}
\unboldmath
\subsubsection{Basic Formulae}
It has been pointed out in \cite{Buchalla:2000sk} that the exclusive decays 
$B_d\to K^*\mu^+\mu^-$ and their inclusive counterparts 
$B\to X_s\mu^+\mu^-$ offer excellent means to probe enhanced 
$Z^0$ penguins and, in particular, their complex weak phases.

During the last two years significant progress for these transitions 
has been made both by  experimentalists and theorists.
On the experimental side, the  Belle and BaBar collaborations 
\cite{Kaneko:2002mr}
reported the observation of 
this decay and of the $B\to X_se^+e^-$ channel. 
The $90\%$ C.L. ranges extracted from these papers \cite{HIKR} 
read as follows:
\be\label{data1}
3.5\times 10^{-6}\le \mbox{BR}(B\to X_s\mu^+\mu^-)\le 10.4\times 10^{-6}
\ee
\be\label{data2}
2.8\times 10^{-6}\le \mbox{BR}(B\to X_s e^+e^-)\le 8.8\times 10^{-6}.
\ee
On the theoretical side, important NNLO corrections have been calculated in 
\cite{NNLO1}--\cite{NNLO3}.
The most recent reviews summarizing the theoretical status can be found 
in \cite{HURTH,Ali:2002jg}. The NNLO formulae are very complicated and it is 
not the purpose of our paper to present a detailed NNLO analysis here. 
For this reason, we give below NLO formulae that show transparently the 
size of various effects that we could expect in our scenario.

Let us recall that these  decays are dominated by the operators
\begin{equation}\label{Q9V}
Q_{9V}    = (\bar{s} b)_{V-A}  (\bar{\mu}\mu)_V,\quad
Q_{10A}  =  (\bar{s} b)_{V-A}  (\bar{\mu}\mu)_A,
\end{equation}
which are generated through EW penguin diagrams.  At low
\begin{equation} \label{invleptmass}
\hat s = \frac{(p_{\mu^+} + p_{\mu^-})^2}{m_b^2},
\end{equation}
also the magnetic operator $Q_{7\gamma}$ plays a significant r\^ole.

At the NLO level \cite{Mis:94,BuMu:94}, the invariant dilepton mass spectrum
in the inclusive decay $B\to X_s\mu^+\mu^-$ is given by
\be \label{rateee}
\frac{{d}/{d\hat s} \, 
\Gamma (b \to s \mu^+\mu^-)}{\Gamma
(b \to c e\bar\nu)} = \frac{\alpha^2}{4\pi^2}
\left|\frac{V_{ts}}{V_{cb}}\right|^2 \frac{(1-\hat s)^2}{f(z)\kappa(z)}
U(\hat s),
\ee
where
\be\label{US} 
U(\hat s)=
(1+2\hat s)\left(|\Ctilde_9^{\rm eff}(\hat s)|^2 + |\Ctilde_{10}|^2\right) + 
4 \left( 1 + \frac{2}{\hat s}\right) |C_{7\gamma}^{(0){\rm eff}}|^2 + 12
C_{7\gamma}^{(0){\rm eff}} \ \RE\,\Ctilde_9^{\rm eff}(\hat s), 
\ee
and $\Ctilde_9^{\rm eff}(\hat s)$ is a function of $\hat s$ that depends on
the Wilson coefficient $\Ctilde_9$ and includes also contributions 
from four-quark operators. $C_{7\gamma}^{(0){\rm eff}}$ is the Wilson
coefficient of the magnetic operator $Q_{7\gamma}$.
Explicit formulae can be found in
\cite{Mis:94,BuMu:94}.
The Wilson coefficients $\Ctilde_9$ and  $\Ctilde_{10}$ are given as follows:
\begin{equation}\label{C9tilde}
\Ctilde_9(v)  =  
P_0 + \frac{Y(v)}{\sin^2\theta_{\rm w}} -4 Z(v) +
P_E E(v),\quad
\tilde C_{10}(v) = - \frac{Y(v)}{\sin^2\theta_{\rm w}},
\end{equation}
with
$P_0 = 2.60\pm 0.25$ in the NDR scheme, $P_E=\ord(10^{-2})$ and $v$ denoting 
collectively the parameters involved. 
$\tilde C_{9}$ and $\tilde C_{10}$ are defined by 
\begin{equation} \label{C10}
C_{9V}(v) = \frac{\alpha}{2\pi} \tilde C_9(v),\quad
C_{10A}(v) =  \frac{\alpha}{2\pi} \tilde C_{10}(v).
\ee

Of particular interest is the forward--backward asymmetry in 
$B\to X_s\mu^+\mu^-$ \cite{AMM}. 
It becomes non-zero only at the NLO level. In our scenario it
is given -- in this approximation -- as follows: 
\be\label{ABF}
A_{\rm FB}(\hat s)= -3\RE\left[ \tilde C^*_{10}
\frac{\hat s \Ctilde_9^{\rm eff}(\hat s)
+2 C_{7\gamma}^{(0){\rm eff}}}{U(\hat s)}\right],
\ee
where $U(\hat s)$ is given in (\ref{US}). The expression for the 
corresponding asymmetry in the 
exclusive decay $B_d\to K^*\mu^+\mu^-$ can be found in \cite{Buchalla:2000sk}.
Both asymmetries vanish at a certain $\hat s=\hat s_0$ \cite{Burdman}, which
is determined in the case of the inclusive decay considered here through 
\be\label{ZERO}
\hat s_0 \RE\,\Ctilde_9^{\rm eff}(\hat s_0)+2 C_{7\gamma}^{(0){\rm eff}}=0.
\ee
As in our scenario $\Ctilde_9^{\rm eff}(\hat s_0)$ and 
$C_{7\gamma}^{(0){\rm eff}}$ are assumed to be marginally affected 
by NP contributions, the value of $\hat s_0$ is bound to be very 
close to the one predicted in the SM
\cite{NNLO1,NNLO2}:  $\hat s_0=0.162\pm 0.008$ after the inclusion of 
NNLO corrections.
On the other hand, as we will see below, 
the magnitude of $A_{\rm FB}(\hat s)$ and its sign 
could be strongly affected by the size of  
$\tilde C_{10}$ and its phase $\theta_Y$. 

Equally interesting is the forward--backward CP asymmetry introduced 
for $B_d\to K^*\mu^+\mu^-$ in \cite{Buchalla:2000sk}, with the explicit 
expression given in equation (63) of that paper. 
This asymmetry can be substantially different from zero above the $c\bar c$ 
threshold if $\tilde C_{10}$ contains a large weak phase.
The integrated asymmetry is 
found then to be \cite{Buchalla:2000sk}
\be
\Delta A^{\rm CP}_{\rm FB}=\int^{0.69}_{0.52}~d\hat s
~A^{\rm CP}_{\rm FB}(\hat s)=(0.03\pm0.01)\times
\frac{\IM \tilde C_{10}}{\RE \tilde C_{10}},
\ee
where the range of the integration is chosen to decrease the effect of
non-perturbative uncertainties.

\subsubsection{Numerical Results}
Using the NNLO formulae of \cite{Ali:2002jg} and comparing with 
the Belle and BaBar data for
the branching ratios in (\ref{data1}) and (\ref{data2}) allows us to obtain a
conservative upper bound 
\be
|Y| \le 2.2,
\ee
which we already used in our previous numerical analysis \cite{BFRS-I}. 
Taking this upper 
bound into account and the values of $|Y|$ and $\theta_Y$ in (\ref{rY}), 
we find 
the following results that were anticipated in
\cite{Buchalla:2000sk} in the case of a large CP-violating weak phase 
in $\tilde
C_{10}$.  Our scenario is a concrete realization of such a situation. 
The most important findings are the following:
\begin{itemize}
\item
As
\be
{\rm sgn} (\RE \tilde C_{10})=-{\rm sgn} (\RE Y)= -{\rm sgn} (\cos \theta_Y) 
\ee
governs the sign of $A_{\rm FB}(\hat s)$, and $\theta_Y=0^\circ$ in the SM, in 
our scenario with $\theta_Y>90^\circ$, the sign of 
$A_{\rm FB}(\hat s)$ is {\it opposite} to the one in the SM: the asymmetry 
is positive for $\hat s < \hat s_0$ and negative for
$\hat s > \hat s_0$. The value of $\hat s_0$ is essentially unaffected.
\item
However,
\be
\frac{\RE \tilde C_{10}}{(\RE \tilde C_{10})_{\rm SM}}=
\frac{|Y|}{|Y|_{\rm SM}}\cos\theta_Y\approx -0.5,
\ee
which through (\ref{ABF}) implies the suppression of
$A_{\rm FB}(\hat s)$ in our scenario with respect to the SM case.
\item
On the other hand,
\begin{equation}\label{FBCP}
\Delta A^{\rm CP}_{\rm FB}=(0.03\pm0.01)\times \tan\theta_Y
 =\left\{ \begin{array}{ll}
-0.06 \pm 0.02 & \quad(\theta_Y=115^\circ) \\
-0.11 \pm 0.04 & \quad(\theta_Y=105^\circ) \\
-0.34 \pm 0.11 & \quad(\theta_Y=95^\circ)
\end{array} \right.
\end{equation}
is still another spectacular effect of the large weak phase $\phi$ implied 
by the $B\to\pi K$ data. The large sensitivity of $\Delta A^{\rm CP}_{\rm FB}$ 
to $\theta_Y$ for $\theta_Y=\ord(100^\circ)$ offers a useful method for the
determination of this phase.
\end{itemize}

\boldmath
\subsection{$K_{\rm L}\to\pi^0 e^+ e^-$}
\unboldmath 
The rare decay $K_{\rm L}\to\pi^0 e^+e^-$ is dominated by CP-violating 
contributions. 
It has recently been reconsidered within the SM \cite{BI03} in view 
of the most recent NA48 data  on $K_{\rm S}\to\pi^0 e^+e^-$ and 
$K_{\rm L}\to \pi^0\gamma\gamma$ \cite{NA48KL}, which allow a much better 
evaluation of the CP-conserving and indirectly (mixing) CP-violating    
contributions. The CP-conserving part is found to be below 
$3\times 10^{-12}$. Moreover, in the SM the indirectly (mixing) CP-violating
contribution and the interference of both CP-violating contributions 
dominate the branching ratio in question, while the directly CP-violating 
contribution alone is significantly smaller and in the ballpark of 
$4\times 10^{-12}$. In our scenario, this pattern is significantly 
changed, the latter part becoming the dominant contribution. 
Indeed, similar to $\mbox{BR}(\klpn)$, the directly CP-violating 
contribution to $\mbox{BR}(K_{\rm L}\to\pi^0 e^+ e^-)$ is enhanced by 
more than one order of magnitude. 

Generalizing formula (33) in \cite{BI03} to our scenario, we find 
\begin{equation}\label{9a}
\mbox{BR}(K_{\rm L} \to \pi^0 e^+ e^-)_{\rm CPV} =
10^{-12}\times\left[C_{\rm mix}+
\bar C_{\rm int}\left(\frac{\kappa}{3\times 10^{-4}}\right) +
\bar C_{\rm dir}\left(\frac{\kappa}{3\times 10^{-4}}\right)^2\right],
\ee
where 
\be
\kappa=\frac{\IM\lambda_t}{\sin(\beta-\beta_s)}=\tilde r A^2\lambda^5 R_t, 
\qquad \tilde r=\left|\frac{V_{ts}}{V_{cb}}\right|\approx 0.98,
\ee
\be
C_{\rm mix}=(15.7\pm 0.3)|a_s|^2, \qquad |a_s|=1.08^{+0.26}_{-0.21},
\ee
\be
\bar C_{\rm int}=1.02\, \hat y_{7V}\,\sqrt{C_{\rm mix}},\qquad
\bar C_{\rm dir}=0.56\,(\hat y_{7A}^2 + \hat y_{7V}^2).
\end{equation}
Here
\begin{equation}\label{y7vpbe}
\hat{y}_{7V} =
[P_0 +P_E E(v)]\sin(\beta-\beta_s)+ 
\frac{|Y(v)|}{\sin^2\theta_{\rm w}}\sin\beta_Y - 4 |Z(v)|\sin\beta_Z
\ee
\be\label{y7apbe}
\hat{y}_{7A}=-\frac{|Y(v)|}{\sin^2\theta_{\rm w}} \sin\beta_Y,
\end{equation}
where $P_0=2.89\pm0.06$  \cite{BLMM} and 
$P_E$ is $\ord(10^{-2})$. The effect of the NP contributions is mainly 
felt in $\hat{y}_{7A}$, as the corresponding contributions in 
$\hat{y}_{7V}$ cancel each other to a large extent.

The present experimental bound from KTeV \cite{KTEVKL},
\be\label{ktev1} 
\mbox{BR}(K_{\rm L} \to \pi^0 e^+ e^-)< 
2.8\times 10^{-10}\quad (\mbox{90\% {\rm C.L.}}),
\ee
should be compared with the SM prediction \cite{BI03},
\be
\mbox{BR}(K_{\rm L} \to \pi^0 e^+ e^-)_{\rm SM}= 
(3.2^{+1.2}_{-0.8})\times 10^{-11}.
\ee
The enhancement of $\sin\beta_Y$ over $\sin(\beta-\beta_s)$ and of $|Y(v)|$
over $Y$ makes the directly CP-violating contribution the dominant part of
the branching ratio, for which we find
\be
\mbox{BR}(K_{\rm L} \to \pi^0 e^+ e^-)_{\rm CPV}= (7.8\pm 1.6)
\times 10^{-11},
\ee
which is lower than the upper bound in (\ref{ktev1}) by only a factor of 3.
\boldmath
\subsection{$\epe$}
\unboldmath
The formula for the CP-violating ratio $\epe$ of \cite{BJ03} generalizes 
for the NP scenario considered here as follows: 
\be \frac{\varepsilon'}{\varepsilon}= \tilde r A^2R_t\lambda^5
\times \tilde F_{\varepsilon'}(v),
\label{epeth}
\ee
with
\be
\tilde F_{\varepsilon'}(v) =[P_0 +P_E \, E(v)]\sin(\beta-\beta_s) 
+ P_X \, |X(v)|\sin\beta_X + 
P_Y \, |Y(v)|\sin\beta_Y
 + P_Z \, |Z(v)|\sin\beta_Z.
\label{FE}
\ee
Here the $P_i$ encode the information about the physics at scales 
$\mu \le\ord(m_t, M_W)$, and are given in terms of 
the hadronic parameters
\be\label{RS}
R_6\equiv \bsi\left[ \frac{121\mev}{m_s(m_c)+m_d(m_c)} \right]^2,\quad
R_8\equiv \bei\left[ \frac{121\mev}{m_s(m_c)+m_d(m_c)} \right]^2
\ee 
as follows:
\begin{equation}
P_i = r_i^{(0)} + 
r_i^{(6)} R_6 + r_i^{(8)} R_8 .
\label{eq:pbePi}
\end{equation}
The coefficients $r_i^{(0)}$, $r_i^{(6)}$ and $r_i^{(8)}$ comprise
information on the Wilson-coefficient functions of the $\Delta S=1$ weak
effective Hamiltonian at the next-to-leading order
\cite{Buchalla:1995vs}; their numerical values for
different choices of $\Lms^{(4)}$ at $\mu=m_c$ in the NDR renormalization
scheme can be found in \cite{BJ03}.
The hadronic parameters $\bsi$ and $\bei$  represent 
the matrix elements of the dominant QCD penguin operator $Q_6$ and the 
dominant EW penguin operator $Q_8$, respectively. The numerical values 
of the $P_i$ are sensitive functions of $R_6$ and $R_8$, as well as of
$\alpha_s$.

On the experimental side, the world average based on the latest results 
from NA48 \cite{NA48} and KTeV
\cite{KTeV} and the previous results from NA31 and E731 reads 
\begin{equation}\label{eps}
\epe=(16.6\pm 1.6) \times 10^{-4}.
\end{equation}
While several analyses made in recent years within the SM found results 
that are compatible with (\ref{eps}), it is fair to say that the 
large hadronic uncertainties  in the coefficients $P_i$ still allow for 
sizeable NP contributions. The relevant list of references can be
found in~\cite{BJ03}.

In the SM, $\theta_X=\theta_Y=\theta_Z=0^\circ$ and therefore
$\beta_X=\beta_Y=
\beta_Z=\beta-\beta_s$. We then find \cite{BJ03} that with 
\be\label{PI1}
P_0=19.5,\quad P_X=0.6,\quad P_Y=0.5,\quad P_Z=-12.4,\quad P_E=-1.6,
\ee
corresponding to $R_6=1.2$, $R_8=1.0$ and $\alpha_s(M_Z)=0.119$,
an agreement with the experimental data can be obtained. 

In our scenario, the first term in (\ref{FE}) involving
$P_0$ and originating dominantly in the matrix elements of the 
QCD penguin operator $Q_6$ does not contain any NP contributions, 
while the important negative last term involving $P_Z$ and being related 
to the EW penguin operator $Q_8$ is strongly enhanced.
With the values in (\ref{PI1}), a negative $\epe$ is then obtained.
Thus for our scenario to be consistent with the data, the hadronic 
matrix element of $Q_6$ or equivalently $R_6$ must be significantly 
enhanced over $R_8$. For instance for 
$\bsi=2.0$, $\bei=0.62$ and $m_s(m_c)+m_s(m_c)=106~{\rm MeV}$, corresponding 
to $R_6=2.6$ and $R_8=0.81$,
one finds for
 $\alpha_s(M_Z)=0.121$:
\be\label{PI2}
P_0=48.0, \qquad P_X=0.65, \qquad P_Y=0.73, \qquad P_Z=-10.4, \qquad P_E=-4.9.
\ee
Setting $\beta_X$, $\beta_Y$, $\beta_Z$ at their central values, we obtain
\be
\epe=15.2 \times 10^{-4},
\end{equation}
which is consistent with the experimental data but the result for $\epe$ is 
very sensitive to the actual values of the coefficients $P_i$ and the angles 
$\beta_i$.
As reviewed in \cite{BJ03}, $R_8=0.8$ used here is consistent with the 
most recent
estimates that give $R_8=1.0\pm0.2$. 
Values for $R_6$ as high as needed here have been reported in \cite{B6A,B6B}.
In particular the values for $R_6=2.2\pm 0.4$ and 
$R_8=1.1\pm 0.3$ found in \cite{B6A}, that within 
the SM would give $\epe$ substantially higher than the experimental data, 
would be very welcome within the scenario considered here.

\subsection{Summary}
In this section, we have demonstrated that the sizeably enhanced EW 
penguins with their large CP-violating NP phase -- as implied by the 
$B\to\pi K$ data -- have important implications for rare $K$ and $B$ 
decays, as well as for $\epe$.
We find several predictions that differ significantly from the SM 
expectations and could easily be identified once the data improve. 
The most interesting results of this study can be summarized as 
follows: 
\begin{itemize}
\item
An enhancement of $\mbox{BR}(\klpn)$ by one order of magnitude 
without any significant change in $\mbox{BR}(\kpn)$, implying 
$\mbox{BR}(\klpn)$ to be close  to its absolute
upper bound derived in \cite{GRNR}.
\item
A spectacular  violation of 
$(\sin 2 \beta)_{\pi \nu\bar\nu}=(\sin 2 \beta)_{\psi K_{\rm S}}$ 
\cite{BBSIN}, which is valid in the SM and any model with minimal
flavour violation.
\item
A large branching ratio 
$\mbox{BR}(K_{\rm L}\to\pi^0e^+e^-)= (7.8\pm 1.6)\times 10^{-11}$,
which is governed by direct CP violation in this scenario, as opposed to 
the SM, where indirect CP violation dominates \cite{BI03}.
\item
A strong enhancement of the integrated forward--backward CP 
asymmetry for $B_d\to K^*\mu^+\mu^-$. 
\item
Enhancements
of $\mbox{BR}(B\to X_{s,d}\nu\bar\nu)$ and 
$\mbox{BR}(B_{s,d}\to \mu^+\mu^-)$  by factors of $2$ and $5$, 
respectively. 
\item
As far as $\epe$ is concerned, the enhanced EW penguins with their 
large CP-violating NP phase, as suggested by the $B\to\pi K$ analysis, 
require a significant enhancement of the relevant hadronic matrix 
elements of the QCD penguin operators with respect to the one of the 
EW penguin operator $Q_8$ to be consistent with the $\epe$ data. 
\end{itemize}

\section{\boldmath Other Prominent Non-Leptonic $B$-Meson 
Decays \unboldmath}\label{sec:B-prom}
\setcounter{equation}{0}
As the last element of our analysis, we explore the implications of
our NP scenario of enhanced EW penguins with a new CP-violating weak
phase for other prominent non-leptonic $B$-meson decays, which play 
another key r\^ole in the physics programme of the $B$ factories.
\boldmath
\subsection{$B\to \phi K$}\label{ssec:BphiK}
\unboldmath
In the SM, the $B\to\phi K$ system is governed by QCD penguins 
\cite{BphiK-old} and receives sizeable contributions from EW penguin 
topologies \cite{RF-EWP,DH-PhiK}. Consequently, these modes offer
an interesting tool to search for signals of NP, and may
of course also be affected within our specific scenario. 
\subsubsection{Observables}
In order to address this exciting issue, we follow \cite{FM-BphiK}, and
consider first 
\begin{equation}\label{Bcal-def}
{\cal B}_{\phi K}\equiv\frac{1-{\cal A}_{\phi K}}{1+{\cal A}_{\phi K}},
\end{equation}
with
\begin{equation}
{\cal A}_{\phi K}\equiv\left[\frac{\mbox{BR}(B^+\to\phi K^+)+
\mbox{BR}(B^-\to\phi K^-)}{\mbox{BR}(B_d^0\to\phi K^0)+
\mbox{BR}(\bar B_d^0\to\phi \bar K^0)}\right]
\left[\frac{\tau_{B^0_d}}{\tau_{B^+}}\right],
\end{equation}
and
\begin{eqnarray}
{\cal D}^-_{\phi K}&\equiv&\frac{1}{2}\left[{\cal A}_{\rm CP}^{\rm dir}
(B_d\to\phi K_{\rm S})-{\cal A}_{\rm CP}^{\rm dir}
(B^\pm\to\phi K^\pm)\right]\label{Dcal-def}\\
{\cal D}^+_{\phi K}&\equiv&\frac{1}{2}\left[{\cal A}_{\rm CP}^{\rm dir}
(B_d\to\phi K_{\rm S})+{\cal A}_{\rm CP}^{\rm dir}
(B^\pm\to\phi K^\pm)\right].\label{Scal-def}
\end{eqnarray}
If we use the experimental results given in \cite{HFAG} and add the
errors in quadrature, we obtain
\begin{eqnarray}
{\cal B}_{\phi K}&=&
\left\{\begin{array}{ll}
-0.05\pm0.10 &\mbox{(BaBar)}\\
+0.06\pm0.13 &\mbox{(Belle)}
\end{array}\right.\\
{\cal D}^-_{\phi K}&=&
\left\{\begin{array}{ll}
-0.17\pm0.20 &\mbox{(BaBar)}\\
+0.08\pm0.16 &\mbox{(Belle)}
\end{array}\right.\\
{\cal D}^+_{\phi K}&=&
\left\{\begin{array}{ll}
-0.21\pm0.20 &\mbox{(BaBar)}\\
+0.07\pm0.16 &\mbox{(Belle),}
\end{array}\right.
\end{eqnarray}
whereas the $B$-factory averages correspond to 
\begin{equation}\label{BDS-BphiK-averages}
{\cal B}_{\phi K}=0.00\pm0.08,\quad
{\cal D}^-_{\phi K}=-0.01\pm0.13,\quad
{\cal D}^+_{\phi K}=-0.04\pm0.13.\quad
\end{equation}
We note that these observables are consistent with zero and hence do
not indicate any deviation from the SM picture. In particular,
as discussed in detail in \cite{FM-BphiK}, ${\cal B}_{\phi K}$ and 
${\cal D}^-_{\phi K}$ are sensitive to the $I=1$ isospin sector, thereby
indicating that the corresponding amplitude $v_1$ is in fact suppressed, 
as is expected on the basis of plausible general arguments. On the other 
hand, ${\cal D}^+_{\phi K}$ is sensitive to NP in the $I=0$ isospin sector,
which involves an amplitude $v_0$ with the sine of a CP-conserving 
strong phase $\Delta_0$. The same contribution governs also the difference 
of ${\cal A}_{\rm CP}^{\rm mix}(B_d\to \phi K_{\rm S})$ and
${\cal A}_{\rm CP}^{\rm mix}(B_d\to J/\psi K_{\rm S})$, where it
enters, however, with the cosine of $\Delta_0$.
Within the SM, this difference is expected to vanish to a good approximation 
\cite{growo}--\cite{GLNQ}, whereas the current $B$-factory data
in (\ref{aCP-Bd-phiK-mix}) and (\ref{ACPmix-gol}) yield
\begin{equation}
{\cal A}_{\rm CP}^{\rm mix}(B_d\to \phi K_{\rm S})-
{\cal A}_{\rm CP}^{\rm mix}(B_d\to J/\psi K_{\rm S})=
\left\{\begin{array}{ll}
+0.3\pm0.4 & \mbox{(BaBar)}\\
+1.7\pm0.5 & \mbox{(Belle).}
\end{array}\right.
\end{equation}
Consequently, we may well arrive at a discrepancy with the SM in the 
future, although the experimental picture is unclear at the moment. 
Should such a discrepancy really emerge and ${\cal D}^+_{\phi K}$  
still be consistent with zero -- in addition to ${\cal B}_{\phi K}$ and 
${\cal D}^-_{\phi K}$, we would have an indication for NP in the 
$I=0$ isospin sector, with a small CP-conserving strong phase $\Delta_0$
relative to the SM contribution. 

\subsubsection{NP Analysis}
\begin{figure}
\vspace*{0.3truecm}
\begin{center}
\psfrag{Amix}{${\cal A}_{\rm CP}^{\rm mix}(B_d\to\phi K_{\rm S})-
{\cal A}_{\rm CP}^{\rm mix}(B_d\to J/\psi K_{\rm S})$}
\psfrag{SphiK}{${\cal D}^+_{\phi K}$}
\psfrag{v0=.1}{$v_0=0.1$}\psfrag{v0=.4}{$v_0=0.4$}
\psfrag{d0=0}{$\Delta_0\!=\!0^\circ$}\psfrag{d0=90}{$90^\circ$}
\psfrag{d0=180}{$180^\circ$}\psfrag{d0=270}{$270^\circ$}
\psfrag{BaBar}{BaBar}\psfrag{Belle}{Belle}
\includegraphics[width=10cm]{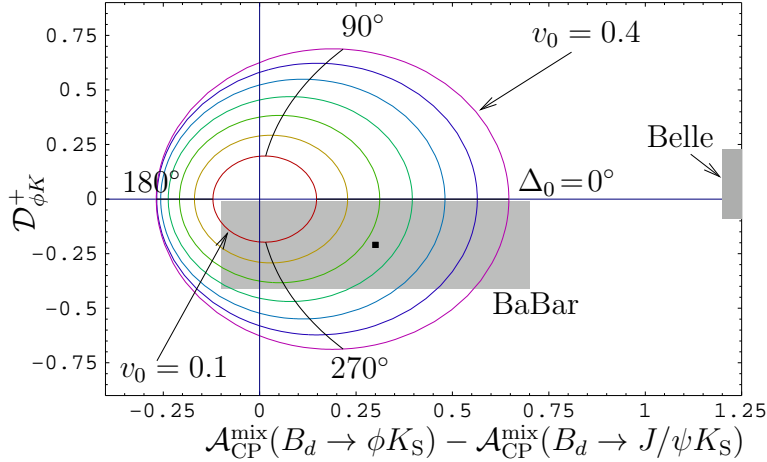}
\end{center}
\caption{The contours in the plane of 
${\cal A}_{\rm CP}^{\rm mix}(B_d\to\phi K_{\rm S})-{\cal A}_{\rm 
CP}^{\rm mix}(B_d\to J/\psi K_{\rm S})$ and
${\cal D}^+_{\phi K}$ for various values of $v_0$ with
$\Delta_0\in[0^\circ,360^\circ]$ and $\phi=-85^\circ$, corresponding to
the central value of (\ref{qnew}).
}\label{fig:BphiK}
\end{figure}

Since our scenario of enhanced EW penguins with a CP-violating weak 
phase $\phi$ belongs to this category of NP, let us explore its 
implications in more detail. To this end, we neglect the amplitude 
$v_1$ of the $I=1$ isospin sector, which corresponds to 
${\cal B}_{\phi K}={\cal D}^-_{\phi K}=0$, and write \cite{FM-BphiK}
\begin{eqnarray}
A(\bar B_d^0\to\phi \bar K^0)&=&A_0\left[1+v_0e^{i(\Delta_0-\phi)}
\right]=A(B^-\to\phi K^-)\label{Bphi-K-ampl}\\
A(B_d^0\to\phi K^0)&=&A_0\left[1+v_0e^{i(\Delta_0+\phi)}
\right]=A(B^+\to\phi K^+)\label{Bphi-K-ampl-CP},
\end{eqnarray}
where the CP-conserving strong amplitude $A_0$ describes the 
QCD penguin contributions, $v_0$ measures the strength of the 
EW penguins with respect to the QCD penguins, which are governed by
the $(\bar s s)(\bar b s)$ quark-flavour structures having $I=0$, and 
$\Delta_0$ is the CP-conserving strong phase mentioned above. Applying 
now once more the standard formalism for the evaluation of the CP-violating
observables, (\ref{Bphi-K-ampl}) and (\ref{Bphi-K-ampl-CP}) yield
\begin{equation}\label{Adir-BphiK}
{\cal A}_{\rm CP}^{\rm dir}(B_d\to\phi K_{\rm S})=
-\left[\frac{2v_0\sin\Delta_0\sin\phi}{1+2v_0\cos\Delta_0\cos\phi+v_0^2}
\right]={\cal A}_{\rm CP}^{\rm dir}(B^\pm\to\phi K^\pm)
\end{equation}
\begin{equation}\label{Amix-BphiK}
{\cal A}_{\rm CP}^{\rm mix}(B_d\to\phi K_{\rm S})=
-\left[\frac{\sin\phi_d+2v_0\cos\Delta_0\sin(\phi_d+\phi)+v_0^2
\sin(\phi_d+2\phi)}{1+2v_0\cos\Delta_0\cos\phi+v_0^2}\right].
\end{equation}
For a simple order-of-magnitude estimate of $v_0 e^{i\Delta_0}$ within
the SM, where $\phi=0^\circ$, we assume that $A_0$ is dominated by 
internal top-quark exchanges, and use na\"\i ve factorization
with the leading-order Wilson coefficients given in \cite{Buras-Les-Houches}
for $\Lambda_{{\overline{\rm MS}}}^{(5)}=225$ $\mbox{MeV}$. Following
these lines, we arrive at
\begin{eqnarray}
\lefteqn{\left.v_0 e^{i\Delta_0}\right|_{\rm fact}^{\rm SM}\approx
-\left[\frac{2(C_9(m_b)+C_{10}(m_b))}{4(C_3(m_b)+C_4(m_b))+
3C_5(m_b)+C_6(m_b)}\right]}\nonumber\\
&&=-\left[\frac{2\times(-1.280+0.328)}{4\times(0.014-0.030)+
3\times0.009-0.038}\right]\times\frac{1}{128}=-0.20,
\end{eqnarray}
which corresponds to $\Delta_0|_{\rm fact}\approx 180^\circ$; for a more
refined treatment addressing also QCD penguins with internal up-
and charm-quark exchanges, see \cite{RF-EWP}. In our scenario of NP,
the enhancement of $C_9+C_{10}$ is the same as that of our EW
penguin parameter $q$. Taking into account also the constraints from the
rare-decay analysis in Section~\ref{sec:rare},  
we conclude that this enhancement may be at most $\sim 1.3$ with
respect to the SM, corresponding to values of $v_0$ at the 0.25 level. 
We may then expand the CP-violating observables in $v_0$, which gives 
\begin{equation}\label{Adir-BphiK-expand}
{\cal D}^+_{\phi K}=-2v_0\sin\Delta_0\sin\phi+{\cal O}(v_0^2)
\end{equation}
\begin{equation}\label{Amix-BphiK-expand}
{\cal A}_{\rm CP}^{\rm mix}(B_d\to \phi K_{\rm S})-
{\cal A}_{\rm CP}^{\rm mix}(B_d\to J/\psi K_{\rm S})=
-2v_0\cos\Delta_0\sin\phi\cos\phi_d+{\cal O}(v_0^2).
\end{equation}
If we use $\cos\phi_d=0.68$, $v_0\approx0.25$, which suffers from large 
hadronic uncertainties, and take into account that our $B\to\pi K$ 
analysis points towards values of $\phi$ around $-90^\circ$, corresponding to 
$\sin\phi\approx-1$, we obtain
\begin{equation}\label{S-approx}
{\cal D}^+_{\phi K}\approx 0.5\times \sin\Delta_0
\end{equation}
\begin{equation}\label{Amix-diff}
{\cal A}_{\rm CP}^{\rm mix}(B_d\to\phi K_{\rm S})-
{\cal A}_{\rm CP}^{\rm mix}(B_d\to J/\psi K_{\rm S})\approx
0.3\times\cos\Delta_0.
\end{equation}
Consequently, the experimental value of ${\cal D}^+_{\phi K}$
in (\ref{BDS-BphiK-averages}) favours a small value of $\sin\Delta_0$. 
If we assume that the sign of $\cos\Delta_0$ is {\it negative}, 
as in factorization, we conclude that the difference in (\ref{Amix-diff}) 
is {\it negative} as well. Since in our notation both mixing-induced CP 
asymmetries would equal $-\sin2\beta\approx-0.7$ in the SM, we generically
arrive at 
\begin{equation}
(\sin 2\beta)_{\phi K_{\rm S}}>(\sin 2\beta)_{\psi K_{\rm S}},
\end{equation}
where $(\sin 2\beta)_{\phi K_{\rm S}}$ around $+1$ may well be possible, 
whereas the current result of the Belle collaboration in 
(\ref{aCP-Bd-phiK-mix}) points to a value with the {\it opposite}\, 
sign around $-1$. From the theoretical 
point of view, there are the following two possible loopholes:
\begin{itemize}
\item[i)] In the case of the unconventional solution $\phi_d=133^\circ$, 
the sign of $\cos\phi_d$ would be negative \cite{Fl-Ma,FIM}. However, 
as we have noted in Subsection~\ref{ssec:Bpipi-gam-det}, this possibility 
appears to be disfavoured. 
\item[ii)] We may encounter large non-factorizable effects, although it
does not look likely that they may flip the sign of $\cos\Delta_0$, which 
would require $\Delta_0$ around $0^\circ$. On the other hand, for 
$\Delta_0$ around $\pm90^\circ$, the impact of NP on 
${\cal A}_{\rm CP}^{\rm mix}(B_d\to\phi K_{\rm S})$ would be very
small, whereas ${\cal D}^+_{\phi K}$ would be large, in conflict with the
data.
\end{itemize}
It is useful to consider the plane of 
${\cal A}_{\rm CP}^{\rm mix}(B_d\to\phi K_{\rm S})-
{\cal A}_{\rm CP}^{\rm mix}(B_d\to J/\psi K_{\rm S})$ and 
${\cal D}^+_{\phi K}$, as done in Fig.~\ref{fig:BphiK}, where we
show also the current experimental results of the BaBar and Belle
collaborations. Applying (\ref{Adir-BphiK}) and (\ref{Amix-BphiK}), we 
may calculate contours for different values of $v_0$, where each 
point is parametrized by the strong phase $\Delta_0$. As far as the 
NP phase $\phi$ is concerned, we use $\phi=-85^\circ$, which is the 
central value in (\ref{qnew}). 
In the future, plots of this kind will allow us to read off 
the hadronic parameters $v_0$ and $\Delta_0$ for our NP scenario easily from 
the improved $B\to\phi K$ data.

\boldmath
\subsection{$B\to J/\psi K$}\label{ssec:BpsiK}
\unboldmath
As emphasized in \cite{RF-rev}, EW penguins may have a non-negligible
impact on the $B\to J/\psi K$ system. Let us, therefore, also address 
these effects in a more quantitative manner. The starting point is
a systematic search for signals of NP arising at the decay-amplitude 
level of these modes \cite{FM-BpsiK}, which is -- from a formal point 
of view -- completely analogous to the $B\to\phi K$ system discussed in
Subsection~\ref{ssec:BphiK}. Using the data for the CP-averaged
$B\to J/\psi K$ branching ratios and the direct CP asymmetries 
compiled in \cite{HFAG} and \cite{PDG}, respectively, 
we arrive at the following picture of the $B\to J/\psi K$ counterparts 
of (\ref{Bcal-def}), (\ref{Dcal-def}) and (\ref{Scal-def}):
\begin{equation}\label{BDS-BpsiK-averages}
{\cal B}_{\psi K}=-0.04\pm0.04,\quad
{\cal D}^-_{\psi K}=+0.023\pm0.025,\quad
{\cal D}^+_{\psi K}=+0.030\pm0.025.\quad
\end{equation}
Consequently, these three observables do not indicate any deviation from
the SM. 

Within our scenario of NP, we may write the $B\to J/\psi K$ amplitudes
in the same form as (\ref{Bphi-K-ampl}) and (\ref{Bphi-K-ampl-CP}), with
expressions for the direct and mixing-induced CP asymmetries as given
in (\ref{Adir-BphiK}) and (\ref{Amix-BphiK}), respectively. 
Using again the na\"\i ve factorization approach to estimate the relevant 
EW penguin parameter, we obtain 
\begin{eqnarray}
\lefteqn{\left.(v_0 e^{i\Delta_0})_{\psi K}\right|_{\rm fact}^{\rm SM}
\approx\frac{C_9(m_b)+C_{10}(m_b)-(1-1/N_{\rm C})C_{10}(m_b)}{C_1(m_b)
+C_2(m_b)/N_{\rm C}}}\nonumber\\
&&\approx\frac{C_9(m_b)+C_{10}(m_b)}{a_2^{\psi K}}\approx
\left[\frac{-1.280+0.328}{0.25}\right]\times\frac{1}{128}\approx-0.03.
\end{eqnarray}
It should be emphasized that this expression suffers from large hadronic
uncertainties. In particular, it is very challenging to estimate the
effects of QCD penguins to the whole amplitude, which we may
consider to be ``effectively'' included in the phenomenological 
$B\to J/\psi K$ colour-suppression factor $a_2^{\psi K}$.
Nevertheless, it is instructive to consider again an
enhancement of $C_9+C_{10}$ of 1.3 and a NP phase $\phi$ around
$-90^\circ$, which yields 
\begin{equation}
{\cal D}^+_{\psi K}\approx0.08\times\sin(\Delta_0)_{\psi K}
\end{equation}
\begin{equation}
{\cal A}_{\rm CP}^{\rm mix}(B_d\to J/\psi K_{\rm S})\approx-\sin2\beta
+0.05\times \cos(\Delta_0)_{\psi K},
\end{equation}
which should be compared with (\ref{S-approx}) and (\ref{Amix-diff}). 
Consequently, we may encounter a NP correction to the determination
of $\sin2\beta$ from the golden mode $B_d\to J/\psi K_{\rm S}$
at the 0.05 level in our scenario, which corresponds to a shift of at most
$\pm 2^\circ$ in $\beta$. Such small effects are still beyond the 
present experimental and theoretical accuracy but could be reinvestigated in
the LHC era. Moreover the comparison with similar effects in $\Delta
M_{s,d}$ could, in the future, shed some light on the size of
the parameter $(v_0 e^{i\Delta_0})_{\psi K}$.
It should also be noted
that this parameter is expected to be different for higher $J/\psi$
resonances, so that the NP effects may largely cancel in the
averages over such modes, which are usually formed by the BaBar and
Belle collaborations and are also at the basis of (\ref{ACPmix-gol}).

In order to search for NP effects in the ``golden'' mode 
$B_d\to J/\psi K_{\rm S}$, decays of the kind $B_d\to D\pi^0,
D\rho^0, ...$\ are particularly interesting. If the neutral $D$
mesons are observed through their decays into CP eigenstates $D_\pm$, 
these pure ``tree'' decays, which do not receive any penguin contributions,
allow extremely clean determinations of $\sin2\beta$ \cite{RF-BDpi}. 
Consequently, a future tiny discrepancy with $(\sin2\beta)_{\psi K_{\rm S}}$
could also provide valuable insights into our specific scenario of NP.

\boldmath
\subsection{Summary}
\unboldmath
The main results of our analysis of the $B\to\phi K$ 
and $B\to J/\psi K$ systems can be summarized as follows:
\begin{itemize}
\item Within our NP scenario, we expect generically 
$(\sin 2\beta)_{\phi K_{\rm S}}>(\sin 2\beta)_{\psi K_{\rm S}}$,
where $(\sin 2\beta)_{\phi K_{\rm S}}\sim+1$ may well be possible. 
This pattern is qualitatively different from the current $B$-factory 
data, which are, however, not yet conclusive. On the other hand, 
a future confirmation of this pattern would be another signal of 
enhanced CP-violating EW penguins at work. 
\item The measurement of $\sin2\beta$ through the golden mode
$B_d\to J/\psi K_{\rm S}$ may receive NP corrections as large
as 0.05, which are, however, affected by large -- essentially 
unknown -- hadronic effects. The current $B$-factory data disfavour
an enhancement of the relevant NP parameter through such effects.   
\end{itemize}

%
%
%
\section{Conclusions}\label{sec:concl}
\setcounter{equation}{0}
%
%
%
In this paper, we have developed a strategy that allows us to study
simultaneously non-leptonic two-body $B$ decays and rare $K$ and $B$ decays
within the SM and its simple extension in which the dominant NP effects come
from modified $Z^0$-penguin contributions with a new CP-violating weak 
phase. This simple scenario is parametrized by only two variables 
$(q,\phi)$ that take the values $q=0.69$ and $\phi=0^\circ$ in the SM.

The aim of our study was not only to learn about  possible NP
effects in the processes in question but also to determine the angle $\gamma$
and to gain some insight into the hadron dynamics, which plays an important
r\^ole in non-leptonic $B$ decays and $\epe$,  is less important in decays 
such as $B\to X_s\mu^+\mu^-$, and is essentially irrelevant in 
theoretically clean processes such as $K\to\pi\nu\bar\nu$.

If we had at our disposal precise experimental data for all processes 
considered here, the most efficient strategy would be to choose decays 
that are theoretically clean and free from NP contributions to determine 
the angle $\gamma$, use then this value in clean rare decays sensitive to 
NP in order to determine $(q,\phi)$, and finally use $\gamma$ and 
$(q,\phi)$ in processes sensitive to hadron dynamics with the aim to obtain
insight into the latter. In order to implement this procedure, clean
tree-level strategies for the determination of $\gamma$ 
(see \cite{RF-Phys-Rep,LHC-BOOK,RF-BDpi} and references therein), 
the rare decays $K\to\pi\nu\bar\nu$, and non-leptonic decays such as 
$B\to\pi\pi$ and $B\to\pi K$ could be used, respectively.

Unfortunately, at present, the first two steps of this strategy cannot be
made in view of the lack of data required for the clean strategies for 
$\gamma$ in question and in view of the insufficient experimental information 
on $K\to\pi\nu\bar\nu$ decays. On the other hand, data on $B\to\pi\pi$ and
$B\to\pi K$ are already available. Even though they are not yet very
accurate, the progress expected in the coming years in measuring the 
relevant observables at the $B$ factories allows us to expect that 
these non-leptonic decays will be known at an acceptable precision well 
before $K\to\pi\nu\bar\nu$ decays -- in particular 
$K_{\rm L}\to\pi^0\nu\bar\nu$ -- and clean tree-level strategies 
for $\gamma$ will become useful.

In view of this situation, it is necessary to concentrate first on 
the usual UT fits for $\gamma$,
$B\to\pi\pi$ and $B\to\pi K$ decays, determine $\gamma$, $q$, $\phi$ and
hadronic parameters by using appropriate observables and 
$SU(3)$ flavour-symmetry and plausible dynamical
assumptions, and subsequently make predictions for rare decays. The three-step
procedure for achieving this goal was presented in Section 1 and
executed in the rest of the paper.

Our analysis has been summarized briefly in \cite{BFRS-II} and presented 
in detail in Sections~\ref{sec:Bpipi}--\ref{sec:B-prom} of the present work.
A list of results can be found at the end of each section. Here we recall 
the most important findings of our ``anatomy'':
\begin{itemize}
\item
Using the available data for $B\to\pi\pi$ decays and neglecting small EW
penguin contributions to the relevant decay amplitudes, we were able to
determine four hadronic parameters using only the isospin symmetry of
strong interactions and the information for the angle $\gamma$ from the UT
fits. The result is given in (\ref{d-det}) and (\ref{x-det}). This 
determination is essentially free from
theoretical uncertainties and the large errors in (\ref{d-det})
and (\ref{x-det}), which reflect the large experimental uncertainties in 
$R_{+-}^{\pi\pi}$, $R_{00}^{\pi\pi}$, 
${\cal A}_{\rm CP}^{\rm dir}(B_d\to \pi^+\pi^-)$ and 
${\cal A}_{\rm CP}^{\rm mix}(B_d\to \pi^+\pi^-)$, should be significantly
reduced in the coming years. Contour plots in Figs.~\ref{fig:theta-d}   
and \ref{fig:Delta-x} will allow a monitoring of these improvements.
\item
Having the determined hadronic parameters of the $B\to\pi\pi$ system at hand,
we may predict the CP-violating observables of the $B_d\to\pi^0\pi^0$ 
channel as given in (\ref{detpipi-dir}) and (\ref{detpipi-mix}), with
the interesting perspective of having large direct and mixing-induced
CP violation in this decay. Moreover, if at least one of these observables 
is measured, $\gamma$ can be determined in a theoretically clean way.   
\item If we make the plausible assumption that penguin annihilation
and exchange topologies play a negligible r\^ole and employ the
$SU(3)$ flavour symmetry of strong interactions, we may complement the
CP-violating observables of $B_d\to\pi^+\pi^+$ in a variety of ways
with the data provided by the $B_d\to\pi^\mp K^\pm$ modes, which 
are only insignificantly affected by EW penguin contributions. 
Following these lines, we may extract $\gamma$, as given in (\ref{gam-H}). 
This determination can be refined with the help of other $B\to \pi K$ 
observables, with the result in (\ref{gam-fin}), which is in very good 
agreement with the UT fits, as can be seen in Fig.~\ref{fig:ut-compare}.
The remarkably consistent overall picture of our analysis also indicates 
that non-factorizable $SU(3)$-breaking effects are moderate and that 
our other dynamical assumptions are justified. 
\item
Using the hadronic $B\to\pi\pi$ parameters determined in the first step 
of our strategy, we may fix their $B\to\pi K$ counterparts with the help of
the $SU(3)$ flavour symmetry and plausible dynamical assumptions (see
previous item), allowing
us to analyse the $B\to\pi K$ system. Interestingly, we find that the 
SM cannot properly describe those $B\to \pi K$ observables that are 
sensitive to EW penguin contributions. This is in particular the case 
for $R_{\rm c}$ and $R_{\rm n}$, for which we obtain $R_{\rm c}\sim 1.14$ 
and $R_{\rm n}\sim 1.11$ in the SM, whereas experiments give 
$R_{\rm c}\sim 1.17$ and $R_{\rm n}\sim0.76$. On the other hand, the 
pattern of the $B$-factory data for those $B\to\pi K$ observables that 
are only insignificantly affected by EW penguins does not show any 
anomalous behaviour, i.e.\ is in accordance with the SM picture. 
\item
We have demonstrated that all features of 
the present $B\to\pi K$ data can be described within a NP scenario,
where EW penguin topologies are moderatly enhanced and carry
a large CP-violating NP phase $\phi$ around $-90^\circ$, thereby requiring
new sources for CP violation that lie beyond the KM mechanism. 
Of particular interest are our predictions for 
${\cal A}_{\rm CP}^{\rm dir}(B_d\!\to\!\pi^0 K_{\rm S})$ and 
${\cal A}_{\rm CP}^{\rm mix}(B_d\!\to\!\pi^0 K_{\rm S})$, and the suggestion 
to use these observables in the future for the determination of the 
parameters of the EW penguin sector. Moreover, also studies of 
$B_s\to K^+K^-$ and $B_s\to \pi^\pm K^\mp$ decays, which are very 
accessible at LHCb, complement our strategy nicely. 
We made also predictions for the corresponding $B_s$-decay observables.
\item Restricting ourselves to a more specific NP scenario with
enhanced $Z^0$ penguins, we may explore the implications of our
$B\to\pi K$ analysis for rare $K$ and $B$ decays, where in particular
the CP-violating NP phase $\phi$ has important consequences. In turn, the
currently available rare-decay data have already some impact on the 
allowed ranges for the $B\to\pi K$ observables as summarized in 
Table~\ref{tab:with-withoutconstraints}. A detailed analysis of
these effects has been presented in Section~\ref{sec:rare}. Possibly the most
interesting effects are found in the $K\to\pi\nu\bar\nu$ system, where the
enhancement of $\mbox{BR}({\klpn})$ by one order of magnitude and a strong
violation of the MFV relation $(\sin 2\beta)_{\pi\nu\bar\nu}=(\sin
2\beta)_{\psi K_{\rm S}}$ are very spectacular.
Similarly to the asymmetries
${\cal A}_{\rm CP}^{\rm dir}(B_d\!\to\!\pi^0 K_{\rm S})$ and 
${\cal A}_{\rm CP}^{\rm mix}(B_d\!\to\!\pi^0 K_{\rm S})$, the branching
ratios for $\kpn$ and $\klpn$ will allow a useful determination of the 
EW penguin parameters and the comparison of these two very different 
determinations will be a very important test of the NP scenario considered 
here.
\item
Finally, as discussed in Section~\ref{sec:B-prom}, we expect a drastic 
modification of the Belle result 
$(\sin 2\beta)_{\psi K_{\rm S}}\gg (\sin 2\beta)_{\phi K_{\rm S}}$,
in that we find
$(\sin 2\beta)_{\phi K_{\rm S}}\gsim (\sin 2\beta)_{\psi K_{\rm S}}$.
Consequently, within the NP scenario considered here,
we find $(\sin 2\beta)_{\phi K_{\rm S}}$ to be of the same magnitude as the 
central value found by the Belle collaboration but of {\it opposite} 
sign!
\end{itemize}  

As we have seen in our analysis, studies of  $B\to\pi\pi$ and $B\to\pi K$ 
decays are not only interesting in the context of the exploration of
CP violation and the search for NP, but also to obtain valuable insights
into hadron physics. Consequently, improved measurements of the 
corresponding observables are also very important in order to see whether 
the theoretical approaches like QCDF \cite{BBNS1}, PQCD \cite{PQCD} and 
SCET \cite{SCET}, in addition to their interesting theoretical structures, 
are also phenomenologically useful. Independently of the outcome of these 
measurements, the phenomenological strategy presented here will be very 
useful in correlating the experimental results for $B\to\pi\pi$ and 
$B\to\pi K$ with those for rare $K$ and $B$ decays, $B_d\to\phi K_{\rm S}$ 
and $\epe$.

Assuming that future, more accurate $B\to\pi\pi,\pi K$ data 
will not modify significantly the currently observed pattern in
these decays, the scenario of enhanced $Z^0$ penguins with a large
NP phase will remain an attractive possibility. While the 
enhancement of $\mbox{BR}(\klpn)$ by one order of magnitude would be very 
welcome to our experimental colleagues and $(\sin2\beta)_{\pi\nu\bar\nu}<0$
would be a very spectacular signal of NP, even more moderate departures 
of this sort from the SM and the MFV expectations could be easily 
identified in the very clean $K\to\pi\nu\bar\nu$ decays as clear signals 
of NP.

\vspace*{0.5truecm}

\noindent
{\bf Acknowledgements}\\
\noindent
The work presented here was supported in part by the German 
Bundesministerium f\"ur
Bildung und Forschung under the contract 05HT1WOA3 and the 
DFG Project Bu.\ 706/1-2.

\newpage

\begin{appendix}
\section{Compendium}\label{sec:comp}
\setcounter{equation}{0}
\boldmath
\subsection{The $B\to\pi\pi$ System}
\unboldmath

\begin{equation}\label{Rpm-expr-C}
R_{+-}^{\pi\pi}=\frac{1+2x\cos\Delta+x^2}{1-2d\cos\theta\cos\gamma+d^2}
\end{equation}
\begin{equation}\label{R00-expr-C}
R_{00}^{\pi\pi}=\frac{d^2+2dx\cos(\Delta-\theta)\cos\gamma+
x^2}{1-2d\cos\theta\cos\gamma+d^2}
\end{equation}
\begin{equation}\label{Adir-Bpipi-C}
{\cal A}_{\rm CP}^{\rm dir}(B_d\to \pi^+\pi^-)
=-\left[\frac{2d\sin\theta\sin\gamma}{1-
2d\cos\theta\cos\gamma+d^2}\right]
\end{equation}
\begin{equation}\label{Amix-Bpipi-C}
{\cal A}_{\rm CP}^{\rm mix}(B_d\to \pi^+\pi^-)
=\frac{\sin(\phi_d+2\gamma)-2d\cos\theta
\sin(\phi_d+\gamma)+d^2\sin\phi_d}{1-2d\cos\theta\cos\gamma+d^2}
\end{equation}
\begin{equation}
{\cal A}_{\rm CP}^{\rm dir}(B_d\to\pi^0\pi^0)=\frac{2dx\sin(\theta-\Delta)
\sin\gamma}{d^2+2dx\cos(\theta-\Delta)\cos\gamma+x^2}
\end{equation}
\begin{equation}
{\cal A}_{\rm CP}^{\rm mix}(B_d\to\pi^0\pi^0)=\frac{d^2\sin\phi_d+
2dx\cos(\theta-\Delta)\sin(\phi_d+\gamma)+x^2\sin(\phi_d+2\gamma)}{d^2+
2dx\cos(\theta-\Delta)\cos\gamma+x^2}.
\end{equation}

\boldmath
\subsection{The $B\to\pi K$ System}
\unboldmath

\begin{equation}\label{R-expr-C}
R=1-2r\cos\delta\cos\gamma+r^2
\end{equation}
\begin{eqnarray}\label{Rc-expr-C}
R_{\rm c}&=&1-2r_{\rm c}\cos\delta_{\rm c}\cos\gamma+r_{\rm c}^2\nonumber\\
&&+q r_{\rm c}\left[2\left\{\cos(\delta_{\rm c}+\omega)\cos\phi
-r_{\rm c}\cos\omega\cos(\gamma-\phi)\right\}+qr_{\rm c}\right]
\end{eqnarray}
\begin{equation}\label{Rn-expr-C}
R_{\rm n}=\frac{1}{b}\left[1-2r\cos\delta\cos\gamma+r^2\right]
\end{equation}
\begin{eqnarray}
b\equiv \frac{R}{R_{\rm n}}&=&1-2qr_{\rm c}\cos(\delta_{\rm c}+\omega)\cos\phi
+q^2r_{\rm c}^2\nonumber\\
&&+2\rho_{\rm n}\left[\cos\theta_{\rm n}\cos\gamma-
qr_{\rm c}\cos(\theta_{\rm n}-\delta_{\rm c}-\omega)\cos(\gamma-\phi)
\right]+\rho_{\rm n}^2\label{b-expr-C}
\end{eqnarray}
\begin{equation}\label{ACPdir-C}
{\cal A}_{\rm CP}^{\rm dir}(B_d\to\pi^\mp K^\pm)=
\frac{2r\sin\delta\sin\gamma}{1-2r\cos\delta\cos\gamma+r^2}
\end{equation}
\begin{equation}
{\cal A}_{\rm CP}^{\rm dir}(B^\pm\to\pi^\pm K)
=-\left[\frac{2\rho_{\rm c}\sin\theta_{\rm c}\sin\gamma}{1+2\rho_{\rm c}
\cos\theta_{\rm c}\cos\gamma+\rho_{\rm c}^2}\right]\label{ACP-B+pi+K-C}
\end{equation}
\begin{equation}
{\cal A}_{\rm CP}^{\rm dir}(B^\pm\to\pi^0 K^\pm)=
\frac{2}{R_{\rm c}}\left[r_{\rm c}\sin\delta_{\rm c}\sin\gamma
-qr_{\rm c}\left\{\sin(\delta_{\rm c}+\omega)\sin\phi+r_{\rm c}
\sin\omega\sin(\gamma-\phi)\right\}\right]
\end{equation}
\begin{eqnarray}
\lefteqn{{\cal A}_{\rm CP}^{\rm dir}(B_d\to\pi^0 K_{\rm S})=
\frac{2}{b}\Bigl[qr_{\rm c}\sin(\delta_{\rm c}+\omega)\sin\phi}\nonumber\\
&&-\rho_{\rm n}\left\{\sin\theta_{\rm n}\sin\gamma-qr_{\rm c}
\sin(\theta_{\rm n}-\delta_{\rm c}-\omega)
\sin(\gamma-\phi)\right\}\Bigr]
\end{eqnarray}
\begin{eqnarray}
\lefteqn{{\cal A}_{\rm CP}^{\rm mix}(B_d\to\pi^0 K_{\rm S})=
-\frac{1}{b}\Bigl[\sin\phi_d-2qr_{\rm c}\cos(\delta_{\rm c}+\omega)
\sin(\phi_d+\phi)+q^2r_{\rm c}^2\sin(\phi_d+2\phi)}\quad\\
&&+2\rho_{\rm n}\left\{\cos\theta_{\rm n}\sin(\phi_d+\gamma)-
qr_{\rm c}\cos(\theta_{\rm n}-\delta_{\rm c}-\omega)
\sin(\phi_d+\gamma+\phi)\right\}+\rho_{\rm n}^2\sin(\phi_d+2\gamma)\Bigr].
\nonumber
\end{eqnarray}

\boldmath
\subsection{The $B_s\to K^+K^-$, $B_s\to \pi^\pm K^\mp$ System}
\unboldmath
Using the $U$-spin flavour symmetry of strong interactions, we
may express the corresponding CP-violating observables in terms of 
the hadronic $B_d\to\pi^+\pi^-$ and $B_d\to\pi^\mp K^\pm$ 
parameters as follows:

\begin{equation}\label{ACPdir-BsKK-C}
{\cal A}_{\rm CP}^{\rm dir}(B_s\to K^+K^-)=\frac{2\epsilon d
\sin\theta\sin\gamma}{\epsilon^2+2\epsilon d\cos\theta\cos\gamma+d^2}
\end{equation}
\begin{equation}\label{ACPmix-BsKK-C}
{\cal A}_{\rm CP}^{\rm mix}(B_s\to K^+K^-)=
\frac{\epsilon^2\sin(\phi_s+2\gamma)+2\epsilon d\cos\theta\sin(\phi_s+\gamma)
+d^2\sin\phi_s}{\epsilon^2+2\epsilon d\cos\theta\cos\gamma+d^2}
\end{equation}
\begin{equation}
{\cal A}_{\rm CP}^{\rm dir}(B_s\to \pi^\pm K^\mp)=
-\left[\frac{2\epsilon r\sin\delta\sin\gamma}{\epsilon^2+2\epsilon r
\cos\delta\cos\gamma+r^2}\right].
\end{equation}

\section{Comment on the New Belle \boldmath$B_d\to\pi^+\pi^-$ \unboldmath
Results}\label{app:belle-new}
\setcounter{equation}{0}
During the final stages of this work, the Belle collaboration 
announced the following update of the results for the 
CP-violating observables of the $B_d\to\pi^+\pi^-$ channel 
\cite{Belle-new}:
\begin{eqnarray}
{\cal A}_{\rm CP}^{\rm dir}(B_d\to\pi^+\pi^-)&=&
-0.58\pm0.15\pm0.07\label{Bpipi-dir-new}\\ 
{\cal A}_{\rm CP}^{\rm mix}(B_d\to\pi^+\pi^-)&=&
+1.00\pm0.21\pm0.07.\label{Bpipi-mix-new}
\end{eqnarray}
Using these new data, the averages in (\ref{Bpipi-CP-averages-dir}) and
(\ref{Bpipi-CP-averages-mix}) change to
\begin{eqnarray}
{\cal A}_{\rm CP}^{\rm dir}(B_d\to\pi^+\pi^-)&=&
-0.42\pm0.13\label{Bpipi-CP-averages-dir-new}\\ 
{\cal A}_{\rm CP}^{\rm mix}(B_d\to\pi^+\pi^-)&=&
+0.70\pm0.19.\label{Bpipi-CP-averages-mix-new}
\end{eqnarray}
Whereas the global picture of the analysis presented in this paper 
is not affected by these new numbers, some numerical results change. 
In particular, instead of (\ref{gam-H}), the averages in 
(\ref{Bpipi-CP-averages-dir-new}) and (\ref{Bpipi-CP-averages-mix-new})
would correspond to the following smaller values of $\gamma$:
\begin{equation}\label{gam-H-new}
\gamma=(38.6^{+6.1}_{-7.2})^\circ \, \lor \, (55.6^{+7.0}_{-8.1})^\circ,
\end{equation}
which would also be in accordance with UT fits, although a bit on
the lower side. The contour corresponding to the new values was added to 
Fig.~\ref{fig:ut-compare}, and the central values in (\ref{alpha-beta})
change to $\alpha=103.3^\circ$ and $\beta=21.1^\circ$.

\section{Error Treatment}\label{app:error-treatment}
\setcounter{equation}{0}
Unless otherwise stated (e.g.\ in the determination of $\gamma$ in
Subsection~\ref{ssec:Bpipi-gam-det}), we treat the errors in the 
following way: all predicted quantities depend on the hadronic parameters
$d$, $\theta$, $x$ and $\Delta$, and on some other input parameters.
To take into account the fact that dependences on the hadronic
parameters can cancel out, we do not vary the hadronic parameters
inside their error bands, but rather vary the input parameters that 
are used to obtain them, i.e.\ ${\cal A}_{\rm CP}^{\rm dir}(B_d\to\pi^+\pi^-)$,
${\cal A}_{\rm CP}^{\rm mix}(B_d\to\pi^+\pi^-)$, $R_{+-}^{\pi\pi}$ and 
$R_{00}^{\pi\pi}$. Other input parameters that contribute to the errors 
are $\gamma$ and $\phi_d$, as well as $A$ and additional hadronic parameters 
appearing in the context of Section~\ref{sec:rare} (some quantities  
involve also $R_b$). The individual errors associated
with the uncertainty of a specific input parameter are found by varying 
the corresponding parameter within its 1$\sigma$ band, while keeping 
the other parameters fixed at their central values. These errors are then 
added up in quadrature to obtain the total error for each observable.

\section{\boldmath Colour-Suppressed EW Penguins in $B\to\pi K$ \unboldmath}
\label{APP:BpiK-CSEWP}
\setcounter{equation}{0}
\subsection{General Structure}
In order to discuss the EW penguin contributions to the decays 
$B^+\to\pi^+K^0$ and $B_d^0\to\pi^-K^+$, which are
usually referred to as ``colour-suppressed'', we have to look at
the EW penguin operators, exhibiting the following generic flavour
structure:
\begin{equation}
{\cal Q}_{\rm EW}^{\rm pen}\sim\frac{1}{2}\left[2(\bar c c)-(\bar s s)
-(\bar b b)+\left\{2(\bar u u)-(\bar dd)\right\}\right](\bar b s).
\end{equation}
Using then the well-known isospin decomposition
\begin{equation}
(\bar u u) = \frac{1}{2}\underbrace{\left[(\bar u u)+(\bar dd)\right]}_{I=0} 
+ \frac{1}{2}\underbrace{\left[(\bar u u)-(\bar dd)\right]}_{I=1},
\end{equation}
we may decompose the EW penguin operators into isospin singlet and
triplet pieces as follows:
\begin{equation}
({\cal Q}_{\rm EW}^{\rm pen})_{I=0}
\sim \frac{1}{2}\left[2(\bar c c)-(\bar s s)-(\bar b b)+
\frac{1}{2}\left\{(\bar u u)+(\bar dd)\right\}\right](\bar b s)
\end{equation}
\begin{equation}
({\cal Q}_{\rm EW}^{\rm pen})_{I=1}
\sim \frac{3}{4}\left[(\bar u u)-(\bar dd)\right](\bar b s).
\end{equation}
If we then apply the isospin flavour symmetry of strong interactions,
as discussed in detail in \cite{BFM}, we obtain\footnote{For simplicity, 
we suppress in the following discussion
the primes introduced in Section~\ref{sec:BpiK} 
to distinguish the $B\to\pi K$ amplitudes from 
their $B\to\pi\pi$ counterparts.}
\begin{eqnarray}
A(B^+\to\pi^+K^0)_{\rm EW}&=&+\frac{1}{2}\left[P_{\rm EW}^{{\rm C}(0)}-
P_{\rm EW}^{{\rm C}(1)}\right]\equiv -\frac{1}{3} P_{\rm EW}^{\rm C-}\\
A(B^0_d\to\pi^-K^+)_{\rm EW}&=&-\frac{1}{2}\left[P_{\rm EW}^{{\rm C}(0)}+
P_{\rm EW}^{{\rm C}(1)}\right]\equiv -\frac{2}{3}P_{\rm EW}^{\rm C+},
\end{eqnarray}
with
\begin{equation}\label{PEW-0}
P_{\rm EW}^{{\rm C}(0)}=-\frac{G_{\rm F}}{\sqrt{2}}\lambda^2A\, 
\frac{1}{2} C \langle K^+\pi^-| \left[4(\bar c c)-2(\bar s s)-2(\bar b b)+
\left\{(\bar u u)+(\bar dd)\right\}\right](\bar b s)|B^0_d\rangle
\end{equation}
\begin{equation}\label{PEW-1}
P_{\rm EW}^{{\rm C}(1)}=-\frac{G_{\rm F}}{\sqrt{2}}\lambda^2A\,
\frac{3}{2} C \langle K^+\pi^-| 
\left[(\bar u u)-(\bar dd)\right](\bar b s) |B^0_d\rangle,
\end{equation}
summarizing the isospin singlet and triplet pieces, respectively.
Here the combination of the generic Wilson coefficient $C$ with
the four-quark operators denotes symbolically the sum over the relevant
EW penguin operators. We observe that
\begin{equation}
A(B^+\to\pi^+K^0)+A(B^0_d\to\pi^-K^+)=-P_{\rm EW}^{{\rm C}(1)},
\end{equation}
where the expression for $P_{\rm EW}^{{\rm C}(1)}$ in (\ref{PEW-1})
agrees with the one for the colour-suppressed EW penguin amplitude
introduced in \cite{defan}. 

It is instructive to consider the tree-diagram-like matrix elements 
entering (\ref{PEW-0}) and (\ref{PEW-1}), which read as follows:
\begin{equation}\label{PEW-0-T}
\left[P_{\rm EW}^{{\rm C}(0)}\right]_{\rm T}
=-\frac{G_{\rm F}}{\sqrt{2}}\lambda^2A\, 
\frac{1}{2} C \langle K^+\pi^-|  (\bar u u)(\bar b s)|B^0_d\rangle_{\rm T}
\end{equation}
\begin{equation}\label{PEW-1-T}
\left[P_{\rm EW}^{{\rm C}(1)}\right]_{\rm T}
=-\frac{G_{\rm F}}{\sqrt{2}}\lambda^2A\,
\frac{3}{2} C \langle K^+\pi^-| (\bar u u)(\bar b s) |B^0_d\rangle_{\rm T},
\end{equation}
and imply
\begin{equation}\label{EWP-T-Rel}
\left[ P_{\rm EW}^{\rm C-}\right]_{\rm T}
=\Bigl[P_{\rm EW}^{\rm C+}\Bigr]_{\rm T}
=\left[P_{\rm EW}^{{\rm C}(1)}\right]_{\rm T}.
\end{equation}
The EW penguin operator $Q_9$ with the largest Wilson coefficient 
has the colour structure $(\bar u_\alpha u_\alpha)(\bar b_\beta s_\beta)$
and its matrix elements (\ref{PEW-0-T}) and (\ref{PEW-1-T}) are 
consequently colour-suppressed. On the other hand,
$Q_{10}\sim(\bar u_\alpha u_\beta)(\bar b_\beta s_\alpha)$ has
a significantly smaller Wilson coefficient. The coefficients of the 
$(V-A)\otimes(V+A)$ EW penguin operators $Q_7$ and 
$Q_8$ are even further suppressed, so that these operators can be neglected.
Consequently, the hadronic matrix elements in (\ref{PEW-0-T}) and 
(\ref{PEW-1-T}) are colour-suppressed, and (\ref{EWP-T-Rel}) represents
the picture of the colour-suppressed EW penguins that is usually adopted 
in the literature. 

In our analysis of the $B\to\pi\pi$ system in Section~\ref{sec:Bpipi},
we have seen that penguin topologies with internal up- and charm-quark 
exchanges, which correspond to matrix elements with penguin-like
contractions of $(\bar u u)(\bar b s)$ and $(\bar c c)(\bar b s)$ 
operators with $u$ and $c$ quarks running in the loops \cite{BFM}, 
play an important r\^ole. Consequently, 
the standard picture of colour-suppressed EW penguins in $B\to\pi K$ 
could also be affected through these penguin topologies. 
Moreover, 
the parameters
\begin{equation}\label{aC-BpiK-tilde}
\tilde a_{\rm C}e^{i\tilde \Delta_{\rm C}}\equiv
\frac{P_{\rm EW}^{\rm C-}}{P_{\rm EW}}=
\frac{P_{\rm EW}^{\rm C-}}{P_{\rm EW}^{\rm C-}+
\tilde P_{\rm EW}^{\rm A}}
\end{equation}
\begin{equation}\label{aC-BpiK}
a_{\rm C}e^{i\Delta_{\rm C}}\equiv
\frac{P_{\rm EW}^{\rm C+}}{P_{\rm EW}}=
\frac{P_{\rm EW}^{\rm C+}}{P_{\rm EW}^{\rm C+}+
P_{\rm EW}^{\rm A}},
\end{equation}
where $\tilde P_{\rm EW}^{\rm A}$ and $P_{\rm EW}^{\rm A}$ are the 
colour-allowed EW penguin amplitudes contributing to  
$\sqrt{2}A(B^+\to\pi^0K^+)$ and $\sqrt{2}A(B^0_d\to\pi^0K^0)$, respectively,
may not be as small as na\"\i vely expected. In (\ref{aC-BpiK-tilde})
and (\ref{aC-BpiK}), we have used that the isospin flavour symmetry implies
\begin{equation}\label{EWP-iso}
P_{\rm EW}= P_{\rm EW}^{\rm C-}+\tilde P_{\rm EW}^{\rm A}=
P_{\rm EW}^{\rm C+}+P_{\rm EW}^{\rm A},
\end{equation}
where $P_{\rm EW}$ enters the EW penguin parameter $q e^{i\phi}e^{i\omega}$. 
As in Subsection~\ref{ssec:EWP-REL}, we define the weak phases of  
$ P_{\rm EW}^{\rm C-}$ and $P_{\rm EW}^{\rm C+}$ as the one 
associated with the combination $C_9+C_{10}$ of Wilson coefficients,
which can be done approximately. 

In the remainder of this appendix, we discuss in detail the possible impact 
of the parameters in (\ref{aC-BpiK-tilde}) and (\ref{aC-BpiK}) on the 
$B\to\pi K$ analysis performed in Section~\ref{sec:BpiK}, and propose 
strategies to search for indications of their effects in the 
corresponding data.

\boldmath
\subsection{Generalization of the Decay Amplitudes}
\unboldmath
The generalization of the decay amplitudes in 
(\ref{B+pi+K0})--(\ref{B0pi0K0}), taking the colour-suppressed EW 
penguin topologies into account, is given by
\begin{eqnarray}
A(B^+\to\pi^+K^0)&=&-P'\left[1+\rho_{\rm c}e^{i\theta_{\rm c}}e^{i\gamma}
-\frac{1}{3}\tilde a_{\rm EW}^{\rm C}e^{i\tilde \psi_{\rm C}}e^{i\phi}
\right]\label{B+pi+K0-CS}\\
\sqrt{2}A(B^+\to\pi^0K^+)&=&
P'\left[1+\rho_{\rm c}e^{i\theta_{\rm c}}e^{i\gamma}
-\left\{e^{i\gamma}-\left(1-\frac{1}{3}\tilde a_{\rm C}
e^{i\tilde \Delta_{\rm C}}\right)
qe^{i\phi}e^{i\omega}\right\}r_{\rm c}e^{i\delta_{\rm c}}
\right]\qquad\mbox{}\label{B+pi0K+-CS}\\
A(B^0_d\to\pi^-K^+)&=&P'\left[1+\frac{2}{3}a_{\rm EW}^{\rm C}
e^{i\psi_{\rm C}}e^{i\phi}-re^{i\delta}e^{i\gamma}\right]\label{B0pi-K+-CS}\\
\sqrt{2}A(B^0_d\to\pi^0K^0)&=&-P'\left[1+\rho_{\rm n}e^{i\theta_{\rm n}}
e^{i\gamma}-\left(1-\frac{2}{3}a_{\rm C}e^{i\Delta_{\rm C}}\right)
qe^{i\phi}e^{i\omega}r_{\rm c}e^{i\delta_{\rm c}}\right],\label{B0pi0K0-CS}
\end{eqnarray}
with
\begin{eqnarray}
\tilde a_{\rm EW}^{\rm C}e^{i\tilde \psi_{\rm C}}&\equiv&
qe^{i\omega}r_{\rm c}e^{i\delta_{\rm c}} \tilde a_{\rm C}
e^{i\tilde\Delta_{\rm C}}\\
a_{\rm EW}^{\rm C}e^{i\psi_{\rm C}}&\equiv&
qe^{i\omega}r_{\rm c}e^{i\delta_{\rm c}} a_{\rm C}e^{i\Delta_{\rm C}}.
\end{eqnarray}
Because of the isospin relation in (\ref{EWP-iso}), the terms proportional 
to $\tilde a_{\rm C}$ and $a_{\rm C}$ have to cancel in the sums of the 
$A(B^+\to\pi^+K^0)$, $\sqrt{2}A(B^+\to\pi^0K^+)$
and $A(B^0_d\to\pi^-K^+)$, $\sqrt{2}A(B^0_d\to\pi^0K^0)$ amplitudes, 
respectively.

\boldmath
\subsection{The Case of $\phi=0^\circ$}
\unboldmath
Let us first consider $\phi=0^\circ$, which applies also to the case of
the SM. Although the analysis of the current $B$-factory data performed 
in Section~\ref{sec:BpiK} favours a large phase of 
$\phi\sim-90^\circ$, it is interesting and 
instructive to have a detailed look at the situation arising for
$\phi=0^\circ$. In this case, we may straightforwardly absorb 
the terms in (\ref{B0pi-K+-CS}) and (\ref{B0pi0K0-CS}) proportional to
$a_{\rm C}e^{i\Delta_{\rm C}}$ in the amplitude $P'$, yielding
\begin{eqnarray}
A(B^0_d\to\pi^-K^+)&=&P'\left[1-re^{i\delta}e^{i\gamma}
\right]\label{B0pi-K+-CS2}\\
\sqrt{2}A(B^0_d\to\pi^0K^0)&=&-P'\left[1+\rho_{\rm n}e^{i\theta_{\rm n}}
e^{i\gamma}-qe^{i\omega}r_{\rm c}e^{i\delta_{\rm c}}
\right].\label{B0pi0K0-CS2} 
\end{eqnarray}
As far as (\ref{B+pi+K0-CS}) and (\ref{B+pi0K+-CS}) are concerned, we 
obtain amplitudes of the structure
\begin{eqnarray}
A(B^+\to\pi^+K^0)&=&-P'\left[1+\rho_{\rm c}e^{i\theta_{\rm c}}e^{i\gamma}
-a_{\rm EW}^{{\rm C}(1)}e^{i\psi_{\rm C}^{(1)}}\right]\label{B+pi+K0-CS2}\\
\sqrt{2}A(B^+\to\pi^0K^+)&=&P'\left[1+\rho_{\rm c}
e^{i\theta_{\rm c}}e^{i\gamma}-\left\{e^{i\gamma}-
\left(1-a_{\rm C}^{(1)}e^{i\Delta_{\rm C}^{(1)}}\right)qe^{i\omega}\right\}
r_{\rm c}e^{i\delta_{\rm c}}\right],\qquad\mbox{}\label{B+pi0K+-CS2}
\end{eqnarray}
with
\begin{eqnarray}
a_{\rm C}^{(1)}e^{i\Delta_{\rm C}^{(1)}}&\equiv&
\frac{P_{\rm EW}^{{\rm C}(1)}}{P_{\rm EW}}\label{aC1-def}\\
a_{\rm EW}^{{\rm C}(1)}e^{i\psi_{\rm C}^{(1)}}&\equiv&
qe^{i\omega}r_{\rm c}e^{i\delta_{\rm c}} 
a_{\rm C}^{(1)}e^{i\Delta_{\rm C}^{(1)}},
\end{eqnarray}
where $P_{\rm EW}^{{\rm C}(1)}$ is given
in (\ref{PEW-1}). It should be noted that $P'$, $re^{i\delta}$, 
$\rho_{\rm n}e^{i\theta_{\rm n}}$, $r_{\rm c}e^{i\delta_{\rm c}}$
and $\rho_{\rm c}e^{i\theta_{\rm c}}$ now contain also contributions from
colour-suppressed EW penguins. The expressions in (\ref{B0pi-K+-CS2})
and (\ref{B0pi0K0-CS2}) are the counterparts of (\ref{Bdpi+pi-EWP}) and 
(\ref{Bdpi0pi0EWP}), respectively. 

We observe that the colour-suppressed EW penguins do {\it not} explicitly 
affect the analysis of the $B\to\pi\pi$ modes and the {\it neutral} 
$B\to\pi K$ decays. Because of this feature, $B^0_d\to\pi^-K^+$ is 
actually the ``natural'' partner of $B^0_d\to\pi^+\pi^-$ to deal with 
the famous penguin problem and not the $B^+\to\pi^+K^0$ channel, as is
sometimes done in the literature. If we use the $B\to\pi\pi$ data as 
described above to fix the hadronic $B\to\pi K$ parameters, we may
extract $q$ and $\omega$ from $R_{\rm n}$ and one of the CP-violating 
$B_d\to\pi^0K_{\rm S}$ observables; the remaining CP asymmetry provides
an important consistency check of the $\phi=0^\circ$ scenario, 
in particular to see whether this weak phase actually vanishes. Interestingly, 
enhanced colour-suppressed EW penguins would yield a difference between 
the values of $r_{\rm c}$ following from (\ref{rc-simple-expr}) and 
(\ref{rc-alt-det}), which is {\it not} indicated by the data. 

If we fix the hadronic $B\to\pi K$ parameters through
the $B\to \pi\pi$ system, the analysis of the {\it charged} 
$B\to\pi K$ modes may be affected both by the colour-suppressed EW 
penguins and by the $\rho_{\rm c}$ parameter. Following the avenue 
described above, we may predict the corresponding observables, and 
may check whether we obtain agreement with the experimental picture. 
Alternatively, we may use the value of $r_{\rm c}$ following from
(\ref{rc-alt-det}), and may shift the terms in (\ref{B+pi+K0-CS2}) and 
(\ref{B+pi0K+-CS2}) proportional to $a_{\rm C}^{(1)}e^{i\Delta_{\rm C}^{(1)}}$ 
into $P'$. We may then also analyse the observables of the charged 
$B\to\pi K$ system in a manner that is not affected by the 
colour-suppressed EW penguins, whereas $\rho_{\rm c}$ may still enter.

\boldmath
\subsection{The Case of $\phi\not=0^\circ$}
\unboldmath
In the case of $\phi\not=0^\circ$, we could of course still absorb
colour-suppressed EW penguin contributions in the amplitude $P'$. However, 
as these terms now contain a CP-violating weak phase, we have no longer 
the simple amplitude structure of $P'=|P'|e^{i\delta_{P'}}$, where
$\delta_{P'}$ is a CP-conserving strong phase. 

If we look at (\ref{B+pi+K0-CS}) and (\ref{B0pi-K+-CS}), we see
that the observable $R$ and the direct CP asymmetries of the 
$B^\pm\to\pi^\pm K$ and $B_d\to\pi^\mp K^\pm$ modes would be 
significantly affected by the terms proportional to $\tilde a_{\rm C}
e^{i\tilde\Delta_{\rm C}}$ and $a_{\rm C}e^{i\Delta_{\rm C}}$ should
these coefficients not be small quantities, i.e.\ should the 
colour suppression of the EW penguins not be effective. However,
as discussed in Subsection~\ref{ssec:BpiK-mixed}, these observables
do not indicate any anomalous behaviour. Moreover, as we have already
noted above, also the agreement between the values of $r_{\rm c}$
in (\ref{r-c-det-Bpipi}) and (\ref{rc-alt-det}) does not favour
an enhancement of the colour-suppressed EW penguins. In
this context, it is also important to note that the $B\to\pi K$ data 
do not require a dramatic enhancement of the parameter $q$. Moreover,
the analysis of the rare decays in Section~\ref{sec:rare} favours 
$q$ to be even smaller than the central value given in (\ref{q-det}). 

In order to deal further with the colour-suppressed EW penguins
in the $\phi\not=0^\circ$ case, it is useful to introduce
\begin{equation}\label{q-eff}
\langle q\rangle e^{i\langle\omega\rangle}\equiv 
\left[1-\frac{1}{2}a_{\rm C}^{(1)}e^{i\Delta_{\rm C}^{(1)}}\right]qe^{i\omega},
\end{equation}
so that
\begin{eqnarray}
\left[1-\frac{1}{3}\tilde a_{\rm C}e^{i\tilde\Delta_{\rm C}}\right]
qe^{i\omega}&=&\langle q\rangle e^{i\langle\omega\rangle}+
\frac{1}{2}a_{\rm C}^{(0)}e^{i\Delta_{\rm C}^{(0)}}qe^{i\omega}\\
\left[1-\frac{2}{3}a_{\rm C}e^{i\Delta_{\rm C}}\right]qe^{i\omega}
&=&\langle q\rangle e^{i\langle\omega\rangle}-
\frac{1}{2}a_{\rm C}^{(0)}e^{i\Delta_{\rm C}^{(0)}}qe^{i\omega},
\end{eqnarray}
with
\begin{equation}
a_{\rm C}^{(0)}e^{i\Delta_{\rm C}^{(0)}}\equiv
\frac{P_{\rm EW}^{{\rm C}(0)}}{P_{\rm EW}},
\end{equation}
in analogy to (\ref{aC1-def}). Considering only tree-diagram-like
matrix elements, as in (\ref{PEW-0-T}) and (\ref{PEW-1-T}), yields
\begin{equation}
\left[\frac{a_{\rm C}^{(0)}}{a_{\rm C}^{(1)}}\right]_{\rm T}=
\frac{1}{3}.
\end{equation}
If we assume that also $a_{\rm C}^{(0)}$ is suppressed with respect to 
$a_{\rm C}^{(1)}$, we may identify the terms proportional to $qe^{i\omega}$ 
in (\ref{B+pi0K+-CS}) and (\ref{B0pi0K0-CS}) simply with the ``effective'' 
EW penguin parameter $\langle q\rangle e^{i\langle\omega\rangle}$. 
Interestingly, the strong phase $\Delta_{\rm C}^{(1)}$ may induce a sizeable
value of $\langle\omega\rangle$, although $\omega$ could still be tiny. 
If we complement then 
\begin{equation}
\left[\frac{\mbox{BR}(B^+\to\pi^0 K^+)+
\mbox{BR}(\bar B^-\to\pi^0 K^-)}{\mbox{BR}(B^0_d\to\pi^0 K^0)+
\mbox{BR}(\bar B^0_d\to\pi^0 \bar K^0)}\right]\frac{\tau_{B^0_d}}{\tau_{B^+}}
\equiv \frac{R_{\rm c}}{b}=0.99\pm0.15
\end{equation}
with two of the three CP-violating observables provided by the 
$B^\pm\to\pi^0 K^\pm$, $B_d\to\pi^0 K_{\rm S}$ modes, we may
determine $\langle q\rangle$, $\langle\omega\rangle$ and $\phi$;
the remaining third CP asymmetry can be predicted and allows
a crucial consistency check.

To conclude, let us emphasize once more that our analysis 
of the current $B$-factory data for the $B\to\pi K$ modes 
performed in Section~\ref{sec:BpiK} points towards a consistent 
overall picture. In particular, $R$ and the direct CP asymmetries of the 
$B_d\to\pi^\mp K^\pm$, $B^\pm\to\pi^\pm K$ modes do not show any 
anomalous behaviour. Consequently, we have no experimental indications 
for an enhancement of the colour-suppressed EW penguins, which have
actually a very complicated internal structure.  In the 
future, the strategies discussed in this appendix will allow us to explore 
these contributions in a more stringent manner with the help of improved 
experimental data.

\end{appendix}
\end{document}